\documentclass[a4paper,11pt]{article}

\usepackage{geometry}
\usepackage[utf8]{inputenc}     
\usepackage{amsmath, amssymb}   
\usepackage{csquotes}           
\usepackage{eurosym}            
\usepackage{xcolor}             
\usepackage{bm}                 
\usepackage{hyperref}           
\usepackage[titletoc,page]{appendix}
\usepackage{authblk}            
\usepackage[bottom]{footmisc}   
\usepackage{float}
\usepackage{graphicx}           
\graphicspath{{./Figures/}}     
\usepackage{wrapfig}
\usepackage{subcaption}
\usepackage{booktabs}           
\usepackage{multirow}           
\usepackage{tikz}               
\usetikzlibrary{shapes,arrows,positioning}
\usetikzlibrary{positioning, arrows.meta}
\usepackage[backend=biber,style=numeric,sorting=nyt]{biblatex}
\title{\textbf{Learning the Exact SABR Model}}
\author[1]{Giorgia Rensi}
\author[1,2,*]{Pietro Rossi}
\author[1,3]{Marco Bianchetti}
\affil[1]{Department of Statistical Sciences \textquote{Paolo Fortunati}, University of Bologna, Italy}
\affil[2]{Prometeia S.p.a., Bologna, Italy}
\affil[3]{Market and Financial Risk Management, Intesa Sanpaolo, Milan, Italy}
\affil[*]{Corresponding author}
\date{\today}
\begin{document}
\maketitle

\begin{abstract}
The SABR model is a cornerstone of interest rate volatility modeling, but its practical application relies heavily on the analytical approximation by Hagan et al., whose accuracy deteriorates for high volatility, long maturities, and out-of-the-money options, admitting arbitrage. While machine learning approaches have been proposed to overcome these limitations, they have often been limited by simplified SABR dynamics or a lack of systematic validation against the full spectrum of market conditions.
\par 
We develop a novel SABR DNN, a specialized Artificial Deep Neural Network (DNN) architecture that learns the true SABR stochastic dynamics using an unprecedented large training dataset of interest rate Cap/Floor volatility surfaces (more than 200 million points), including very long maturities (30Y) and extreme strikes consistently with market quotations. Our dataset is obtained via high-precision unbiased Monte Carlo simulation of a scaled shifted-SABR stochastic dynamics, which allows dimensional reduction without any loss of generality.
\par
Our SABR DNN provides arbitrage-free calibration of real market volatility surfaces and Cap/Floor prices for any maturity and strike with negligible computational effort and without retraining across business dates. 
Our results fully address the gaps in the previous machine learning SABR literature in a systematic and self-consistent way, and can be extended to cover any interest rate European options in different rate tenors and currencies, thus establishing a comprehensive functional SABR framework that can be adopted for daily trading and risk management activities. 
\end{abstract}

\vfill 
\noindent \textbf{Classifications}: JEL: C45, C63, G12, G13; MSC: 91G60 (Primary) 60H35, 68T07, 65C05 (Secondary); ACM: G.1.6, I.2.6, I.5.1, J.4.\par 
\vspace{0.25cm} \noindent \textbf{Keywords}: stochastic volatility; SABR, Hagan, model calibration; volatility surface; interest rate derivatives; Cap; Floor; Monte Carlo simulation; machine learning; deep neural network.
\par 
\vspace{0.25cm} \noindent \textbf{Acknowledgements}: M.B. acknowledges fruitful discussions with many colleagues at international conferences and at Risk Management, Financial Engineering and Trading Desks of Intesa Sanpaolo.\par 
\vspace{0.25cm} \noindent \textbf{Disclaimer}: the views expressed here are those of the authors and do not represent the opinions of their employers. They are not responsible for any use that may be made of these contents.

\newpage
\tableofcontents
\newpage

\section{Introduction}
\label{sec:Intro}

\subsection{Machine Learning in Pricing Model Calibration}
\label{sec:MLinFinance}
The daily market risk management routine of banks dealing with large portfolios of financial instruments requires substantial computational resources to run a number of tasks: model calibration, instrument pricing and hedging, risk measurement, risk analyses, and reporting. 
All the pricing measures used in this context depend on market variables and must be recalculated once the market moves significantly, even in real-time in the most complex situations. 
Risk measures add another layer of complexity, since they are typically based on risk scenarios which require multiple model calibrations and portfolio valuations.
Hence, the quantitative finance community is continuously developing, on the one side, sophisticated models to properly manage complex portfolios and, on the other side, efficient algorithms to contain the required computational effort. The trade-off between these two competing objectives drives model selection, typically limiting the choice to less sophisticated models and approximate numerical techniques compatible with the available computational budget.
\par 
However, the repeated calculations described above are typically quite self-similar: most of the time large portfolios have a limited daily turnover and market data inputs change smoothly. Furthermore, the most relevant inputs are typically limited to a subset of market points (e.g. forward and volatility), trade parameters (e.g. maturity and strike), pricing model parameters (e.g. 1 for Black-Scholes, 4 for SABR, 5 for Heston, etc.) and risk model parameters (e.g. market risk historical series).
Clearly, this context offers an ideal setting for supervised machine learning techniques, where all the necessary input and output information can be encoded into a labeled dataset, which is used to train appropriate algorithms and learn the relevant pricing or risk function.
In particular, deep neural networks (DNNs) are designed to approximate unknown functions $y=f(x)$ that are available only through sample pairs of given input and output data $\{x,y\}$ using a mapping $\hat{y}=\hat{f}(x,w)$ where the parameters $w$ are calibrated to minimize some distance $\|\hat{y}-y\|$. This task is feasible thanks to the universal approximation theorem \cite{Cybenko1989}, \cite{Hornik1989NN}, also including function derivatives \cite{Hornik1990NN} (provided that the activation function is smooth), with a maximum of three node layers \cite{Eldan2016}. See the classic textbook \cite{Goodfellow2016} for more details about Deep (Artificial) Neural Networks.
\par 
These interesting properties of DNNs are particularly useful in the case of \emph{pricing model calibration}, which can be described as a \textit{inverse optimization problem}: given a set of $n_{mkt}$ quoted plain-vanilla financial instruments, each characterized by a set of \textit{contract parameters} $\theta^\mathcal{C}\in\mathbb{R}^n$ and market price $V^\textit{mkt}(\theta^\mathcal{C})$ (e.g. European options characterized by maturity and strike, $\theta^\mathcal{C}=\{T,K\}\in\mathbb{R}^2$), and a pricing model $\mathcal{M}$ depending on a set of \textit{model parameters} $\theta^\mathcal{M}\in\mathbb{R}^m$ which produces model prices $V^\mathcal{M}(\theta^\mathcal{C},\theta^\mathcal{M})$, the model calibration amounts to estimate the model parameters $\hat{\theta}^\mathcal{M}$ such that model prices match market prices, $V^\mathcal{M}(\theta^\mathcal{C},\hat{\theta}^\mathcal{M}) \simeq V^\textit{mkt}(\theta^\mathcal{C})$. 
Both model and market prices are typically expressed as discrete two-dimensional implied volatility surfaces of maturities and strikes\footnote{In the case of interest rate Swaptions we have a three-dimensional cube of underlying swap tenors, maturities and strikes, $\theta^\mathcal{C}\in\mathbb{R}^3$. Also interest rate Caps/Floors written on different IBOR tenors may be viewed as a volatility cube. Since Caplets/Floorlets may be viewed as Swaptions on a single-period Swap, the two cubes are interconnected. Volatilities are conventionally implied using normal or lognormal models, a.k.a. Bachelier or Black models, respectively, and appropriate numerical procedures. See e.g. \cite{Choi2022} for further details. 
Note that we use the same notation $\theta^\mathcal{C}$ to denote contract parameters both in case of a single contract, with $\theta^\mathcal{C}\in\mathbb{R}^n$, and in case of an entire market dataset of $n_{mkt}$ instruments, trusting on the clarity of the context.}, $\sigma^\mathcal{M}(\theta^\mathcal{C},\theta^\mathcal{M}),\sigma^\textit{mkt}(\theta^\mathcal{C})$.
The numerical estimation procedure requires, in principle, a $m-$dimensional global optimization algorithm which computes $n_{mkt}$ prices in $n_s$ iterations until some convergence criteria is met, thus calling the $\mathcal{M}-$model pricing function at least $n_{mkt} \times n_s$ times\footnote{Actually, population-based global optimization algorithms, such as Genetic Algorithms or Particle Swarm Optimization, consider multiple possible solutions $n_r\in\mathbb{N}$ at each iteration, leading to $n_{mkt} \times n_s \times n_r$ calls of the pricing function.}.
Once the model is calibrated, it may be used to price other financial instruments consistently with the market plain-vanillas, which work as the natural hedges. 
In principle, the model must be recalibrated after each significant market movement for real-time pricing applications, or to compute scenario-based risk measures, e.g. historical Value at Risk. 
Clearly, this approach is feasible only when the pricing function 
$V^\mathcal{M}(\theta^\mathcal{C},\theta^\mathcal{M})$ 
can be computed using exact or approximated (semi)analytical formulas, a case occurring only for the simplest pricing models $\mathcal{M}$. If this condition is not met, one must resort to expensive numerical methods such as Monte Carlo simulation or PDE solution, which typically make the problem unfeasible in practice. This is known as the \textit{model calibration bottleneck}.
\par
A relatively recent stream of research (see e.g. \cite{Baschetti2024QF} and refs. therein) explores a DNN-based approach to model calibration, where the \textquote{online} calibration illustrated above is substituted by an \textquote{offline} pre-processing procedure, in which an appropriate DNN is trained on a large precomputed set of model prices $V^\mathcal{M}(\bm{\theta}^\mathcal{C},\bm{\theta}^\mathcal{M})$ 
obtained via some numerical procedure (e.g. Monte Carlo simulation) using a large set of contract and model parameters $\{\bm{\theta}^\mathcal{C},\bm{\theta}^\mathcal{M}\}$ selected within appropriate domains.
In this way, the DNN \textquote{learns} the exact model pricing function 
$V^\textit{NN}(\bm{\theta}^\mathcal{C},\bm{\theta}^\mathcal{M},w) \simeq
V^\mathcal{M}(\bm{\theta}^\mathcal{C},\bm{\theta}^\mathcal{M})$,
where $w$ are the network weights, and the corresponding implied volatilities 
$\sigma^\textit{NN}(\bm{\theta}^\mathcal{C},\bm{\theta}^\mathcal{M},w) \simeq
\sigma^\mathcal{M}(\bm{\theta}^\mathcal{C},\bm{\theta}^\mathcal{M})$, 
in a wide variety of market situations, corresponding to the many different model and contract parameter values used to generate  the training set. 
Once the offline step is executed, the online step amounts to calibrate the model $\mathcal{M}$ to market prices using the DNN, i.e. to estimate the model parameters $\hat{\theta}^\mathcal{M}$ such that $V^\textit{NN}(\theta^\mathcal{C},\hat{\theta}^\mathcal{M},w) \simeq V^\textit{mkt}(\theta^\mathcal{C})$, 
a task that typically takes very short times thanks to the pure algebraic mathematical structure of the DNN.
Clearly, the production of the training set and the offline DNN training procedure is computationally expensive, but, provided that the training set is large enough to represent a sufficient variety of market volatility surfaces, it is not necessary to update it, except in the event of market regime changes.

\subsection{The SABR Model}
\label{sec:SABR_discussion}
The SABR (Stochastic Alpha Beta Rho) is one of the most popular models in finance. It assumes a CEV (Constant Elasticity of Variance) stochastic dynamics of single forward quantities, i.e. rates or prices, with a correlated lognormal stochastic volatility. It was originally introduced by \cite{Hagan2002} and later extended by \cite{Hagan2016} to accommodate negative forwards. 
The reasons behind the extraordinary success of the SABR model, despite the complexity of its stochastic dynamics, are the following.
\begin{itemize}
\item \textbf{Parsimony}: the shifted-SABR stochastic dynamics depends on 5 parameters with a simple and transparent financial interpretation. In particular, the $\beta$ parameter allows the model to interpolate between normal ($\beta=0$) and lognormal ($\beta=1$) dynamics.
\item \textbf{Analytical approximation}: there exists a simple analytical approximation (a.k.a. Hagan et al. formula) of normal or lognormal SABR volatility implied in European option prices, which was originally derived in \cite{Hagan2002} using singular perturbation techniques, and later improved in \cite{Obloj2008} and \cite{Hagan2016}. This approximation allows to price European options depending on a single forward rate simply using either the Bachelier (normal) or the Black (lognormal) analytical formulas (see e.g. \cite{Choi2022} for further details), even if the SABR dynamics is, in general, neither normal nor lognormal.
\item \textbf{Calibration and performance}: the two previous properties allow fast and precise calibration of the SABR model to fit typical volatility skew and smile shapes encountered on the market. The global optimization problem of volatility skew/smile calibration is, in principle, five-dimensional, but it may be reduced by fixing the shift parameter to some conventional value and by leveraging the $\beta$ parameter redundancy. The resulting three-dimensional problem may be managed by using standard numerical techniques, i.e. smart parameter guess and local Levenberg-Marquardt algorithms, as suggested e.g. by \cite{Gauthier2009}.
\item \textbf{Greeks and hedging}: the Hagan et al. formula also allows to use either analytical formulas for greeks, making hedging and risk management immediate. Furthermore, contrary to local volatility models, the SABR stochastic volatility nature allows to capture the correct smile dynamics\footnote{when the underlying increases/decreases, the smile shifts to higher/lower prices.} and provides more accurate and robust hedging, as discussed in the original paper \cite{Hagan2002} and in more detail in \cite{Hagan2019Bartlett}.
\end{itemize}
Thanks to the properties discussed above, the SABR model \emph{plus} the Hagan et al. approximation became soon the most widely used approach to calibrate the interest rate volatility smile and to price the corresponding European options, i.e. interest rate Caps, Floors, Swaptions. The model is also used to price Constant Maturity Swaps (CMS) and CMS European options, where the convexity adjustment depends on the full volatility smile. 
Since, as shown in tab. \ref{tab:notional_and_gmv} in app. \ref{app:TradingVolumes}, interest rate options are the most traded options on the market, we may conclude that the SABR model is one of the most important model used by global financial markets.
\par
Clearly, the SABR model also has a few limitations, as discussed below.
\begin{itemize}
\item \textbf{Single forward model}: the SABR model refers to the volatility smile of a single forward. For example, each single forward rate entering in a Cap/Floor requires a different SABR model with its own SABR parameters and calibration to the corresponding volatility smile. The European Swaptions cube requires a different SABR model for each ATM swaption.
\item \textbf{Complex dynamics}: the Monte Carlo simulation of the SABR stochastic volatility dynamics is not straightforward; see e.g. \cite{Cai2017OR}. The same happens when solving the corresponding PDE. As a consequence, the model calibration to market implied volatility surface is extremely time consuming and prone to numerical approximations. Even more complex is the calculation of Greeks and hedging.
\item \textbf{Approximation accuracy}: the accuracy of the Hagan et al. analytical approximation deteriorates for high volatilities, long maturities and out-of-the-money options, yielding negative densities in the distribution’s tails for realistic parameter values. As a consequence, the model allow volatility arbitrages and produce inconsistent prices for financial products depending on the wings of the volatility smile, such as deep out-of-the-money options and constant maturity swaps and options. 
\item \textbf{Parameter redundancy}: as discussed in \cite{Hagan2002}, parameters $\beta$ and $\rho$ affect the volatility smile in a similar way, leading to some level of redundancy and to the possibility of fixing one of them to some specific value, e.g. $\beta=0$ in low rates environments, or the Solomonic CIR choice $\beta=0.5$. This is not actually a limitation of the model, since $\beta$ commands the smile wings and can be used to calibrate smile-dependent instruments such as Constant Maturity Swaps and options, as described by \cite{MerPal2006Risk}. A more sophisticated approach is explored in \cite{ZhangFabozzi2016}, which take into account at the same time both the traditional smile calibration and the hedging performance.
\end{itemize}
The limitations of the SABR model discussed above have given rise to a whole strand of literature; see e.g. the introduction in \cite{Hagan2018VolSurfaces} and \cite{Kienitz2022} for excellent reviews.
One possible approach to overcome the shortcoming of Hagan et al. analytical approximation while preserving fast calibration performances consists in using machine learning algorithms based directly on the SABR stochastic dynamics, as discussed in the next section.

\subsection{Machine Learning for SABR}
\label{sec:SABR_ML}
To the best of our knowledge, the first published application of machine learning techniques to the SABR model dates back to \cite{McGhee2021JCF}. In this seminal paper, a lognormal ($\beta=1$) SABR model is used to generate a dataset of exact implied volatility surfaces up to short maturities (2Y) and 10 strikes, on a grid of SABR parameters $\alpha,\rho,\nu$, both randomly and regularly sampled, using either moment-matching analytical integration or numerical PDE solution with finite-difference method (FDM). Then, a single-layer ANN is trained on the dataset and carefully tested, showing good precision and much better performance. 
In \cite{Jeon2022KBS} is taken an approach similar to \cite{McGhee2021JCF}, generating a large dataset of volatility surfaces via GPU-assisted MC simulation instead of FDM\footnote{Even using GPUs, their dataset generation took 1 month.}. They find that DNNs trained on this dataset achieve a very high calibration accuracy and argue that large training sets allow DNNs to filter out  random errors caused by MC, while DNNs fail to rectify systematic biases caused by FDM.
Also \cite{kienitz2020cv} take a similar approach, improving the ANN training by using Hagan et al. approximate volatilities as control variates.
\cite{Funahashi2023QF} extends the control variate approach by using more sophisticated SABR versions from \cite{Antonov2013Risk} for long maturities and from \cite{Antonov2015Risk} for negative forwards.  
\cite{Hoshisashi2024} introduces a Derivative-Constrained Neural Network (DCNN) that incorporates price sensitivities in the objective function and allows to generate smooth volatility surfaces respecting no-arbitrage conditions. The DCNN is trained on sparse volatility surfaces generated by the SABR model using  Hagan et al. approximation with varying parameters. 
Finally, \cite{Su2025MS} solve the backward Kolmogorov equation for the cumulative probability function and train a DNN to learn the corresponding transition probability density function (TPDF). They test different models, including SABR, and benchmark the ANN pricing accuracy against MC simulations using 100 SABR parameter sets for short maturities (0.25,0.5,0.75, 1.0 years, see their tab. 7). 
\par 
Different but related approaches are taken by a number of authors. In particular, we cite \cite{Cuchiero2020Risks}, who calibrate SABR-like Local Stochastic Volatility models using ANNs to identify the leverage function and GANs (Generative Adversarial Networks) to identify the loss function. MC simulation with Euler discretization and Black-Scholes delta hedge variance reduction is used to generate option prices on a fixed grid of 4 short maturities (up to 1Y) and 20 evenly spaced strikes. 
We also cite \cite{Lokvancic2020}, who uses a looking up and interpolation algorithm on a dense dataset of option values computed by Monte Carlo (MC) simulation with Euler discretization on a fixed grid of SABR parameter values (with constant $\beta$). 
\par 
Many other papers do not specifically discuss the SABR model, but take more general approaches that could be applied to SABR, see e.g. the original proposal in \cite{Bayer2018}, further elaborated in \cite{Horvath2020QF}.

\subsection{Our Contribution}
\label{sec:OurContribution}
The literature cited in the previous section \ref{sec:SABR_ML}, presents a number of limitations, which we list below.
i) The original Hagan et al. approximation \cite{Hagan2002} is used instead of the more advanced version by \cite{Hagan2016}, which allows a better performance.
ii) The full shifted-SABR dynamics is reduced to the lognormal case by fixing $\beta=1$ to simplify the numerical solutions.
iii) The volatility surface is limited to short maturities and/or fixed strike ranges, without addressing the regions where the Hagan et al. approximation is less accurate.
iv) Do not systematically consider real market data and/or the most diffused interest rate instruments traded on the market.
v) The datasets used for DNN training and/or the DNN structure and/or the training procedure are not fully specified in all the necessary details, thus preventing a complete understanding and reproducibility of the results.
\par 
Although the limitations listed above are scattered across different papers, to the best of our knowledge there is no single work where all of them are addressed together for the specific SABR model in a systematic and self-consistent way. 
In this paper, we aim to cover these gaps by combining all the key elements and establish a comprehensive and functional SABR framework that practitioners can adopt in their daily trading and risk management activity on interest rate options. 
Specifically, we contribute to this research field with the following elements.

\begin{enumerate}

\item We consider real market data on different business dates for the most traded interest rate derivatives, i.e. EURIBOR Caps and Floors\footnote{As discussed in sec. \ref{sec:Conclusions}, the same approach described in this paper can be applied to Caps/Floors on overnight rates and to Swaptions. We note that Swaptions depend on a single forward swap rate, while Caps/Floors are more complex since each option depends on multiple forward rates, one for each Caplet/Floorlet.}, taking into account the entire volatility surface quoted on the market up to very long maturities (30Y) and extreme strikes (14 points from -1.5\% to 10\%, including the ATM)\footnote{The market strike grid changed in the past to accommodate negative strikes, but remained stable following the return to positive interest rates in 2022.}, and different EURIBOR tenors (i.e. 3M and 6M).

\item We consider the full shifted-SABR stochastic dynamics (without any usage of the Hagan et al. approximation), taking into account possible negative forward rates. In particular, the $\beta$ parameter is not fixed (as done e.g. in \cite{McGhee2021JCF}), but is calibrated together with the other model parameters, preserving the full flexibility of the original SABR formulation. 
Caplet/Floorlet model prices are generated using high-precision unbiased Monte Carlo simulation of the scaled shifted-SABR dynamics $X(t) := \bar{F}(t)/\bar{F}_0$, thus reducing the number of parameters without loss of generality.

\item Since DNN performances depend crucially on data, we deserve particular importance to the initial offline generation of a very large dataset of interest rate volatility surfaces (more than 200 million points) for EUR Caplets/Floorlets consistent with market quotations.
To this scope, we adapt to the SABR model the random grid approach proposed by \cite{Baschetti2024QF}, introducing two differences: first, we consider a different SABR model for each maturity, consistently with the SABR assumptions in the previous section; second, we select a strike grid independent of maturity, in line with EUR market conventions where Caps and Floors are quoted on a fixed strike grid across all maturities. 

\item We develop and optimize a DNN architecture specific for Caps and Floors and we train it on the dataset described above, reaching a very good level of precision. 
We check that our DNNs, with a single training, are able to calibrate different market volatility surfaces quoted on different business dates with the same very good precision and performance. As a consequence, they are robust with respect to evolving markets, and do not need retraining, at least in the absence of market regime shifts leading to configurations significantly different from those included in the training dataset.

\item Using our DNNs, we challenge the latest version of Hagan et al. analytical approximation developed in \cite{Hagan2016} for the Shifted-SABR model, by computing the distance between the implied volatilities obtained via Monte Carlo simulation using the two corresponding calibrated parameters. This approach allows us to precisely measure how much, in real market situations, the Hagan et al. approximation accuracy deteriorates in specific regions of the volatility surface, particularly for medium-long maturities and out-of-the-money strikes.

\end{enumerate}
\par
The rest of the paper is organized as follows. 
Section \ref{sec:Theory} presents the theoretical framework, introducing the financial instruments involved, the SABR model, its scaled Monte Carlo simulation, and the DNN-based SABR calibration approach proposed in this research. Additional details are reported in App. \ref{app:MktData} and \ref{app:SABRdetails}. 
Section \ref{sec:NumericalResults} describes in detail the numerical implementation and results of our methodology, including the generation of the training, validation, and test datasets (sec. \ref{sec:Datasets}), the DNN construction, optimization and training (sec. \ref{sec:DNNSetup}), and the calibration results, comparing the DNN performance with that of the Hagan et al. approximation (sec. \ref{sec:CalibrationResults}). Additional details are reported in app. \ref{app:DatasetDetails}, \ref{app:DNNdetails}, and \ref{app:CalibrationDetails}.
Section \ref{sec:Conclusions} concludes the paper, summarizing the main findings and outlining potential directions for future research.

\section{Theoretical Framework}
\label{sec:Theory}

\subsection{Financial Instruments}
\label{sec:CF}
In this paper we focus on interest rate Cap/Floor European options. In particular, we select Caps/Floors on EURIBOR typical of the EUR market. Nevertheless, our results do not depend on the specific underlying rate or payoff, and could be easily extended to other instruments, e.g. EURIBOR Swaptions and Caps/Floors/Swaptions on compounded overnight rates typical of other markets (e.g. SOFR for USD, SONIA for GBP, etc.). 
\par 
A Cap/Floor is a portfolio of Caplets/Floorlets, i.e. call/put options with payoff at cash flow date $T_i$ given by
\begin{equation}\label{eq:cfPayoff}
\textit{cf}(T_i;T_{i-1},T_i,K,\omega)
=\textit{Max}\left\{\omega\left[L(T_{i-1},T_i)-K\right]\right\}\tau(T_{i-1},T_i),
\end{equation}
where $L(T_{i-1},T_i)$ is a realized IBOR rate fixed at $T_{i-1}$ and referred to the time interval $[T_{i-1},T_i]$, a.k.a. the rate tenor (e.g. 6 months), $\tau(T_{i-1},T_i)$ is the year fraction consistent with the rate tenor, $K$ is the strike, and $\omega=\pm 1$ for Caplets/Floorlets, respectively. 
The undiscounted price at time $t<T_{i-1}$ (before the IBOR fixing date) of Caplets/Floorlets and the price of Caps/Floors is given by the expectations
\begin{align}
\textit{cf}(t;T_{i-1},T_i,K,\omega)
&= \mathbb{E}_t^{Q_d^{T_{i}}} \left\{
\textit{Max}\left\{\omega\left[L(T_{i-1},T_i)-K\right]\right\}
\right\}\tau(T_{i-1},T_i),\label{eq:cfPrice}\\
\textit{CF}(t;\mathbf{T_n},K,\omega) 
&= \sum_{i=i_0(t)}^n P_d(t;T_i)\textit{cf}(t;T_{i-1},T_i,K,\omega)\label{eq:CFPrice},
\end{align}
where $\mathbf{T_n} = \left\{T_0,\cdots T_n\right\}$ is the Cap/Floor schedule\footnote{The Cap/Floor schedule is consistent with the rate tenor, e.g. semi-annual on IBOR6M, quarterly on IBOR3M, etc., otherwise the pricing formulas would include a convexity adjustment, which is not common on the market and beyond the scope of this paper.} for maturity $T_n$, $i_0(t)$ indexes the current time interval such that $t\in\left[T_{i_0(t)-1};T_{i_0(t)}\right]$, the expectation is taken under the forward measure $Q_d^{T_i}$ associated with the discounting numeraire $P_d(t;T_i)$, given the information available at time $t$. Notice that we have shifted the discount factor $P_d(t;T_i)$ from Caplets/Floorlets to Caps/Floors for later convenience, in order to avoid discounting rates in the DNN. 
Real Caps/Floors quoted on the market are further characterized by a number of details that we report in app. \ref{app:CapsFloorsQuotes}.
\par 
The corresponding IBOR forward rate observed at time $t<T_{i-1}$ is defined as
\begin{equation}\label{eq:fwdRate}
F_i(t) := F(t;T_{i-1},T_i) := \mathbb{E}_t^{Q_d^{T_{i}}} \left[ L(T_{i-1},T_i) \right].
\end{equation}
Forward rate models assume that the IBOR forward rate in eq. \eqref{eq:fwdRate} is the fundamental quantity whose stochastic dynamics must be modeled to compute the prices in eqs. \eqref{eq:cfPrice} and \eqref{eq:CFPrice}. This is obviously not the only possible choice; see e.g. \cite{BrigoMercurio2006} for details on interest rate modeling approaches.
\par 
For the purpose of DNN training discussed in the following sec. \ref{sec:LearningSABRwithDNN}, the contract parameters for Caplets/Floorlets are given by $\theta^\textit{CF} = \left\{T,K\right\}$, where $T$ is the contract fixing date and $K$ is the strike.

\subsection{Scaled Shifted-SABR Model}
\label{sec:ScaledSABRmodel}
The SABR\footnote{The name SABR given in the original paper by Hagan et al. \cite{Hagan2002} stands for \textquote{Stochastic Alpha Beta Rho}, where the stochastic forward and volatility processes were denoted by $\hat{F}$ and $\hat{\alpha}$, respectively, and the parameter associated with the initial volatility value was denoted by $\alpha=\hat{\alpha}(0)$. In this paper we adopt a lighter notation, denoting the two stochastic processes simply by $F(t)$ and $\sigma(t)$, and keeping the initial volatility parameter as $\alpha=\sigma(0)$.} stochastic dynamics of the forward rate $F_i(t)$ in eq. \eqref{eq:fwdRate} and of its volatility, denoted by $\sigma_i(t)$, is given, dropping the index $i$, in \cite{Hagan2002} as
\begin{align}
    & d\bar{F}(t) = \sigma(t) \bar{F}^{\beta}(t) dW(t), \hspace{0.5cm} \bar{F}(0) := \bar{F}_0 = F_0 + \lambda, \label{eq:SABRFwdDynamics}\\
    & d\sigma(t) = \nu \sigma(t) d Z(t), \hspace{1.5cm} \sigma(0) = \alpha, \label{eq:SABRVolDynamics}\\
    & dW(t) dZ(t)= \rho dt \label{eq:SABRcorrelation},\\
    &\bar{F}(t) = F(t) + \lambda \label{eq:FwdShifted},\\
    & F_0\in\mathbb{R}^+, \alpha\in\mathbb{R}^+,\beta\in [0;1], \rho\in [-1;1], \nu\in\mathbb{R}^+,\lambda\in\mathbb{R}^+,\label{eq:SABRconditions}
\end{align}
where $\bar{F}(t)$ is the shifted forward rate, and the processes $\{W(t)\}_t$ and $\{Z(t)\}_t$ are two $Q_d^T$ Brownian motions associated to the discounting zero coupon bond $P_d(t;T)$ as numeraire. 
The SABR model, for each forward rate $F(t)$, is characterized by 6 model parameters 
$\theta^\textit{SABR}=\left\{F_0,\alpha,\beta,\rho,\nu,\lambda\right\}$,
i.e., respectively, the initial forward rate level, the initial volatility level, the CEV elasticity, the  correlation between the forward rate and its volatility, the volatility of volatility, and the rate shift, which bounds the forward rate to $-\lambda$, ensuring that $\bar{F}_0\in\mathbb{R}^+$. 
The Hagan et al. analytical approximation of the SABR implied volatility for European options is reported in app. \ref{app:Hagan}.
\par
As discussed in the following sec. \ref{sec:LearningSABRwithDNN}, the SABR model and Caplet/Floorlet contract parameters 
$\left\{\theta^\textit{SABR},\theta^\textit{CF}\right\}$ are used as DNN input variables. 
In order to enhance the network's ability to learn the relationships between the output and the input values, we reduce the number of SABR model parameters.
First, according to common market practice, we set the shift parameter $\lambda$ to a fixed value, large enough to handle negative forward rates. 
Secondly, we rescale the shifted-SABR process to $X(t) := \frac{\bar{F}(t)}{\bar{F}_0}$ and $\hat{\sigma}(t) := \sigma(t) \bar{F}(t)^{\beta -1}$, leading to the scaled shifted-SABR Model
\begin{align}
    & dX(t) = \hat{\sigma}(t) X^{\beta}(t) dW(t), \hspace{0.5cm} X(0) = 1, \label{eq:XSABRFwdDynamics}\\
    & d\hat{\sigma}(t) = \nu \hat{\sigma}(t) d Z(t), 
    \hspace{1.5cm} \hat{\sigma}(0) = \alpha \bar{F}_0^{\beta -1} =: \hat{\alpha}, \label{eq:XSABRVolDynamics}\\
    & dW(t) dZ(t)= \rho dt. \label{eq:XSABRcorrelation}
\end{align}
Consistently, we also rescale the Caplet/Floorlet shifted strike to $\hat{K} = \frac{\bar{K}}{\bar{F}_0}$. 
Finally, we set\footnote{Note that we denote always with $\theta^\textit{SABR},\theta^\textit{CF}$ different combinations of model and contract parameters, e.g. scaled and not-scaled, hoping that the difference will be clear depending on the context.} 
$\theta^\textit{SABR}=\left\{\hat{\alpha},\beta,\rho,\nu\right\}$ and $\theta^\textit{CF} = \left\{T,\hat{K}\right\}$. 
Note that the Caplet/Floorlet implied volatlity is the same, i.e.
$\sigma(\hat{\alpha},\beta,\rho,\nu,T,\hat{K}) = \sigma(\bar{F}_0,\alpha,\beta,\rho,\nu,T,\bar{K})$,
The proof is provided in app. \ref{app:ScaledSABRmodel}. 
\par
We stress that we do not further reduce the model parameters by fixing $\beta=1$ as done in some previous papers, e.g. \cite{McGhee2021JCF}.

\subsection{SABR Monte Carlo}
\label{sec:MC}
A possible approach to circumvent the limitations of \cite{Hagan2016} analytical approximation is to directly simulate the SABR dynamics in eqs. \eqref{eq:SABRFwdDynamics}-\eqref{eq:SABRconditions} using Monte Carlo. This approach is not straightforward, since it requires the joint simulation of the stochastic volatility process \eqref{eq:SABRVolDynamics}, the simulation of the integrated variance (conditional on the volatility process), and the simulation of the underlying CEV process \eqref{eq:SABRFwdDynamics} (conditional on the volatility and integrated variance processes). Furthermore, an absorbing boundary must be included in the MC simulation to deal with the non-null probability of the underlying CEV process to touch zero for $\beta\in (0,1)$. All of these elements lead to computationally expensive simulation that reduces the actual usefulness of the Monte Carlo approach.
\par 
In this paper we adopt a simple log-Euler discretization scheme of the SABR model dynamics. Although more refined discretization schemes exist in the literature, such as the approaches proposed by \cite{Chen2012IJTAF} and \cite{Cai2017OR}, their additional complexity does not offer significant advantages for our purposes. Moreover, our framework is independent of the specific MC discretization scheme, ensuring that our results remain valid and replicable under different discretization schemes.
Since the two Brownian motions $W(t),Z(t)$ in eq. \eqref{eq:SABRFwdDynamics}-\eqref{eq:SABRconditions} are correlated, we introduce the following transformation
\begin{align}
    &dW(t) = \rho d\tilde{Z}(t) + \hat{\rho} d\tilde{W}(t),\\
    &dZ(t) = d\tilde{Z}(t),\\
    &\hat{\rho}=\sqrt{1-\rho^2},
\end{align}
where now $\tilde{W}(t),\tilde{Z}(t)$ are independent Brownian motions. 
Applying the log-transformation to both the forward rate and volatility processes and introducing the Euler exponential discretization, leads to the following shifted-SABR model's discretization scheme, 
\begin{align}
    & \bar{F}(t_{j+1}) = \bar{F}(t_{j}) \exp\left\{- \frac{1}{2} \sigma^2(t_j) \bar{F}^{2\beta-2}(t_j) \Delta + \sigma(t_j) \bar{F}^{\beta-1}(t_j) \sqrt{\Delta}\left[\rho \zeta_{\tilde{Z}} 
    + \hat{\rho} \zeta_{\tilde{W}} \right] \right\}, \label{eq:MCSABRFwdDynamics}\\
    & \sigma(t_{j+1}) =  \sigma(t_j) \exp\left\{-\frac{1}{2} \nu^2 \Delta + \nu \sqrt{\Delta} \zeta_{\tilde{Z}}\right\}, \label{eq:MCSABRVolDynamics}
\end{align}
where $j \in [0,n] \subset \mathbb{N}$, $\{t_0,\dots,t_n\}$ is the regular grid of discretization dates with $t_0=0, t_n = T_{i-1}$ (the underlying rate fixing date), $\Delta = t_{j+1} - t_{j}$ is a constant time step, and $\zeta_{\tilde{Z}}, \zeta_{\tilde{W}} \sim N(0,1)$ are standard normal random variables. 
Finally, scaling the process $\bar{F}(t)$ to $X(t)=\bar{F}(t)/\bar{F}_0$ as discussed in sec. \ref{sec:ScaledSABRmodel} and app. \ref{app:ScaledSABRmodel} lead to the following scaled shifted-SABR discretization scheme,
\begin{align}
    & X(t_{j+1}) = X(t_{j}) \exp\left\{-\frac{1}{2} \hat{\sigma}^2(t_j) X^{2\beta-2}(t_j) \Delta + \hat{\sigma}(t_j) X^{\beta-1}(t_j) \sqrt{\Delta}\left[\rho \zeta_{\tilde{Z}} 
    + \hat{\rho} \zeta_{\tilde{W}} \right] \right\}, \label{eq:MCXSABRFwdDynamics}\\
    & \hat{\sigma}(t_{j+1}) =  \hat{\sigma}(t_j) \exp\left\{-\frac{1}{2} \nu^2 \Delta + \nu \sqrt{\Delta} \zeta_{\tilde{Z}}\right\}, \label{eq:MCXSABRVolDynamics}
\end{align}
which we used in practice to produce the datasets described in the following sec. \ref{sec:Datasets}. 
\par
In order to avoid that the CEV forward rate process reaches zero during the MC simulation, we set an absorbing boundary at $10^{-14}$ to ensure that, in case the shifted forward rate reaches the boundary, it remains there for the rest of the simulation.

\subsection{Learning SABR with DNNs} 
\label{sec:LearningSABRwithDNN}
As discussed in sec. \ref{sec:Intro}, an increasing number of studies explore machine learning algorithms to address the calibration of stochastic volatility models using a DNN approach, which we formalize below in the context of SABR model applied to Caps/Floors. 
Unlike the previous literature, we split and formalize our algorithm into a three-stage approach, to give the appropriate importance to the initial dataset generation stage, matching the general idea that DNN performances depend crucially on data. 

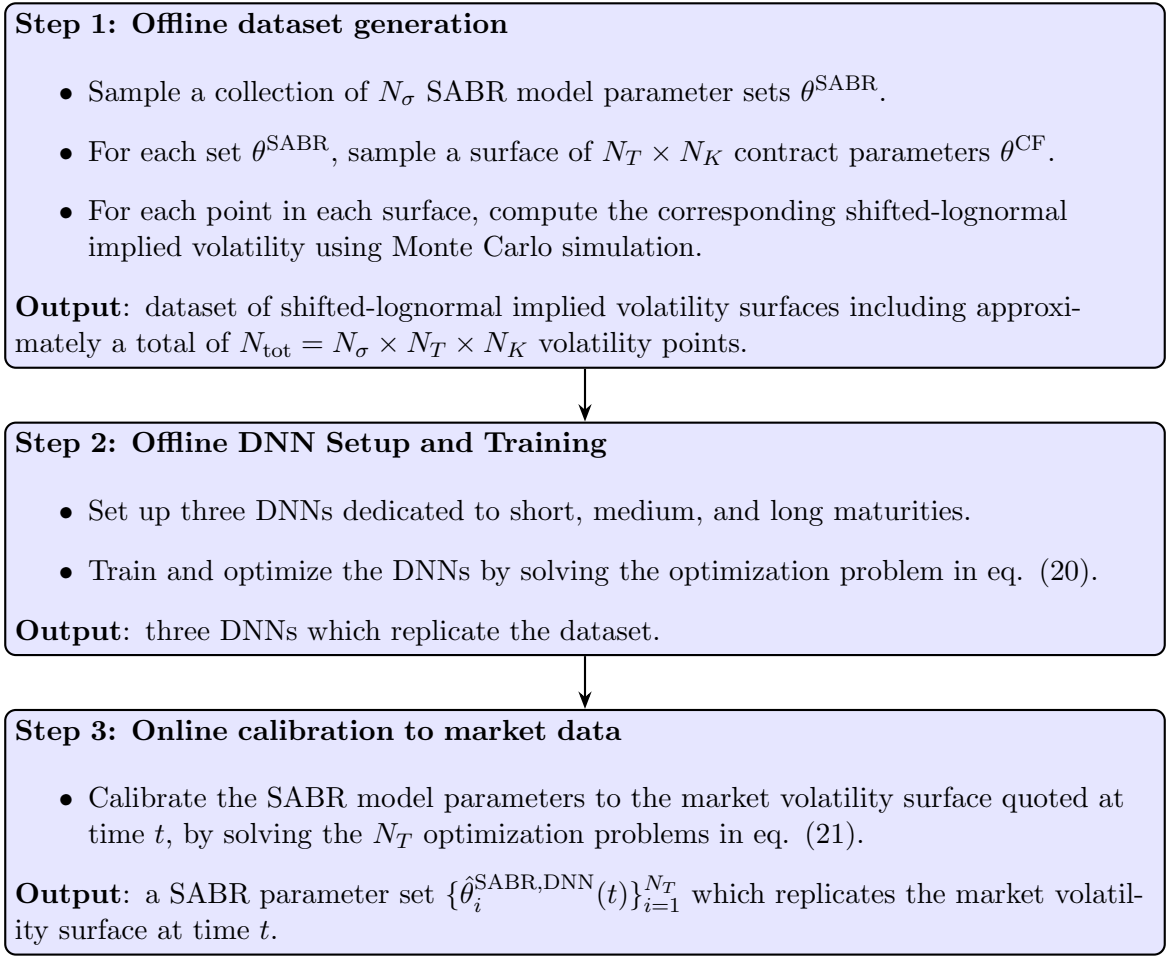
\begin{figure}[H]
\centering
\begin{tikzpicture}[
    block/.style={                    
        rectangle,
        draw,
        thick,                        
        rounded corners,
        fill=blue!10,                 
        text width=15cm,              
        align=left,                   
        minimum height=3em
    },
    connector/.style={                
        -Stealth,                     
        thick                         
    }
]
\node (n1) [block]
{
    \textbf{Step 1: Offline dataset generation}
    \vspace{1mm} 
    \begin{itemize}
        \item Sample a collection of $N_\sigma$ SABR model parameter sets $\theta^{\text{SABR}}$.
        \item For each set $\theta^{\text{SABR}}$, sample a surface of $N_T \times N_K$ contract parameters $\theta^{\text{CF}}$.
        \item For each point in each surface, compute the corresponding shifted-lognormal implied volatility using Monte Carlo simulation.
    \end{itemize}
    \textbf{Output}: dataset of shifted-lognormal implied volatility surfaces including approximately a total of $N_{\text{tot}}=N_\sigma\times N_T\times N_K$ volatility points.
};
\node (n2) [block, below = 0.7cm of n1] 
{
    \textbf{Step 2: Offline DNN Setup and Training}
    \vspace{1mm}
    \begin{itemize}
        \item Set up three DNNs dedicated to short, medium, and long maturities.
        \item Train and optimize the DNNs by solving the optimization problem in eq. \eqref{eq:DNNtraining}.
    \end{itemize}
    \textbf{Output}: three DNNs which replicate the dataset.
};
\node (n3) [block, below = 0.7cm of n2] 
{
    \textbf{Step 3: Online calibration to market data}
    \vspace{1mm}
    \begin{itemize}
        \item Calibrate the SABR model parameters to the market volatility surface
        quoted at time $t$, by solving the $N_T$ optimization problems in eq. \eqref{eq:SABRDNNCalibration}.
    \end{itemize}
    \textbf{Output}: a SABR parameter set $\{\hat{\theta}_i^{\text{SABR,DNN}}(t)\}_{i=1}^{N_T}$ which replicates the market volatility surface at time $t$.
};
\draw [connector] (n1) -- (n2);
\draw [connector] (n2) -- (n3);
\end{tikzpicture}
\caption{Flow chart of the SABR DNN calibration methodology.}
\label{fig:Scheme}
\end{figure}

\begin{enumerate}

\item \textbf{Offline dataset generation}.
We sample a large collection of $N_\sigma$ SABR model parameter sets $\theta^\textit{SABR}=\{F_0,\alpha,\beta,\rho,\nu\}$ using appropriate parameter ranges that cover different market situations. 
For each set $\theta^\textit{SABR}$, we sample a set of $N_T \times N_K$ contract parameters $\theta^\textit{CF}=\{T,K\}$ forming a surface with $N_T$ dates and $N_K$ strikes, consistent with the date and strike ranges observed in market quotations. 
For each point in each surface, we compute the corresponding Caplet/Floorlet price using the scaled shifted-SABR Monte Carlo simulation described in sec. \ref{sec:MC}, and the shifted-lognormal implied volatility 
$\sigma^\textit{MC}(\theta^\textit{SABR},\theta^\textit{CF})$ 
by inverting the shifted-Black formula. 
The results is a large dataset of $N_\sigma$ shifted-lognormal implied volatility surfaces of dimension $N_T\times N_K$ including, in total, $N_{tot}=N_\sigma\times N_T\times N_K$ volatility points, consistent with the shifted-SABR stochastic dynamics in eqs. \eqref{eq:SABRFwdDynamics}--\eqref{eq:SABRconditions} and market quotations.
Note that we do not use the Hagan et al. approximation for dataset construction. 

\item \textbf{Offline DNN Setup and Training}. 
We set up three Deep Neural Networks (DNNs) dedicated to short, medium, and long maturities. 
The DNNs considered here are feed-forward networks, similar to e.g. \cite{Horvath2020QF} and \cite{Baschetti2024QF}. 
The key characteristic of their topology is that each node in a given layer is connected to all nodes in the subsequent layer, whereas vertical connections between nodes within the same layer or backward from a layer to a previous layer are not allowed. In these DNNs data flows uni-directionally, from the input nodes to the output nodes, without any feedback loops. 
Denoting with $\theta_s :=\{\theta_s^\textit{SABR},\theta_s^\textit{CF}\}, s=1,\dots,N_{tot}$ the total set of parameters associated to a single Monte Carlo implied volatility $\sigma^\textit{MC}(\theta_s)$, the DNNs are trained to learn the implied volatility map 
$\theta_s \rightarrow \sigma^\textit{MC}(\theta_s)$ by finding the optimal DNNs' weights $\hat{w}$ that solve the optimization problem
\begin{equation}
\hat{w} = \underset{w}{\arg \min} \sqrt{\frac{1}{N_{tot}}\sum_{s=1}^{N_{tot}}
\left[\sigma^\textit{DNN}(\theta_s, w) - \sigma^\textit{MC}(\theta_s)\right]^2}.
\label{eq:DNNtraining}
\end{equation}

\item \textbf{Online calibration to market data}. 
Given the trained DNN, we can now, at any time $t$, calibrate the shifted-SABR model parameters to the market volatility surface
\begin{align*}\sigma^\textit{Mkt}(t;T_i, K_j),\; i=1,\dots,N_T,j=1,\dots,N_K\quad \textit{(smile by smile)} \end{align*}
quoted at time $t$, considering the trained DNN as a substitute of the actual pricing function, by solving the $N_T$ optimization problems 
\begin{equation}
\hat{\theta}_i^\textit{SABR,DNN}(t) = 
\underset{\theta_i^\textit{SABR}}{\arg \min} \sqrt{\frac{1}{N_K}\sum_{j=1}^{N_K}
\xi_j\left[\sigma^\textit{DNN}(\theta_i^\textit{SABR},T_i,K_j, \hat{w}) - \sigma^\textit{Mkt}(t;T_i, K_j)\right]^2},i=1,\dots,N_T,
\label{eq:SABRDNNCalibration}
\end{equation}
where $\xi_j$ denotes the Black vega sensitivity associated with the $j$-th strike in the smile section, normalized with respect to the total vega sensitivity of the smile section. 

\end{enumerate}
We show in fig. \ref{fig:Scheme} the flowchart corresponding to the three-stage approach described above.

\par
The results of the SABR DNN calibration may be compared with the results of the \textquote{classic} SABR calibration, based on the Hagan et al. approximation discussed in app. \ref{app:Hagan}, obtained by the $N_T$ optimization problems
\begin{equation}
\hat{\theta}_i^\textit{SABR,Hagan}(t) = 
\underset{\theta_i^\textit{SABR}}{\arg \min} \sqrt{\frac{1}{N_K}\sum_{j=1}^{N_K}
\xi_j\left[\sigma^\textit{Hagan}(t;\theta_i^\textit{SABR},T_i,K_j) - \sigma^\textit{Mkt}(t;T_i,K_j)\right]^2}, i=1,\dots,N_T.
\label{eq:SABRHaganCalibration}
\end{equation}
\par 
The massive numerical calculation of shifted-lognormal volatilities implied from Monte Carlo prices of a large dataset including deep in/out of the money options and high uncertainty scenarios requires careful consideration and specific techniques to reduce the Monte Carlo error and avoid numerical problems (see e.g. \cite{Choi2022} sec. 4, and refs. therein). 
In this context, we prefer the bisection algorithm applied to Floorlet prices.
Specifically, we always compute both Caplet and Floorlet MC prices, and we select the option with the smallest MC error; if the option is a Floorlet, we proceed with bisection, else if the option is a Caplet, we compute the price of the corresponding Floorlet via the call-put parity and then use bisection. 
In this way, we ensure consistency in the volatility implication process, as the same inversion formula is always applied. Floorlets are preferred because, as shown in app. \ref{app:MCErrorAnalysis}, in low–uncertainty scenarios, the smallest MC pricing error typically occurs for out–of–the–money options, whereas, in high–uncertainty scenarios, put options tend to have a smaller MC error due to the natural upper bound on their payoff. 
This approach ensures that the implied volatility calculation is always based on the most accurate MC price, thus reducing the implied volatility MC error propagated from the MC price. 
\par 
The generation of a large dataset (we will see in sec. \ref{sec:Datasets} that $N_{tot}$ is huge, more than 200 million volatilities) and the training procedure are time-consuming processes that can take hours or days, depending on the available computational resources and model complexity. However, this process is conducted offline, meaning that it is performed once and for all. Subsequently, the trained DNNs are used to address very fast all necessary calibration tasks, even in real time.

\section{Numerical Results}
\label{sec:NumericalResults}
In this section we report our numerical results following the three-stage approach illustrated in the previous sec. \ref{sec:LearningSABRwithDNN}.

\subsection{Datasets}
\label{sec:Datasets}
We built three distinct datasets spanning short, medium and long Caplet/Floorlet maturities up to 30 years, according to the EUR market quotations, as described below.

\paragraph{\textbf{Training Sets}:} the SABR model parameters $\theta^\textit{SABR}=\{F_0,\alpha, \beta, \rho, \nu\}$ were sampled from a uniform distribution using Latin Hypercube Sampling (LHS) within the predefined ranges reported in tab. \ref{tab:SABRparametersRanges}. 
\begin{table}[H]
\centering
\begin{tabular}{ccccccccc}
\toprule  
Training & Fixing & \multirow{2}*{$F_0$} & \multirow{2}*{$\lambda$} & \multirow{2}*{$\alpha$} & \multirow{2}*{$\beta$} & \multirow{2}*{$\rho$} & \multirow{2}*{$\nu$} & \multirow{2}*{$N_\sigma$} \\
set \# & date (y) & &  &  &  &  &  &  \\
\midrule 
1 & $[0.25,4)$  & $[1\%, 5\%]$ & 3\% & [0.001,0.2] & [0.1,0.9]  & [-0.8,0.6] & [0.05,1.6] & $2^{20}$\\
2 & $[4,10.5)$  & $[1\%, 5\%]$ & 3\% & [0.001,0.2] & [0.1,0.9]  & [-0.8,0.6] & [0.05,1.2] & $2^{18}$\\
3 & $[10.5,30]$ & $[1\%, 5\%]$ & 3\% & [0.001,0.2] & [0.05,0.9] & [-0.8,0.6] & [0.05,1.2] & $2^{18}$\\
\bottomrule  
\end{tabular}
\caption{SABR model parameters' ranges and number of samples for each data subset. For each sample is generated an entire volatility surface, as explained below. The total dataset includes $N^\textit{tot}_\sigma = 2^{20}+2^{18}+2^{18}$ volatility surfaces. }
\label{tab:SABRparametersRanges}
\end{table}
The parameters' ranges were selected so as to generate a training set spanning a sufficiently wide range of values, enabling an effective and robust calibration of the SABR model in different market conditions, while at the same time avoiding excessively broad ranges that could lead to the inclusion of unrealistic parameter values.
The initial forward rate $F_0$ was centered around market-observed values, i.e. EURIBOR6M forward rates, and the range was extended to include all levels the rate may reasonably assume under the prevailing economic conditions. The shift parameter was fixed to $\lambda=3\%$, consistently with market quotations for EUR Caps/Floors which include negative strikes (see app. \ref{app:CapsFloorsQuotes}). 
The parameters $F_0$ and $\alpha$ are then scaled as discussed in sec. \ref{sec:ScaledSABRmodel}.
The ranges for the remaining SABR model parameters reported in tab. \ref{tab:SABRparametersRanges} were determined using the following empirical iterative procedure.
\begin{enumerate}
\item Start with a suitable small but realistic guess of the parameters' ranges.
\item Generate the corresponding dataset as described below in this section.
\item Train, validate and test the DNNs as described in sec. \ref{sec:DNNSetup}.
\item Using the DNNs, calibrate the SABR parameters to market data as described in sec. \ref{sec:CalibrationResults}.
\item Check if the resulting SABR parameters lie close to the boundaries of their respective ranges.
\item Adjust the parameters' ranges and repeat the previous steps until the final result is stable.
\end{enumerate}
\par 
We observe in tab. \ref{tab:SABRparametersRanges} that the number of sampled SABR parameter sets $N_\sigma$ is higher for dataset \#1, which corresponds to shorter maturities. This is because the Monte Carlo shifted-Black implied volatilities for shorter maturities exhibit a greater MC error. Consequently, more data are required in dataset \#1 for DNN training to overcome this noise. The SABR parameters' ranges determined according to the procedure above are consistent with the parameters' term structures reported in app. \ref{app:CalibrationDetails}, a confirmation that the empirical iterative procedure outlined above was effective.
\par
Given the SABR parameter sets $\theta^\textit{SABR}_l, l=1,\dots,N_\sigma$, we must associate the contract parameters 
$\theta^\textit{CF}=\{T,K\}$. We built, for each set $\theta^\textit{SABR}_l$, an entire shifted-Black volatility surface including $N_T$ dates $\times N_K$ strikes. 
The dates $\{T_{l,i}\}_{i=1}^{N_T}$ were sampled uniformly and randomly, with each $T_{l,i}$ selected from a corresponding $i$-th sub-interval of the total time span of each dataset, shown in tab. \ref{tab:times}. 

\begin{table}[H]
\centering
\begin{tabular}{ccccc}
\toprule 
    Training set \# & Fixing date intervals $[T_{min},T_{max}]$ & $N_T$ \\
\midrule
    \multirow{3}{*}{\centering 1} & [2m,5m) $\cup$ [5m,8m) $\cup$ [8m,1y) $\cup$ [1y, 1y4m) $\cup$  & \multirow{3}{*}{\centering 10} \\
    & [1y4m,1y6m) $\cup$ [1y6m,1y11m) $\cup$ [1y11m,2y5m) $\cup$  &  \\
    & [2y5m,2y11m)$\cup$ [2y11m,3y5m)$\cup$ [3y5m,3y11m] &\\
\midrule
    \multirow{3}{*}{\centering 2} & [3y10m,4y6m) $\cup$ [4y6m,5y2m) $\cup$ [5y2m,5y10m) $\cup$ [5y10m,6y6m) $\cup$  & \multirow{3}{*}{\centering 10} \\
    &[6y6m,7y2m) $\cup$[7y2m,7y10m) $\cup$ [7y10m,8y6m)$\cup$ [8y6m,9y2m) $\cup$&\\ 
    & [9y2m,9y10m)$\cup$ [9y10m,10y6m] & \\
\midrule
    \multirow{5}{*}{\centering 3} & [10y5m,11y5m) $\cup$ [11y5m,12y5m) $\cup$ [12y5m, 13y5m) $\cup$ [13y5m,14y5m) $\cup$  & \multirow{5}{*}{\centering 20} \\
    & [14y5m,15y5m) $\cup$ [15y5m,16y5m) $\cup$ [16y5m,17y5m) $\cup$ [17y5m,18y5m) $\cup$ & \\
    & [18y5m,19y5m) $\cup$ [19y5m,20y5m) $\cup$ [20y5m,21y5m) $\cup$ [21y5m,22y5m) $\cup$& \\
    &  [22y5m,23y5m) $\cup$ [23y5m,24y5m) $\cup$ [24y5m,25y5m) $\cup$ [25y5m,26y5m)$\cup$ & \\
    & [26y5m,27y5m) $\cup$ [27y5m,28y5m) $\cup$ [28y5m,29y5m) $\cup$ [29y5m,30y5m] &\\
\bottomrule  
\end{tabular}
\caption{Time sub-intervals for each dataset time span. Dates are expressed in months (m) and years (y). As stated immediately below eq. \eqref{eq:MCSABRVolDynamics} in sec. \ref{sec:MC}, these time intervals refer to fixing dates of the forward rate, the final point where the SABR Monte Carlo simulation stops.  The maximum fixing date (30y5m) is larger than the maximum fixing date quoted on the market (29y6m, see tab. \ref{fig:QuotedCapsFloors}), to facilitate the DNN training.}
\label{tab:times}
\end{table}
We note that, although the sub-intervals were consistent across all parameter sets within a dataset, the actual sampled dates are different, since the selection is performed independently for each set. This procedure ensures that the sample dates are well spread throughout the entire date range.
We also note that $N_T$ is increased to 20 for dataset \#3, to adequately cover its larger time span with respect to datasets  \#1 and \#2, and ensure a consistent training precision of the corresponding DNN, as described in the next section.
\par
The strikes were sampled in a similar way: for each SABR parameter set $\theta^\textit{SABR}_l, l=1,\dots,N_\sigma$ and for each date $\{T_{l,i}\}_{i=1}^{N_T}$, we sampled $N_K$ scaled moneyness $\{\hat{K}_{l,i,j}\}_{j=1}^{N_K}$, where $\hat{K} = \bar{K}/\bar{F}_0 = (K+\lambda)/(F_0 + \lambda)$ as discussed in sec. \ref{sec:ScaledSABRmodel}, uniformly and randomly from three fixed sub-intervals, shared across all datasets, as defined in tab. \ref{tab:moneyness}. The ranges were selected according to both market quotations, which feature a fixed strike grid, and the ranges selected for the initial forward rate $F_0$. Also in this case, the actual strike values differ between pairs, since sampling is performed independently for each pair. 
\begin{table}[H]
\centering
\begin{tabular}{ccc}
\toprule  
Training set \# & $\hat{K}$ interval & \# $N_K$ \\
\midrule
\multirow{3}{*}{\centering Any} & $[0.15, 0.70)$ & 4\\
                                & $[0.70, 1.50)$ & 5\\
                                & $[1.50, 3.50]$ & 4\\
\bottomrule  
\end{tabular}
\caption{Scaled moneyness sub-intervals common to all datasets in tab. \ref{tab:times}, and the corresponding number of sample values. The total number of values sampled for each maturity is thus $N_K=13$.}
\label{tab:moneyness}
\end{table}
\par 
For each point the prices of the corresponding Caplet/Floorlet options were estimated using the Monte Carlo simulation described in sec. \ref{sec:MC}, with MC parameters as in the following tab. \ref{tab:TrainingSet}, where we also summarize the characteristics of the training sets. 
Finally, for each price, its corresponding shifted-Black implied volatility was computed as described in sec. \ref{sec:LearningSABRwithDNN}.
\begin{table}[H]
\centering
\begin{tabular}{cccccccc}
\toprule  
Training & \multirow{2}{*}{$N_\sigma$} & Fixing date  & \multirow{2}{*}{$N_T$} & $\hat{K}$ & \multirow{2}{*}{$N_K$} &  \multirow{2}{*}{$N_{MC}$} & $\Delta_{MC}$ \\
set \#   &                             & interval (y) &                        & interval  &                        &  & (days) \\
\midrule
1     & $2^{20}$ & [0.25,4)  & 10 & [0.15,3.5] & 13  & $2^{18}$ & 0.5 \\
2     & $2^{18}$ & [4,10.5)  & 10 & [0.15,3.5] & 13  & $2^{18}$ &   1 \\
3     & $2^{18}$ & [10.5,30] & 20 & [0.15,3.5] & 13  & $2^{18}$ &   3 \\
\bottomrule  
\end{tabular}
\caption{Summary of parameters used to generate the training sets.}
\label{tab:TrainingSet} 
\end{table}
\par 
We note that while the number of MC paths is the same for each dataset, the MC time step increases for medium and longer maturities (associated to subsets \#2 and \#3), to balance efficiency with precision of the dataset generation process. 
\par 
In conclusion, the total training set consists of $N^\textit{tot}_\sigma = 2^{20} + 2^{18} + 2^{18} = 1,572,864$ random grid surfaces, totaling $2^{20}\times 10\times 13 + 2^{18} \times 10\times 13 + 2^{18} \times 20\times 13 = 238,551,040$ points.
Actually, the number of volatility points available in the dataset is slightly smaller, since we dropped any combination of market and model parameters giving prices with a time value below the threshold of $10^{-13}$. This accounts roughly for $1\%$ of the total training set ($\approx 2.4$ million points excluded). 
Furthermore, 20\% of the training set is used as validation set, see below.

\paragraph{\textbf{Validation Sets}:} regarding the validation sets, they were built by randomly sampling 20\% data points from the corresponding training sets. 
Therefore, the total number of data points in the validation set is approximately 47,710,208.

\paragraph{\textbf{Test Sets}:} in addition to the three training sets described above, it is essential to independently generate their corresponding test sets, which are used to evaluate \textit{a posteriori} the performance of the DNN using out of sample data. 
Accordingly, we generated three test sets in a way similar to the training sets, using the $\theta^\textit{SABR}$ parameters' ranges of tab. \ref{tab:SABRparametersRanges}, setting $N_\sigma=2^{10}$. Unlike the training set, we sampled $N_T$ dates uniformly throughout the entire date range, without using the sub-intervals shown in tab. \ref{tab:times}, and a single strike per date. 
In this way we obtain a random selection of points within a volatility surface, less regular than the training set and well-suited for testing the DNNs completely out of sample.
The total number of data points across the three test sets is $2^{10}\times 10\times 1 + 2^{10} \times 10\times 1 + 2^{10} \times 20\times 1 = 40,960$. Also in this case the actual number of volatility points available for testing is slightly smaller, since we dropped configurations giving prices with a time value below the threshold of $10^{-13}$.

\subsection{DNNs Setup and Training} 
\label{sec:DNNSetup}
We set up three distinct DNNs to cover the entire maturity domain of the Caplets/Floorlets quoted on the market, corresponding to the short (\#1), medium (\#2) and long (\#3) maturity datasets shown in tab. \ref{tab:SABRparametersRanges}. Their detailed structure is reported in the following tab. \ref{tab:DNNparameters}.
\begin{table}[hbt]
\centering
\begin{tabular}{lccc}
\toprule  
Parameter & DNN 1 & DNN 2 & DNN 3\\
\midrule
Training set & \#1 & \#2 & \#3 \\
Test set & \#4 & \#5 & \#6 \\
Maturity range (y) & $[0.25,4)$ & $[4,10.5)$ & $[10.5,30]$ \\
Moneyness range & \multicolumn{3}{c}{$[0.15,3.5]$}  \\
Inputs & \multicolumn{3}{c}{$\{\hat{\alpha},\beta,\rho,\nu,T,\hat{K}\}$} \\
Output & \multicolumn{3}{c}{$\sigma^\textit{DNN}_\textit{SLN}(\hat{\alpha},\beta,\rho,\nu,T,\hat{K}) = \sigma^\textit{DNN}_\textit{SLN}(\bar{F}_0,\alpha,\beta,\rho,\nu,T,\bar{K})$} \\
Input layer nodes & \multicolumn{3}{c}{6} \\
Hidden layers & \multicolumn{3}{c}{5}\\
Hidden nodes per layer & \multicolumn{3}{c}{64} \\
Hidden activation function & \multicolumn{3}{c}{ELU} \\
Output layer nodes & \multicolumn{3}{c}{1} \\
Output activation function & \multicolumn{3}{c}{Linear} \\
Loss function & \multicolumn{3}{c}{RMSE} \\
Optimizer & \multicolumn{3}{c}{ADAM} \\
Max number of epochs & \multicolumn{3}{c}{500} \\
Early stopping patient (epochs)& \multicolumn{3}{c}{50} \\
Mini batches & \multicolumn{3}{c}{YES} \\
\bottomrule 
\end{tabular}
\caption{Summary of DNNs characteristics. DNN parameters following the maturity range (moneyness range and below) are common to all the three DNNs.}
\label{tab:DNNparameters}
\end{table}
\par 
The input layer has 6 nodes to accommodate the scaled shifted-SABR and contract parameters 
$\theta = \{\hat{\alpha},\beta,\rho,\nu,T,\hat{K}\}$ (the DNNs' input features).
The output layer consists of a single node representing the shifted-Black implied volatility $\sigma^\textit{DNN}_\textit{SLN}$ with a shift parameter $\lambda=3\%$ consistent with the input volatility data. 
Note that the DNN works with scaled shifted-SABR parameters but the output implied volatility is the same as the non-scaled model, as discussed in sec. \ref{sec:ScaledSABRmodel}.
The linear activation function is used to get the output, since if the neural network is learning properly, it should naturally produce positive implied volatilities, without the need to explicitly enforce it. 
The DNNs' input data are standardized using the StandardScaler from scikit-learn, which transforms each input feature to have zero mean and unit variance, to allow faster convergence and greater efficiency. 
The DNNs' architecture was optimized empirically, checking that reducing the number of hidden layers or nodes per layer led to a worse performance, while increasing them had little or no effect. 
\par
\begin{figure}[hbt]
\begin{subfigure}{.5\textwidth}
  \centering
  \centerline{\includegraphics[scale=0.6]{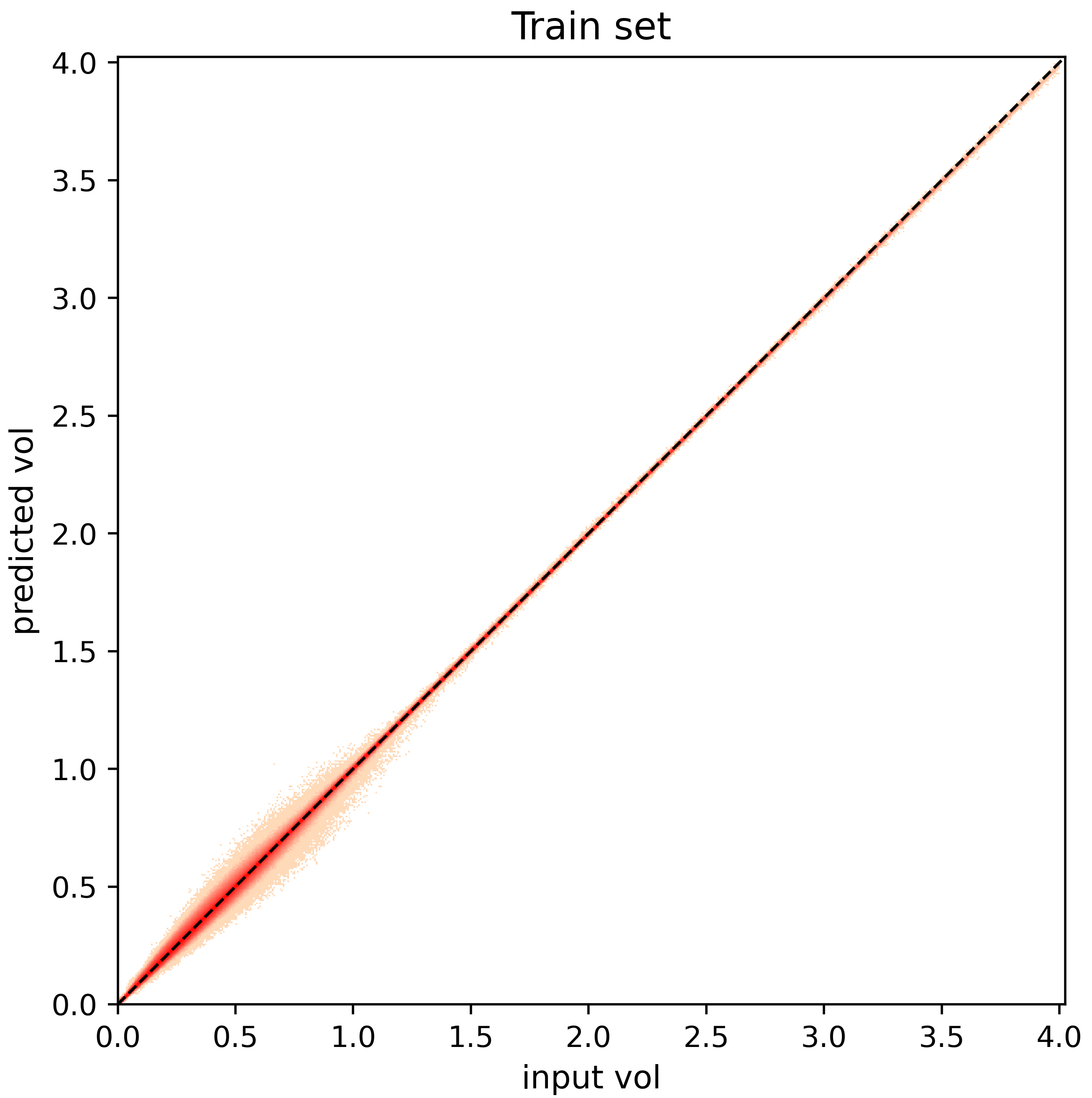}}
\end{subfigure}
\hfill
\begin{subfigure}{.5\textwidth}
  \centering
  \centerline{\includegraphics[scale=0.6]{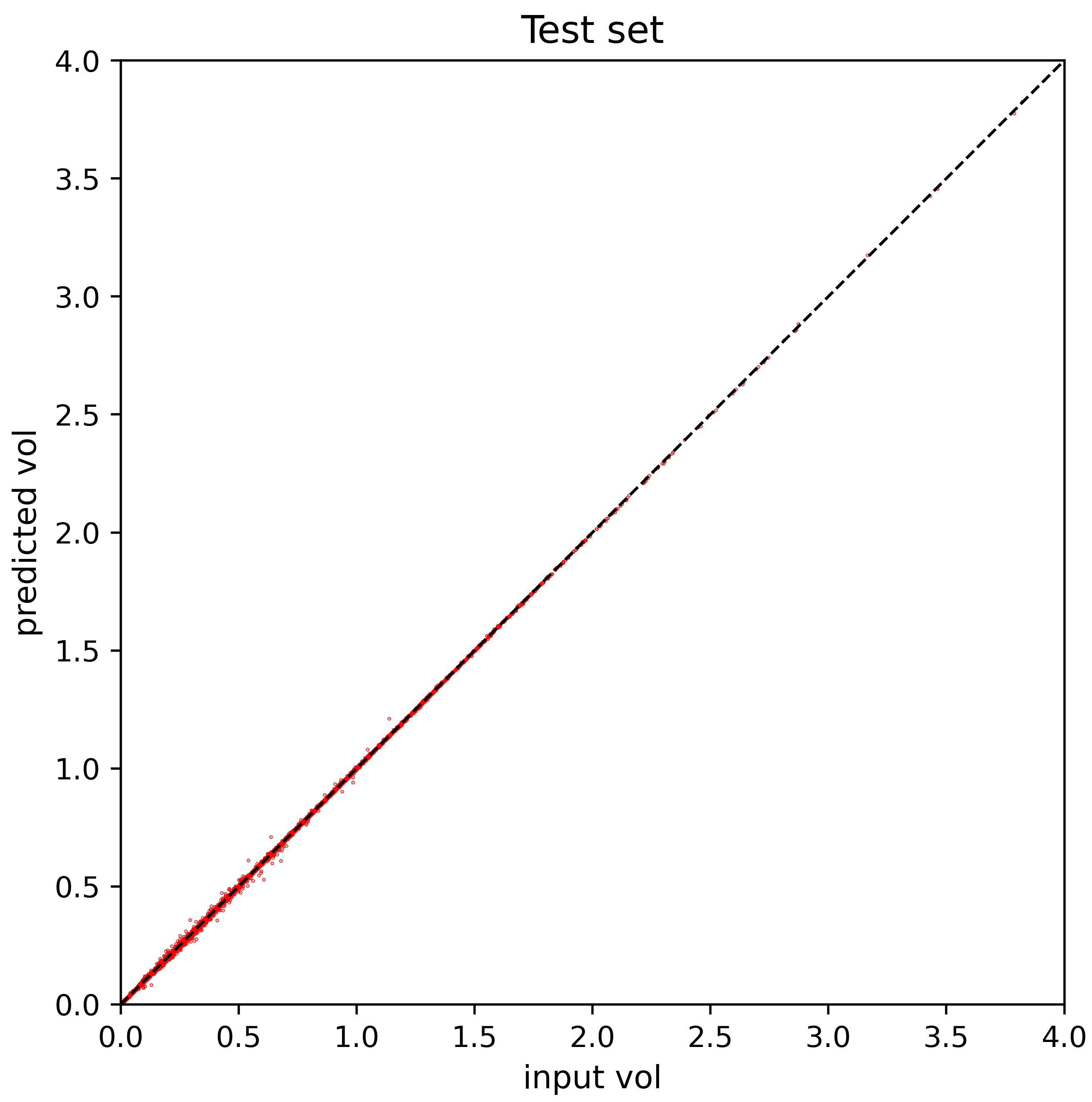}}
\end{subfigure}
\caption{Combined scatter plots for the training (left) and test (right) aggregate results of the three DNNs. More results are reported in app. \ref{app:DNNdetails}. We note in both panels that most of the data lie in the \textquote{reasonable} region below $\sim 75\%$. Few higher data, up to 400\%, occasionally appear because of the random generation of SABR parameters discussed in the previous sec. \ref{sec:Datasets}, which may produce \textquote{strange} combinations leading to extreme volatilities. We did not discarded such data from the training set to let the DNNs learn also extreme combinations, not encountered in calibrations of real market data.}
\label{fig:DNNscatter}
\end{figure}

 
To evaluate the accuracy of the three DNNs to approximate the shifted-SABR model, it is essential to assess their performance on both training, validation, and test sets. 
We show in tab. \ref{tab:TrainingResults} and fig. \ref{fig:DNNscatter} the numerical results of this procedure for our three DNNs together. Specifically, the results reported for the training procedure include their validation using the validation set. 
More details on individual DNNs are reported in app. \ref{app:DNNdetails}. 
\begin{table}[H]
\centering
\begin{tabular}{cccc}
\toprule
\multicolumn{3}{c}{Training Set}                        & Test Set \\
\midrule 
$|\Delta\sigma| >1\%$ & $|\Delta\sigma| >5\%$ & RMSE    & RMSE \\
\midrule 
1\%                   & 0.26\%                & 0.28\%  & 0.25\% \\
\bottomrule
\end{tabular}
\caption{Numerical results of training and test procedure for the three combined DNNs. $\Delta\sigma = \sigma^\textit{DNN}_\textit{SLN} - \sigma^\textit{MC}_\textit{SLN}$ denotes the DNN shifted-Black volatility approximation error.}
\label{tab:TrainingResults}
\end{table}
We observe in tab. \ref{tab:TrainingResults} that both training and test performances in terms of RMSE are quite good with respect to the typical volatility bid-ask spread observed on the EUR Cap/Floor market ($\approx 0.25\%$ for the most liquid options). 
The DNN volatility approximation error $\Delta\sigma$ shown in tab. \ref{tab:TrainingResults} was estimated by a post-training sample analysis. We find that the largest discrepancies are associated with short maturities and extreme strikes, where the MC SABR volatilities show the largest MC errors (defined as the ratio between three MC standard deviations and the shifted-Black vega sensitivity, see app. \ref{app:CalibrationDetails}), and that SABR DNN volatilities fall within the MC errors.
To further check the DNN error, we recalculated the MC SABR volatilities with a larger number of MC scenarios, i.e. $N_{MC} = 2^{20}$, finding lower DNN errors. We conclude that our DNNs consistently learnt the MC SABR volatilities even in the most difficult cases.
\par 
The test plot in fig. \ref{fig:DNNscatter} (right panel) shows that there is no overfitting. The training plot (left panel), however, shows some mismatch between predicted and actual values for the smallest volatilities. This is actually expected, since data in the training and test sets are subject to Monte Carlo error. If the training scatter plots were a perfect line, they would indicate that the DNNs are overfitting the data by fitting even the MC errors. Instead, as observed above, we observe that the DNNs act as filters, extracting true information while discarding noise, which is reflected in the halo observed around the noisier points. 
A closer inspection of these points reveals that they correspond mainly to short maturities and extreme strikes, with the highest MC errors.

\subsection{Volatility Smile Calibration}
\label{sec:CalibrationResults}
Having built the datasets and trained the DNNs, we are now ready to use them to calibrate the SABR model parameters to market quotes, i.e. to solve the optimization problem in eq. \eqref{eq:SABRDNNCalibration}. 
In order to compare the DNN results with those obtained with the \textquote{classic} Hagan et al. approximation, we also solve the corresponding optimization problem in eq. \eqref{eq:SABRHaganCalibration}.
The numerical solution of the calibration problem is influenced by the initial guess of the SABR parameters in eqs. \eqref{eq:SABRDNNCalibration} and  \eqref{eq:SABRHaganCalibration}. We opted for the simple approach of Random Search, considering a large number (300) of initial parameter sets generated randomly and uniformly using Latin Hypercube sampling and the parameters ranges reported in tab. \ref{tab:SABRparametersRanges}. For each initial sample, a local solution of the optimization problem was found using the L-BFGS-B algorithm available in \texttt{scipy.optimize.minimize}, and the best SABR parameter set was identified as the one associated with the minimum value of the objective function. 

\begin{figure}[H]
\begin{subfigure}{0.5\textwidth}
  \centering
  \centerline{\includegraphics[scale=0.45]{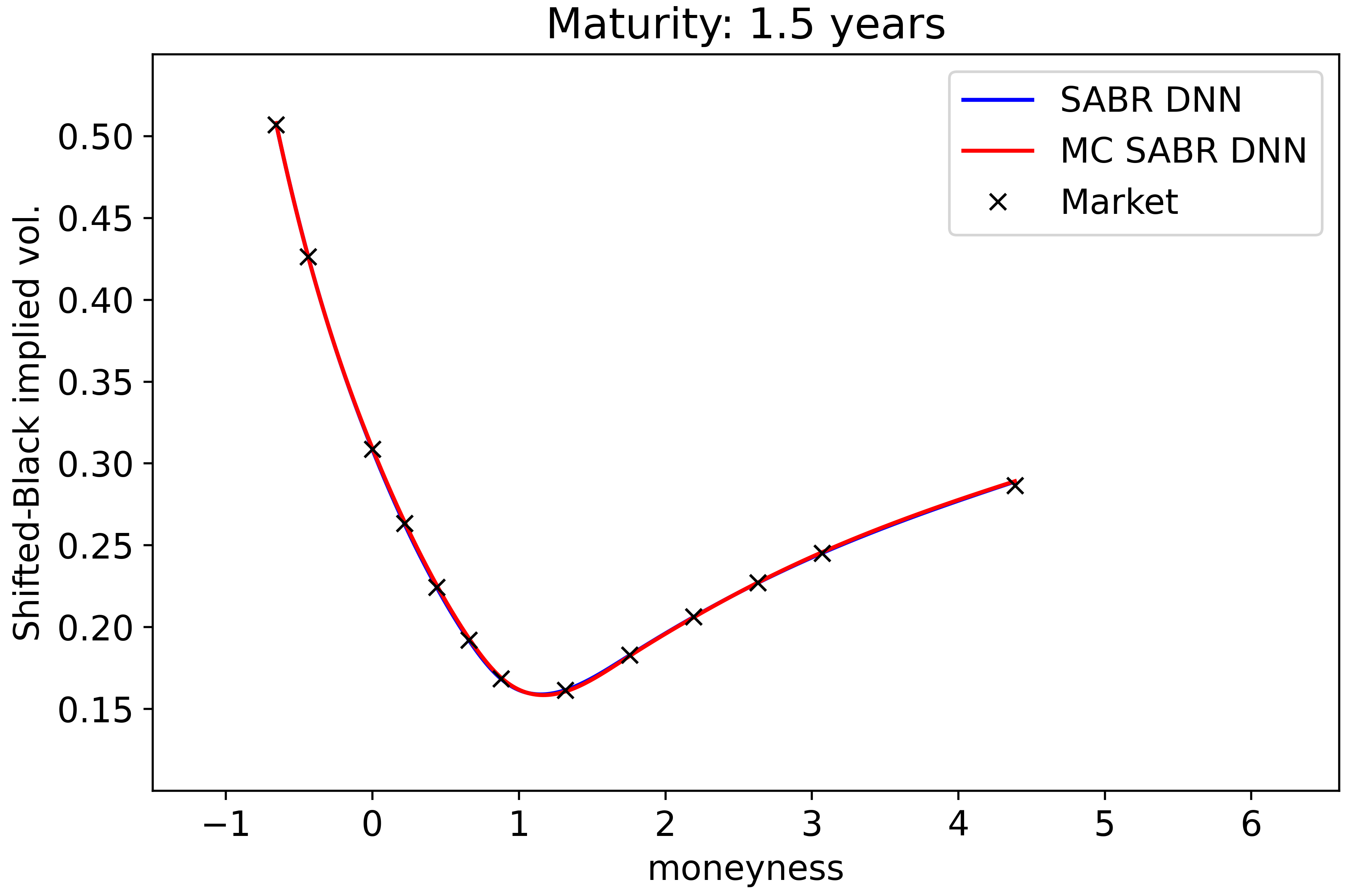}}
\end{subfigure}
\hfill
\begin{subfigure}{0.5\textwidth}
  \centering
\centerline{\includegraphics[scale=0.45]{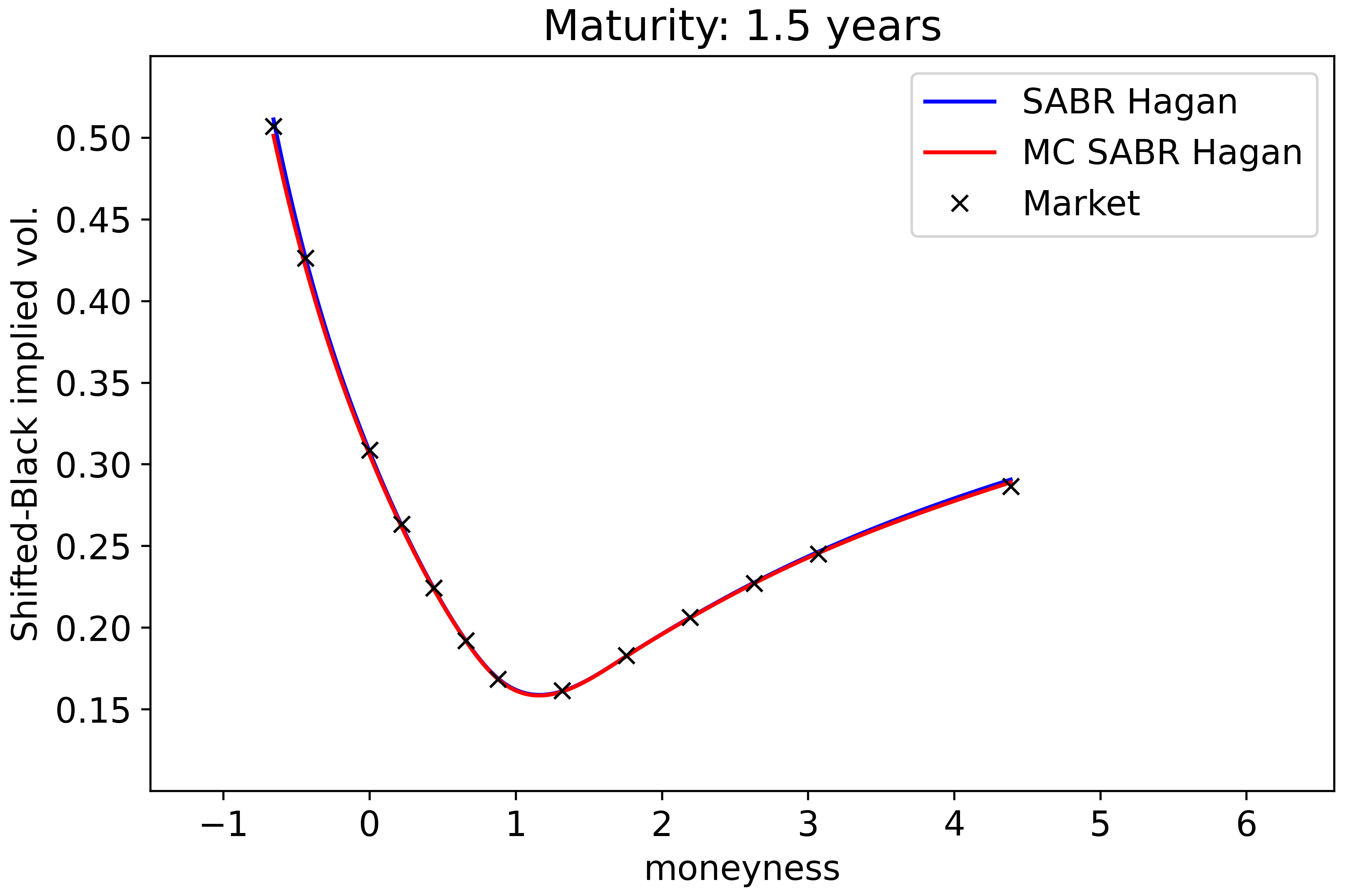}}
\end{subfigure}
\vspace{0.25cm}
\begin{subfigure}{0.5\textwidth}
  \centering
  \centerline{\includegraphics[scale=0.45]{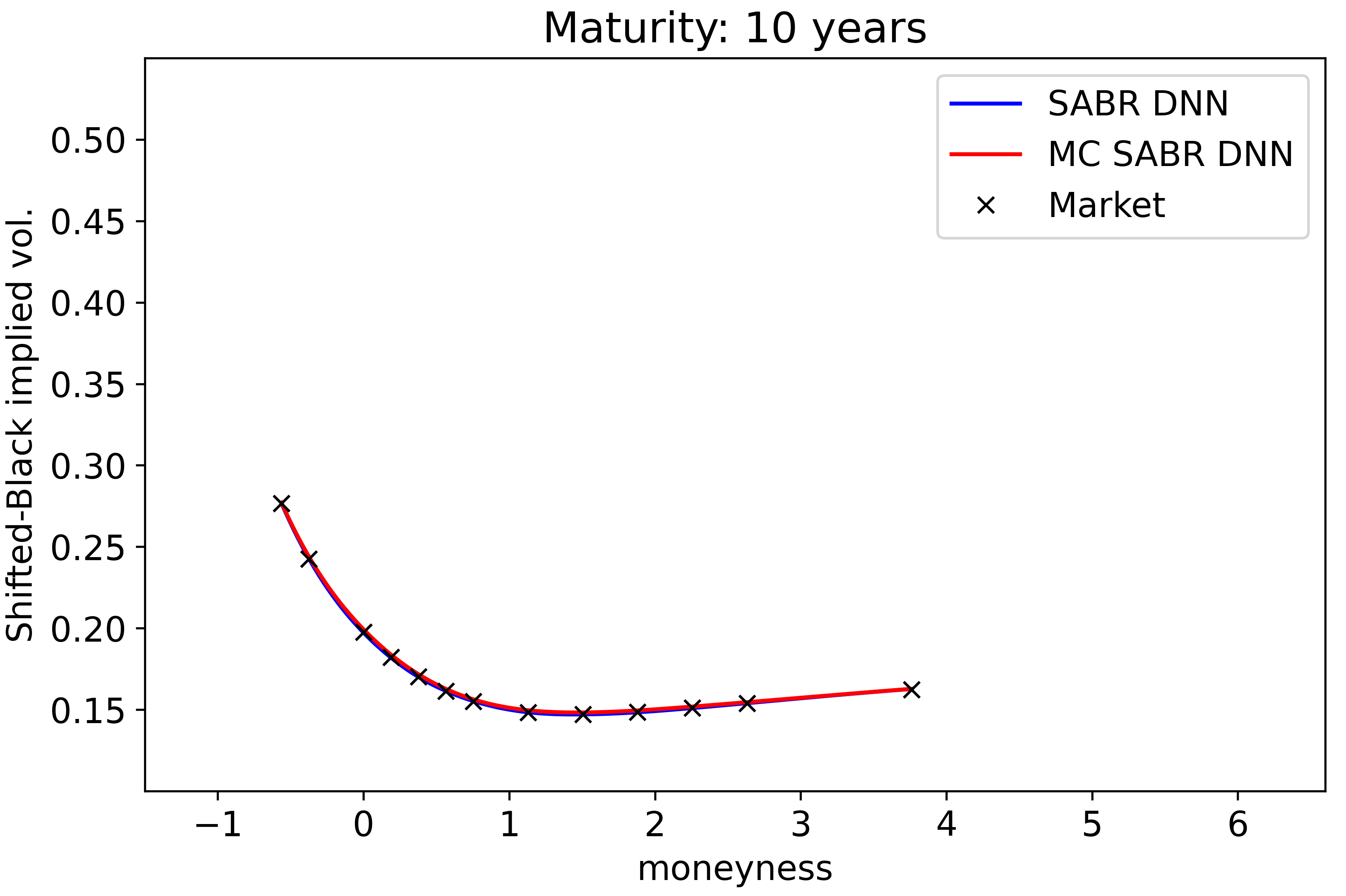}}
\end{subfigure}
\hfill
\begin{subfigure}{0.5\textwidth}
  \centering
\centerline{\includegraphics[scale=0.45]{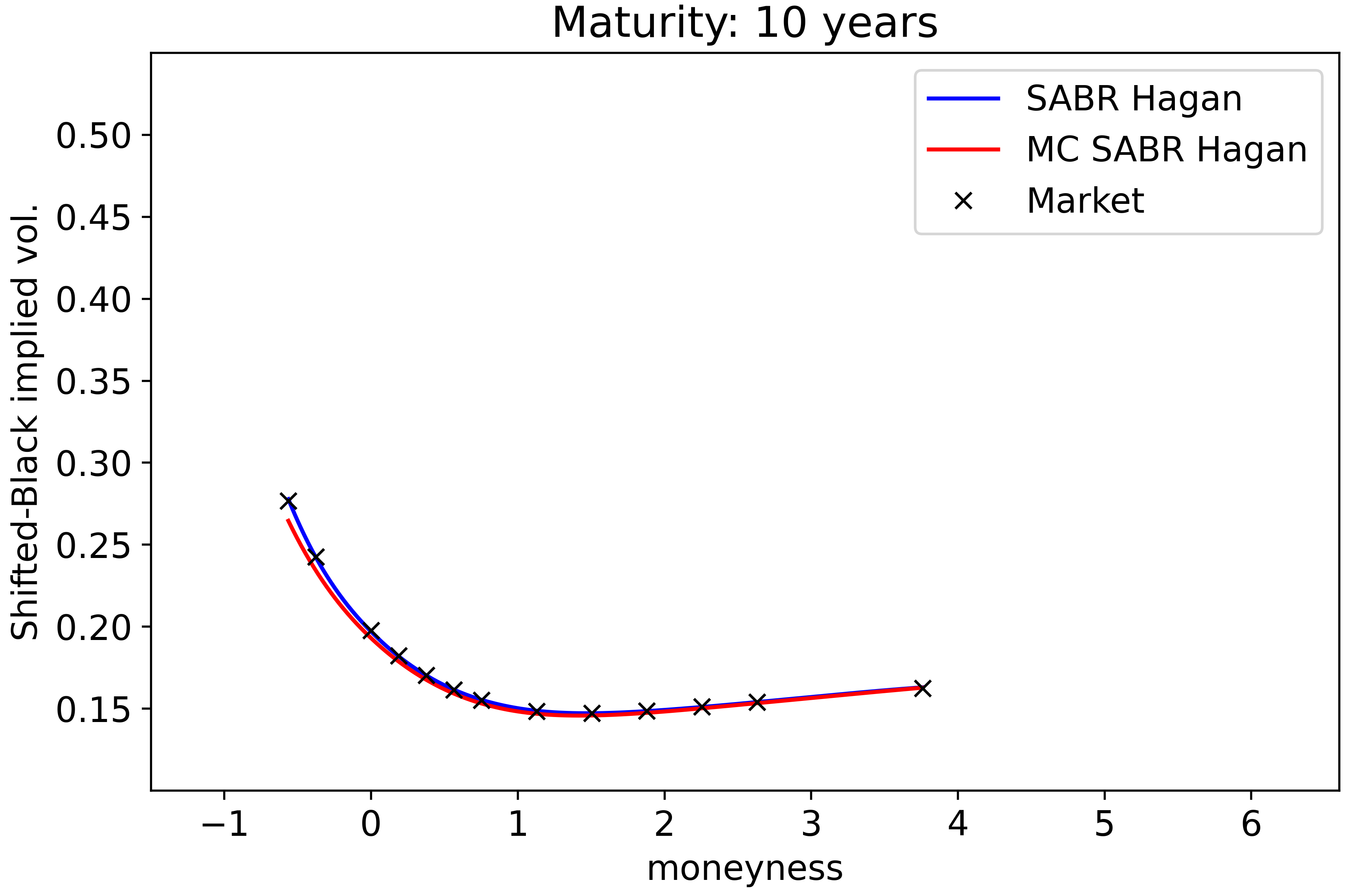}}
\end{subfigure}
\vspace{0.25cm}
\begin{subfigure}{0.5\textwidth}
  \centering
  \centerline{\includegraphics[scale=0.45]{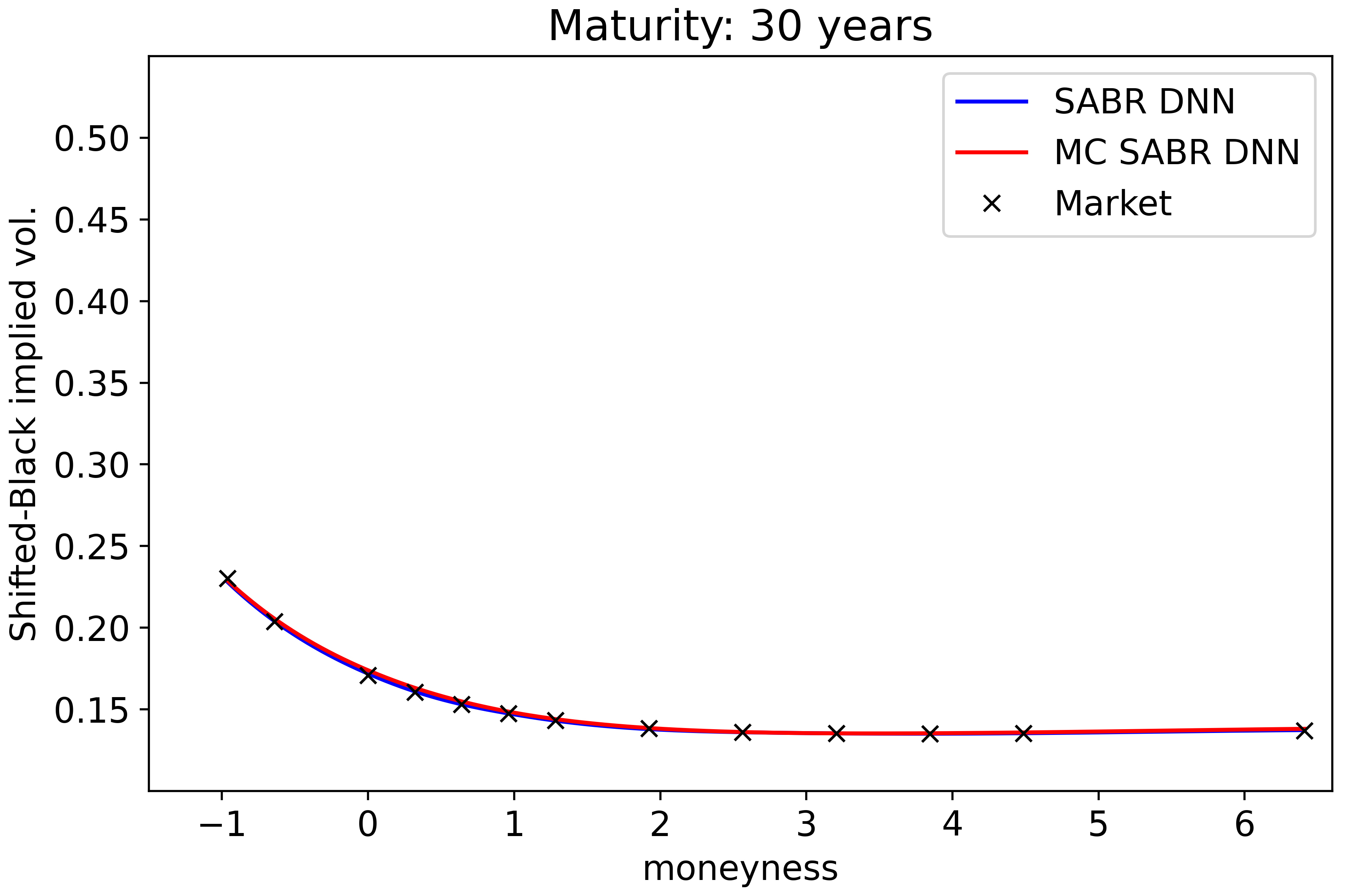}}
\end{subfigure}
\hfill
\begin{subfigure}{0.5\textwidth}
  \centering
\centerline{\includegraphics[scale=0.45]{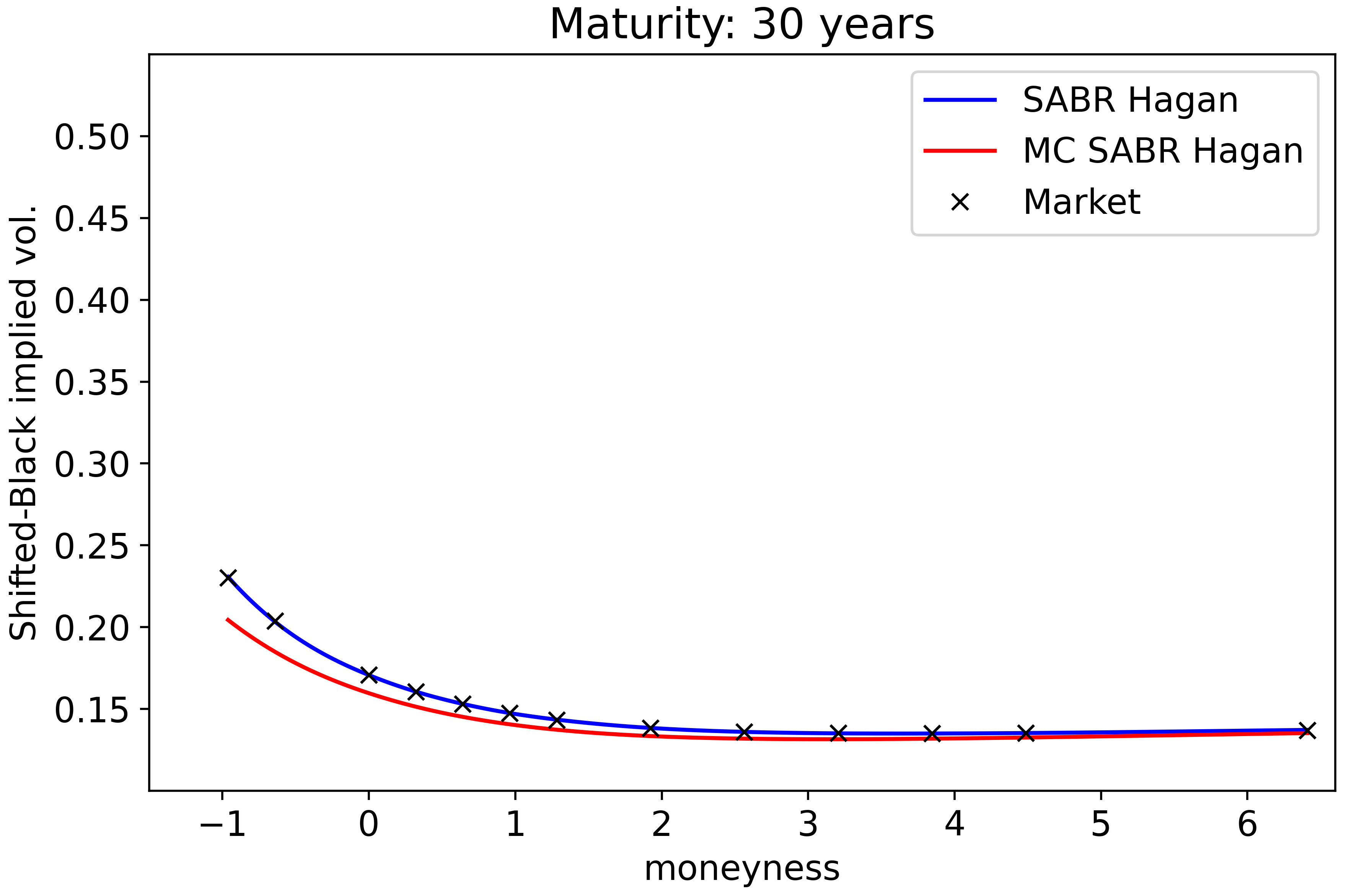}}
\end{subfigure}
\caption{Shifted-SABR model calibrations for short (1.5y, top panel), medium (10y, mid panel) and long (30y, bottom panel) maturities. Black crosses: market data. Blue lines: SABR DNN smile (left) obtained from the SABR parameters calibrated with SABR DNN, and SABR Hagan smile (right) obtained from the SABR parameters calibrated with Hagan et al. approximation.
Red lines: MC SABR DNN smile (left) obtained by MC simulations using the SABR parameters calibrated with SABR DNN, MC SABR Hagan smile (right) obtained by MC simulations using the SABR parameters calibrated with Hagan et al. approximation. High precision MC simulations performed with $2^{24}$ scenarios and time steps as in tab. \ref{tab:TrainingSet}. MC errors not shown being negligible. The same scales are used for all panels to allow direct  comparability. The numerical values corresponding to the smiles depicted in the figure and the respective calibrated parameters are provided in app. \ref{app:CalibrationDetails}.}
\label{fig:smiles}
\end{figure}
\par 
We show in fig. \ref{fig:smiles} the calibration results for three selected short, medium and long maturities, of the Caplet/Floorlet shifted-Black volatility surface derived from EUR Caps/Floors on EURIBOR6M quoted on the over-the-counter (OTC) market as of 30/08/2024 (see app. \ref{app:CapsFloorsQuotes}). 
We observe in fig. \ref{fig:smiles} that both the SABR DNN and SABR Hagan smiles closely match the market smile across all three time horizons. This indicates that both approaches work as effective interpolators of market smiles, i.e. are able to find (different) SABR parameter sets which effectively fit the same market volatilities. 
However, our objective is to assess the DNN performance in calibrating the \textquote{true} model-implied volatility smiles associated with the shifted-SABR model, provided by the unbiased Monte Carlo simulation of the shifted-SABR dynamics. To this end, we compare the SABR DNN and SABR Hagan smiles with those obtained from the MC simulations based on the respective calibrated SABR parameters. The discrepancy between these smiles measures how much the two calibration methodologies are consistent with the true shifted-SABR model. Fig. \ref{fig:smiles} clearly shows that, as the maturity increases, the SABR DNN and MC SABR DNN smiles remain nearly identical, while SABR Hagan smile becomes increasingly inconsistent with the \textquote{true} MC SABR Hagan smile, as expected.
\par 
A more detailed analysis of the volatility smile differences is shown in fig. \ref{RMSD} below, which plots the term structure of the Root Mean Square (relative) Distances (RMSD) 
\begin{equation}
    \textit{RMSD}(T_i) = \sqrt{\frac{1}{N_K}\sum_{j=1}^{N_K}\left[\frac{\sigma(T_i,K_j)}{\sigma^{MC}(T_i,K_j)}-1\right]^2}.
\end{equation}
We observe that, as the maturity increases, the RMSD remains nearly constant for the SABR DNN approach, while diverges for the SABR Hagan approach. More specifically, the right-hand panel shows that the effect is caused essentially by the lowest quoted strike, reaching values higher than those observed in the left-hand panel, while the highest strike works well. This finding confirms that the accuracy of Hagan et al. approximation decreases with increasing maturity and decreasing strikes, as expected.

\begin{figure}[H]
\begin{subfigure}{.5\textwidth}
  \centering
  \centerline{\includegraphics[scale=0.44]{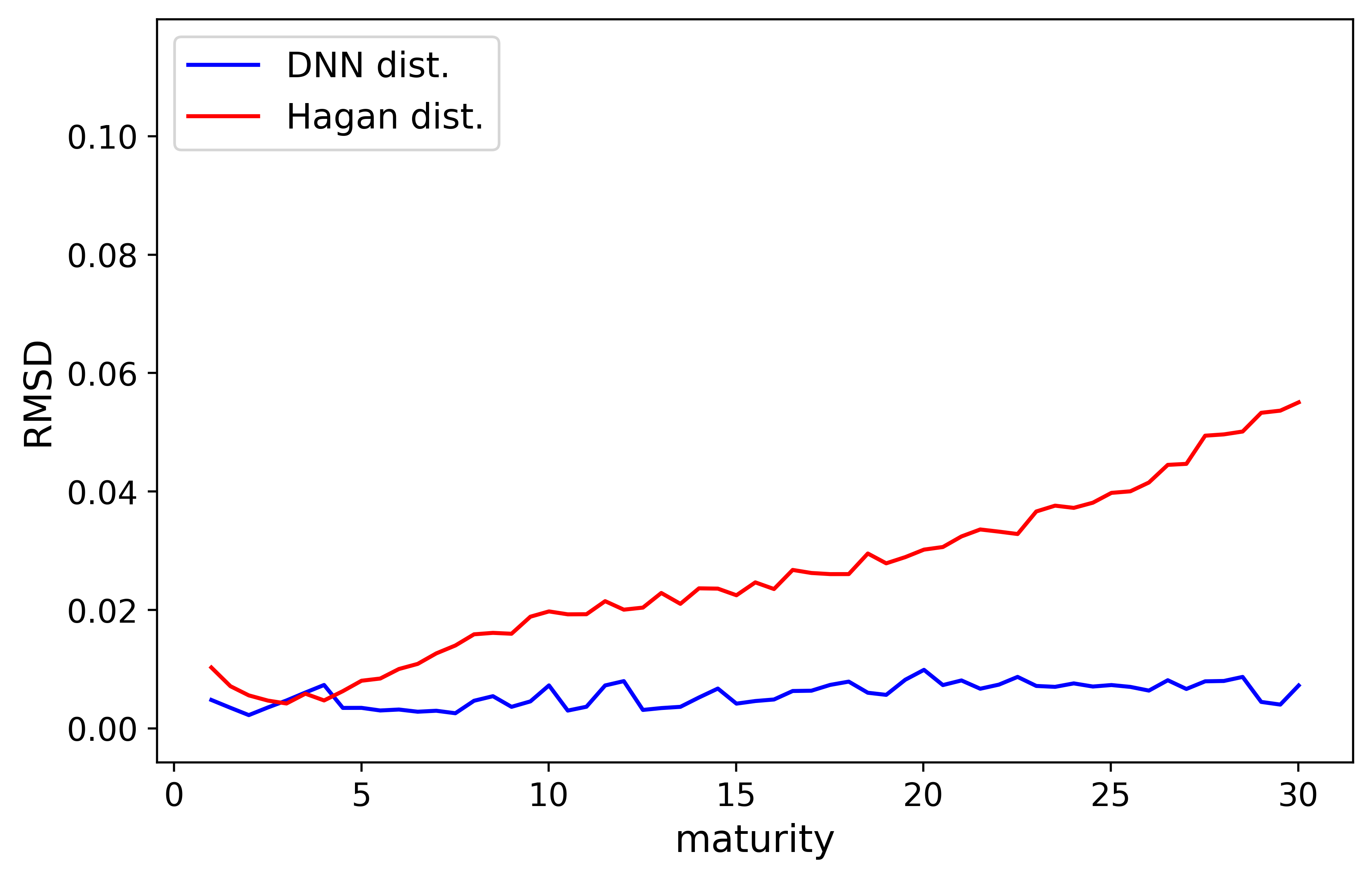}}
\end{subfigure}
\hfill
\begin{subfigure}{.5\textwidth}
  \centering
  \centerline{\includegraphics[scale=0.44]{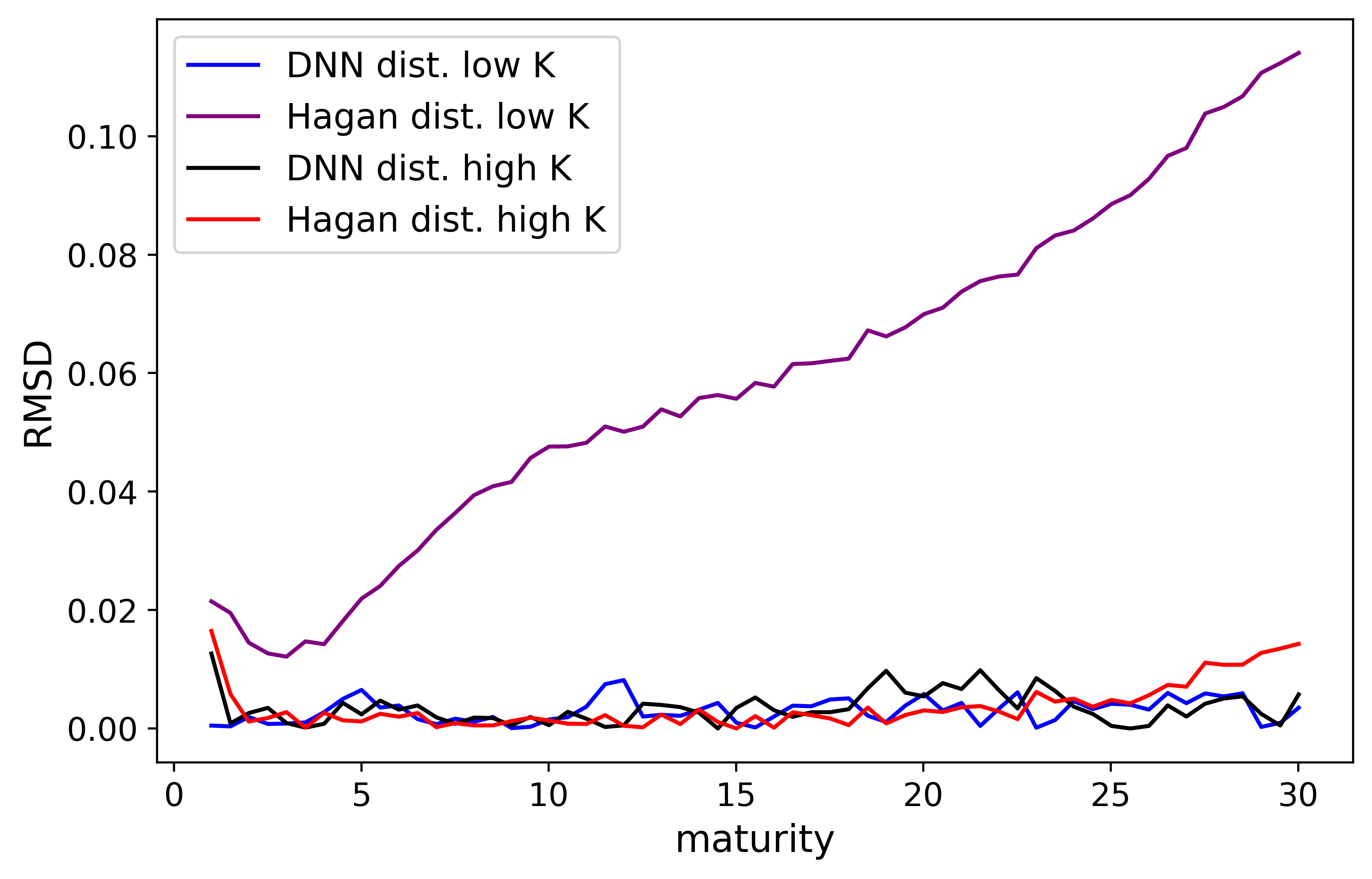}}
\end{subfigure}
\caption{Left panel: root mean square distance (RMSD) plots for both SABR DNN and SABR Hagan smiles as a function of maturity, considering all strikes together. Right panel: focus on the lowest and highest quoted strikes (-1.5\% and 10\%, respectively).}
\label{RMSD}
\end{figure}
\par
\begin{figure}[H]
\begin{subfigure}{.5\textwidth}
    \centering
    \centerline{\includegraphics[scale=0.56]{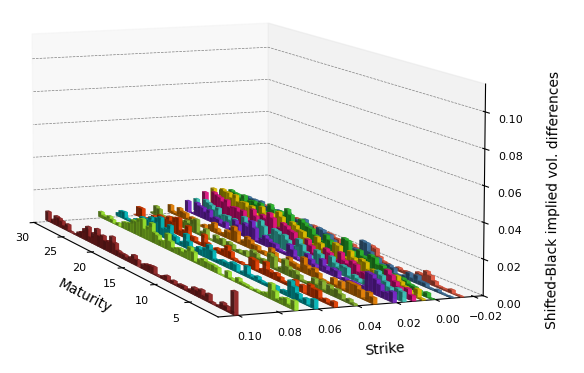}}
    \caption{SABR DNN vs MC SABR DNN.}
    \end{subfigure}
\hfill
\begin{subfigure}{.5\textwidth}
    \centering
    \centerline{\includegraphics[scale=0.55]{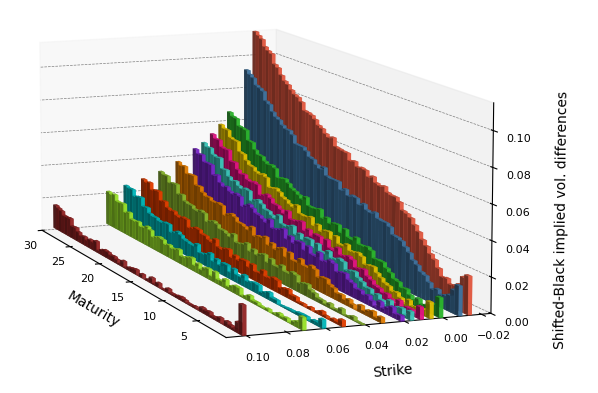}}
    \caption{SABR Hagan vs MC SABR Hagan.}
\end{subfigure}
\caption{3D graphs depicting the absolute relative difference (ARD) between SABR DNN and MC SABR DNN smiles (left-hand panel) and between SABR Hagan and MC SABR Hagan smiles (right-hand panel).}
\label{3D}
\end{figure}

 
Further evidence is shown in fig. \ref{3D}, which reports a granular 3D comparison between SABR DNN and SABR Hagan volatilities as a function of both maturity and strike, using the absolute relative volatility difference (ARD)
\begin{equation}
    \textit{ARD}(T_i, K_j) = \left|\frac{\sigma(T_i,K_j)}{\sigma^{MC}(T_i,K_j)}-1\right|.
\end{equation}
We observe on the left-hand side that the ARD for the SABR DNN approach is nearly constant across maturities and strikes, with a significantly lower magnitude compared to the right-hand side, where the ARD for the SABR Hagan et al. approximation sharply increases with increasing maturity and decreasing strike, reaching approximately $10\%$ in the worst corner (larger maturities, lower strikes).

\par 
The discussion so far regarded the pricing precision of Caplets/Floorlets, whose prices are obtained from Caps/Floors market quotations using the volatility stripping procedure described in app. \ref{app:CapsFloorsQuotes}. We further checked that Caps/Floors, being portfolios of Caplets/Floorlets, are priced within the same precision, once we exclude the further uncertainty introduced by the volatility stripping procedure. We notice that our SABR DNN prices equally well also ATM Caps/Floors, which were not considered in the volatility smile calibration discussed at the beginning of this section.
\par 
We checked our results using other market datasets on different dates, finding consistent results (not reported here to avoid redundancy). Overall, these results confirm that the proposed SABR DNN methodology is consistent with the exact shifted-SABR model, and helps to quantify the increasing inaccuracy of Hagan et al. approximation.

\subsection{Computational Performances}
\label{sec:Performance}
Finally, it is worth to make a remark on the computational time required by the proposed SABR DNN calibration approach. 
\par
The datasets generation discussed in sec. \ref{sec:Datasets} constitutes a high-performance computing task: given the ranges for 
model and contract parameters 
$\left\{\theta^\textit{SABR},\theta^\textit{CF} \right\} = \left\{\hat{\alpha},\beta,\rho,\nu,T,\hat{K}\right\}$ reported in tab. \ref{tab:SABRparametersRanges},
the process of datasets generation required approximately 25,000–30,000 CPU hours, parallelized on 256 CPUs (4 AMD EPYC 7003 processors with 64 cores, 3 GHz each). 
Also the DNN training discussed in sec. \ref{sec:DNNSetup} is not a trivial task, because of the large datasets used. Although the DNNs sizes were contained (see tab. \ref{tab:DNNparameters}), the training process took $\approx 1.5$ GPU hours for each of the three DNNs considered. 
Taking into account also the empirical hyperparameter tuning described in sec. \ref{sec:DNNSetup}, where roughly 3-4 combinations were tested for each network, the total training time for each DNN took $\approx 5$ GPU hours.
However, these processes are conducted offline once and for all, after which the DNNs are ready to be used.
\par 
Contrary to the previous steps, once the datasets are available and the DNNs are trained, the online calibration of the SABR model on the market volatility surface, given the algebraic math of the DNNs, is extremely fast regardless of the hardware used, requiring less than a second per smile.
\par 
In conclusion, the proposed SABR DNN approach allows accurate and fast online calibration of market volatility surfaces while preserving a high computational efficiency.

\subsection{Practical Use of SABR DNN for Caps/Floors}
\label{sec:DNNSABRuse}
The typical usage of the \textquote{classic} SABR model with the Hagan et al. approximation for Caps/Floors is described in app. \ref{app:SABRuse}. 
The SABR DNN discussed in the previous sections allows to completely avoid the Hagan et al. approximation not only in the calibration phase, but also for any pricing purpose. In fact, the SABR DNN allows to compute the shifted-Black implied volatility $\sigma_{SLN}^\textit{DNN}(T,K)$ for any maturity $T$ and strike $K$ using a two-dimensional interpolation scheme based on the SABR DNN in the strike dimension and on interpolation of variances in the maturity dimension using eq. \eqref{eq:VarianceInterpolation}. 
Then, the price of Caplet/Floorlet for maturity $T$ and strike $K$ is computed using the shifted-lognormal (Black) pricing formula, eq.  \eqref{eq:CapletFloorletPriceLogormal}. 
Finally, the Cap/Floor price is obtained as the sum of discounted Caplets/Floorlets as in eq. \eqref{eq:CFPrice}.
We notice that the pricing procedure above is separately applied to each single Caplet/Floorlet included in a Cap/Floor, using the appropriate SABR parameters for each Caplet/Floorlet maturity and the corresponding short, medium, or long maturity DNN.
\par 
Thanks to its pricing and computational performances, the SABR DNN is particularly suitable for situations requiring repeated pricing calls, e.g. pricing of large Caps/Floors portfolios, market and counterparty risk measurement (VaR, stressed VaR, Expected Shortfall, Expected Exposure, etc. even based on Monte Carlo simulations), stress testing, what if analyses, etc.

\section{Conclusions and Directions of Future Work}
\label{sec:Conclusions}

\subsection{Conclusions}
In this paper, we develop and test SABR DNN, a SABR model based on specific Deep Neural Networks (DNNs) trained offline on a very large dataset of interest rate Caplets/Floorlets volatility surfaces consistent with Cap/Floor market quotations. We then use them to calibrate online real market quotations and as a general-purpose Cap/Floor pricer.
\par 
Since DNN performances depend crucially on data, we give particular importance to the initial offline generation of a very large dataset of interest rate volatility surfaces (more than 200 million points) for EUR Caplets/Floorlets including very long maturities (16 points from 1Y to 30Y) and extreme strikes (14 points from -1.5\% to 10\%, including the ATM), consistently with market quotations. 
To this scope, we adapt to the SABR model the random grid approach proposed by \cite{Baschetti2024QF}, taking into account the single-forward nature of SABR (i.e. requiring a different set of SABR parameters for each maturity), and selecting a strike grid independent of maturity. 
\par 
We generate model prices using high-precision unbiased Monte Carlo simulation of the full shifted-SABR stochastic dynamics (without any usage of the Hagan et al. approximation), taking into account possible negative forward rates. In particular, the $\beta$ parameter is not fixed (as done e.g. in \cite{McGhee2021JCF}), but it is calibrated together with the other model parameters, preserving the full flexibility of the original SABR formulation. We still reduce the problem dimensionality recurring to a scaled version of the full SABR dynamics.
\par
We develop, train and optimize offline a SABR DNN architecture specific for Caps and Floors, reaching a very good level of precision. We find that our SABR DNN acts as a filter, extracting true information from the dataset while discarding the noise related to Monte Carlo error.
Once trained, our SABR DNN is able to calibrate online the SABR model parameters to real market volatility surfaces quoted on different business dates with very high precision and computational performance. 
We note that our SABR DNN \textquote{knows} only the forward rate fixing date, not the EURIBOR tenor nor the Caplet/Floorlet maturity date. Hence, once trained, it is able to calibrate Caplet/Floorlet volatilities on any IBOR tenor. 
We conclude that our SABR DNN is robust and can be used in any pricing-intensive situation such as in real-time pricing of large Caps/Floors portfolios, market and counterparty risk measurement, stress testing, what if analyses, etc.
\par 
Furthermore, using our SABR DNN, we challenge the latest version of Hagan et al. analytical approximation \cite{Hagan2016} precisely measuring how much it deteriorates in specific regions of the market volatility surface. The same approach can be adopted to challenge any other SABR approximation. Interestingly, the term structures of the SABR parameters obtained from the SABR DNN calibration display richer shapes than those coming from the Hagan et al. approximation, reflecting the ability of the SABR DNN to learn the \textquote{exact} SABR implied volatility function.
\par 
In conclusion, our results fully address the gaps in the previous machine learning SABR literature in a systematic and self-consistent way, establishing a comprehensive and functional SABR framework that can be adopted by practitioners for daily trading and risk management activities on Cap/Floor options.

\subsection{Directions of Future Work}
\label{sec:FutureWork}
The SABR DNN approach described in this paper can be further developed, experimented, and adapted in a number of ways.
\par 
Regarding the coverage of the interest rate market, first of all one may consider the EUR Swaptions cube, characterized by one additional dimension beyond maturity and strike, i.e. the underlying IRS tenor, which presents a more difficult task for the SABR DNN.
Second, one may consider to join the previous datasets and train the SABR DNN on all possible EUR Caps, Floors and Swaptions together. This step would be very interesting to check the capability of the SABR DNN approach to price consistently any EUR interest rate European option and to detect possible inconsistencies and arbitrage opportunities between the different sections of the market. 
Third, one may consider Caps/Floors/Swaptions on overnight rates, typical of e.g. USD or GBP markets\footnote{After the financial benchmark reform, LIBORs were dismissed and LIBOR markets moved to overnight compounded rates.}. 
The fourth and last step would be to join the previous cases and consider IR options on different rates and currencies.
\par 
The high performance of the SABR DNN calibration enables its test in those cases where multiple SABR model recalibrations are required. A typical example in the context of market risk management would be the calibration of historical or Monte Carlo volatility scenarios encountered in the calculation of market and counterparty risk measurement. 
Further test could be to challenge possible SABR enhancements, such as e.g. those proposed by \cite{Hagan2018VolSurfaces} and \cite{Labordere2010}, using the approach suggested in sec. \ref{sec:CalibrationResults}, fig. \ref{fig:smiles}.
\par 
Regarding the DNNs themselves, one may consider more sophisticated network structures and training strategies. For example, one may introduce control variates in the DNN training, as described e.g. by \cite{kienitz2020cv} and \cite{Funahashi2023QF}. Furthermore, one may consider Derivative-Constrained Neural Networks (DCNNs), including derivatives in the objective function to generate smooth volatility surfaces and incorporate no-arbitrage conditions\footnote{See e.g. \cite{Johnson2009} for no-arbitrage conditions on the Swaption volatility cube.}, as suggested in \cite{Hoshisashi2024}.

\printbibliography[heading=bibintoc,title={References}]

\newpage 
\begin{appendices}

\section{Market Data}
\label{app:MktData}

\subsection{Trading Volumes for OTC Derivatives}
\label{app:TradingVolumes}
We report in tab. \ref{tab:notional_and_gmv} data on trading volumes for OTC derivatives, supporting the SABR model relevance in global financial markets discussed in sec. \ref{sec:SABR_discussion}.
\begin{table}[hbt]
    \centering
    \begin{tabular}{lrrrr}
        \toprule
        & \multicolumn{2}{c}{Notional Amounts} & \multicolumn{2}{c}{Gross Market Value} \\
        \cmidrule{2-5}
        & 2023-S2 & 2024-S2 & 2023-S2 & 2024-S2 \\
        \midrule
        All contracts & 667,058 & 699,476 & 18,122 & 17,615 \\
        \midrule
        Interest rate contracts & 529,813 & 548,341 & 12,783 & 11,364 \\
        \ \ FRAs & 56,023 & 55,105 & 403 & 367 \\
        \ \ Swaps & 425,277 & 446,419 & 11,628 & 10,581 \\
        \ \ Options & 48,288 & 46,156 & 698 & 600 \\
        \ \ Other products & 224 & 205 & 54 & 36 \\
        \midrule
        Foreign exchange contracts & 118,004 & 130,093 & 4,197 & 4,874 \\
        \ \ Outright forwards and forex swaps & 67,797 & 72,827 & 2,196 & 2,553 \\
        \ \ Currency swaps & 36,184 & 38,071 & 1,767 & 1,979 \\
        \ \ Options & 13,999 & 19,171 & 234 & 324 \\
        \ \ Other products & 24 & 24 & - & 18 \\
        \midrule
        Equity-linked contracts & 7,783 & 8,901 & 582 & 662 \\
        \ \ Forwards and swaps & 3,830 & 4,673 & 61 & 68 \\
        \ \ Options & 3,954 & 4,226 & 324 & 332 \\
        \midrule
        Commodity contracts & 2,203 & 2,208 & 301 & 257 \\
        \ \ Forwards and swaps & 1,647 & 1,824 & - & - \\
        \ \ Options & 556 & 383 & - & - \\
        \midrule
        Credit derivatives & 8,708 & 9,229 & 209 & 213 \\
        \midrule
        Other derivatives & 546 & 505 & 50 & 61 \\
        \bottomrule
    \end{tabular}
    \caption{Notional Amounts and Gross Market Values of OTC derivatives traded in the two semesters of 2024. Data in USD billions. Interest rate options result to be the most traded options on the market (600 \$bln in S2-2024). Source: \cite{BISStats}, table DS.1 and DS.2.}
    \label{tab:notional_and_gmv}
\end{table}

\subsection{Caps/Floors Market Quotations}
\label{app:CapsFloorsQuotes}
\begin{figure}[hbt]
    \centering
    \includegraphics[width=1\linewidth]{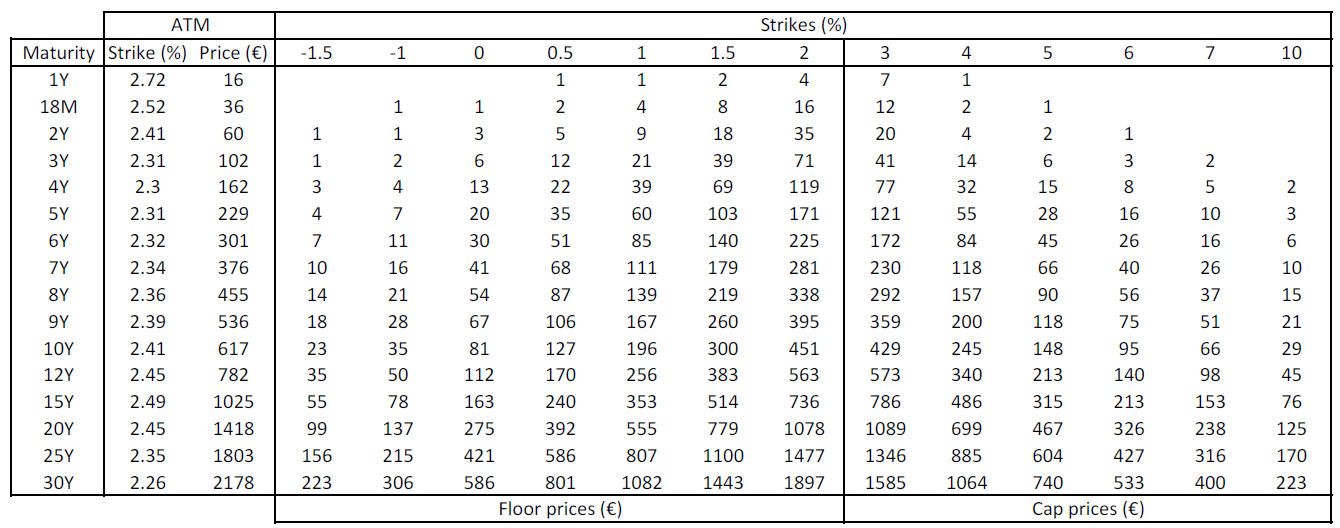}
    \caption{An example of market quotations of EUR Cap/Floor prices as of 30/08/2024. Nominal amount = 10,000\euro. Empty values correspond to negligible prices. Rows for maturities 1Y, 18M, 2Y correspond to options on EURIBOR3M with quarterly cash flows, the other rows to options on EURIBOR6M with semiannual cash flows. The first column reports the option final maturity, i.e. the last cash flow date. At-the-money (ATM) columns reports ATM strikes, lying all between 2\% and 3\%, and prices. The ATM strike is the equilibrium strike which makes Cap and Floor prices equal. Strike columns between -1.5\% and 2\% show out-of-the-money (OTM) Floors, columns between 3\% and 10\% show OTM Caps. Conventionally, the first Caplet/Floorlet is not included in the Cap/Floor, since the underlying rate is already fixed. For example, the Cap 10Y includes 19 Caplets with maturity dates 
    $\{t_1=1Y,t_2=1.5Y,\cdots,t_{19}=10Y\}$. To take into account possible negative rates, the market conventionally assumes a forward rate shift $\lambda=3\%$ to compute shifted-lognormal implied volatilities.}
    \label{fig:QuotedCapsFloors}
\end{figure}

We report here a brief description of over-the-counter (OTC) market quotations for EUR Caps and Floors. As of 30/08/2024, the market quoted a surface of Cap/Floor prices including $N_T = 16$ maturities and $N_K = 13$ strikes (plus one ATM strike) as shown in fig. \ref{fig:QuotedCapsFloors}.
Also other quotations are available, e.g. interest rate Futures Options quoted on exchanges. 
The corresponding Caplet/Floorlet prices can be obtained through a procedure called \textit{volatility stripping} (see e.g. \cite{BBGVolCube2018}), which leverages on the recursive Cap/Floor pricing formula,
\begin{equation}
\textit{CF}(t;\mathbf{T_i},K,\omega) - \textit{CF}(t;\mathbf{T_{i-1}},K,\omega)
= \sum_{t_k>T_{i-1}}^{T_i} P_d(t;t_k) \textit{cf}(t;t_{k-1},t_k,K,\omega),
\end{equation}
where $P_d(t;t_k)$ is the discount factor\footnote{In our definition eq. \eqref{eq:cfPrice} Caplet/Floorlet prices are not discounted.} for date $t_k$, Caplets/Floorlets on the right-hand side cover the time intervals 
$[t_{k-1},t_k]\subset [T_{i-1},T_i]$, and some interpolation scheme is required to fill the gaps. 
A common scheme is linear interpolation of variances, i.e.
\begin{align}
v(T,K) = \frac{t_k-T}{t_k-t_{k-1}}v(t_{k-1},K) + \frac{T-t_{k-1}}{t_k-t_{k-1}}v(t_k,K),\quad
\text{where}\quad v(T,K)=\sigma^2(T,K)(T-t)
\label{eq:VarianceInterpolation}
\end{align}
where $T\in[t_{k-1}, t_k]$ are Caplet/Floorlet fixing (not maturity) dates, and $K\in\{K_1,\cdots, K_{N_K}\}$. 
For example, from the quoted prices of two Caps on EURIBOR6M with maturities 9Y and 10Y, one may obtain the price of two Caplets insisting on [9Y,9.5Y] and [9.5Y,10Y] by interpolating the Caplet variance at 9Y (the fixing date of the first Caplet) between the Caplet variances at 8.5Y (available from the previous stripping step) and 9.5Y (the fixing date of the second Caplet).
The resulting Caplet/Floorlet surface has a semi-annual maturity grid with $N'_T=60$ rows (a.k.a. smile sections), according to the EURIBOR6M tenor, and $N_K=14$ strike columns, including the ATM.
Implied volatility surfaces for other EURIBOR tenors, i.e. 1M, 3M and 12M, require more sophisticated construction algorithms, see e.g. \cite{BrigoMercurio2006}, \cite{Kienitz2013SSRN} for further details.
\par 
We note that the volatility stripping procedure, based on interpolations and other modelling hypotheses (for non-quoted IBOR tenors), introduces a further level of uncertainty on Caplet/Floorlet prices with respect to Cap/Floor quotations. 
Caps/Floors on overnight rates, typical of other markets (e.g. USD on SOFR, GBP on SONIA, etc.) have simpler structures, including one single Caplet/Floorlet per year and one single tenor, leading to less Caplet/Floorlet price uncertainty.

\section{SABR Details} 
\label{app:SABRdetails}
In this appendix we report some details regarding the shifted-SABR model used in this paper.

\subsection{Hagan et al. Approximation}
\label{app:Hagan}
The Hagan et al. analytical approximation of the SABR implied volatility was originally published in \cite{Hagan2002} and improved in \cite{Obloj2008}. 
The most recent version for the shifted-SABR model used in this paper was introduced in \cite{Hagan2016}, where the shifted-SABR implied normal volatility $\sigma_N ^{SABR}$ is given by 
\begin{gather} 
\sigma_N^\textit{SABR}(t;T,\bar{F},\bar{K},\theta) = \nu(\bar{K}-\bar{F}) \frac{Z(z)}{Y(z)},\;
F \neq K,\\
z := \frac{\nu}{\bar{\alpha}} \left\{ 
\begin{array}{ll}
\frac{\bar{K}^{1-\beta} - \bar{F}^{1-\beta}}{1-\beta} & \text{for} \hspace{0.2cm} \beta \neq 1,\\
\mathrm{ln}\frac{\bar{K}}{\bar{F}} & \text{for} \hspace{0.2cm} \beta = 1,\\
\end{array}\right. \hspace{0.5cm} 
Z(z) := \left\{ 
\begin{array}{ll}
1 + \Theta(z) \tau(t,T) & \text{for} \hspace{0.2cm} \Theta(z) \geq 0,\\
\left[1  - \Theta(z) \tau(t,T)\right]^{-1} & \text{for} \hspace{0.2cm} \Theta(z) < 0,\\
\end{array}\right. \\
\Theta(z) := \frac{\nu^2}{24}\left[ -1 +3 \frac{z + \rho -\rho E(z)}{Y(z)E(z)}\right] + \frac{\bar{\alpha}^2 \Delta_0}{6}\left[ 1- \rho^2 + \frac{(z+\rho)E(z) - \rho}{Y(z)}\right],\\
Y(z) := \ln\frac{z + \rho + E(z)}{1+\rho} , \hspace{0.5cm} E(z) := \sqrt{1+2\rho z + z^2},\\
\bar{\alpha} := \alpha \left[ 1+ \frac{1}{4}\alpha\beta\rho\nu \bar{F}^{\beta-1} \tau(t,T) \right], \hspace{0.5cm} 
\Delta_0 := -\frac{\beta(2-\beta)}{8\bar{F}^{2-2\beta}},
\end{gather}
where $t$ is the valuation time, $T$ the forward rate fixing date (\textit{not} the option's maturity date), $\tau(t,T)$ the year fraction for the time interval $[t,T]$ (time to fixing), $\bar{F}:=\bar{F}(t)$ and $\bar{K}$ are the shifted forward rate at pricing time $t$ and the shifted strike, respectively, and $\theta=\{\alpha,\beta,\rho,\nu\}$ stands for the four SABR parameters\footnote{Note that we keep here a generic valuation time $t$ instead of setting $t=0$, consistently with Caplet/Floorlet and Caps/Floors pricing expressions in sec. \ref{sec:CF}, and we denote with $F:=F(t)$ the initial forward rate at time $t<T$. Instead in the SABR dynamics in eqs. \eqref{eq:SABRFwdDynamics} -- \eqref{eq:SABRconditions} we denote the initial forward rate at time $t=0$ by $F(0)=F_0$.}. 
For ATM options with $F = K$ we have
\begin{gather}
\sigma_N^\textit{SABR}(t;T,\bar{F},\bar{F},\theta)
= \bar{\alpha} \bar{F}^\beta Z_\textit{ATM},\quad
\Theta_\textit{ATM} := \frac{\nu^2}{24}(2-3\rho^2) + \frac{\bar{\alpha}^2\Delta_0}{3},\quad F=K.
\end{gather}
\par 
The shifted-SABR implied shifted-lognormal volatility can be implied from shifted-SABR prices using numerical inversion. Alternatively, one can use the approximated conversion formula for $F \neq K$
\begin{equation}\label{hagan}
\sigma_{LN}^\textit{SABR} (t;T,\bar{F},\bar{K},\theta) 
= \sigma_{N}^\textit{SABR}(t;T,\bar{F},\bar{K},\theta) 
\frac{\mathrm{ln}\frac{\bar{F}}{\bar{K}}}{\bar{F} -\bar{K}} 
\left[ 1+ \frac{\sigma^\textit{SABR}_N(t;T,\bar{F},\bar{F},\theta)^2 \tau(t,T) }{24\bar{F} \bar{K}} \right],
\end{equation}
while for ATM options with $F = K $, the approximation remains the one derived in \cite{Hagan2002},
\begin{equation}\label{haganATM}
    \sigma_{LN}^\textit{SABR} (t;T,\bar{F},\bar{F},\theta) = \frac{\alpha}{\bar{F}^{1-\beta}}\left\{ 1+ \left[ \frac{\alpha^2 (1-\beta)^2}{24 \bar{F}^{2-2\beta}} +\frac{\alpha\beta\rho\nu}{4 \bar{F}^{1-\beta}} + \nu^2 \frac{2-3\rho^2}{24}\right]\tau(t,T)\right\}.
\end{equation}
\par
Once the shifted-SABR implied volatility has been obtained using the Hagan et al. approximation, the (undiscounted) price of European Caplet/Floorlet options in eq. \eqref{eq:cfPrice} is simply given by either the shifted-lognormal (Black) or the normal (Bachelier) formulas\footnote{For the sake of simplicity, in this paper we do not use the further corrections proposed in \cite{Hagan2016} (secs. 1.4 and 1.5), but our framework is fully general and they could be easily included.},
\begin{align}
\textit{cf}(t;T_{i-1},T_i,K,\omega)
&= \mathbb{E}_t^{Q_d^{T_{i}}} \left\{\textit{Max}\left\{\omega\left[L(T_{i-1},T_i)-K\right]\right\}\right\}\tau(T_{i-1},T_i)\nonumber\\
&= \textit{Black}\left[\bar{F}_i(t),T_i,\bar{K},v_{LN}^\textit{SABR}(t;T_{i-1}),\omega]\tau(T_{i-1},T_i)\right] \label{eq:CapletFloorletPriceLogormal}\\
&= \textit{Bach}\left[\bar{F}_i(t),T_i,\bar{K},v_{N}^\textit{SABR}(t;T_{i-1}),\omega]\tau(T_{i-1},T_i)\right],
\label{eq:CapletFloorletPriceNormal}
\end{align}
where the SABR variances $v_{x}^\textit{SABR}(t;T)$ are given by
\begin{equation}
v_{x}^\textit{SABR}(t;T) 
= v_{x}^\textit{SABR} (t;T,\bar{F},\bar{K},\theta) 
= \sigma_{x}^\textit{SABR}(t;T,\bar{F},\bar{K},\theta)^2\tau(t,T).
\end{equation}

\subsection{Practical Use of SABR for Caps/Floors}
\label{app:SABRuse}
We describe here the typical application of the SABR model with the Hagan et al. approximation to the Cap/Floor market, based on the following steps.
\begin{enumerate}
\item \textbf{Volatility stripping}: from market prices of Caps/Floors obtain the corresponding Caplet/Floorlet normal or shifted-lognormal implied volatility $N_T \times N_K$ surface using the volatility stripping procedure discussed in app. \ref{app:CapsFloorsQuotes} before.
\item \textbf{Shifted-SABR calibration}: for each smile section (i.e. maturity row $t_k, k=1,\cdots,N_T$) in the Caplet/Floorlet volatility surface, corresponding to forward rate $F_k(t):=F(t;t_{k-1},t_k)$, calibrate a separate shifted-SABR model using the Hagan et al. approximate eq. discussed in app. \ref{app:Hagan} before, and obtain the corresponding SABR parameters $\{\alpha_k,\beta_k,\rho_k,\nu_k\}$.
\item \textbf{Market repricing check}: given the set of SABR parameters $\{\alpha_k,\beta_k,\rho_k,\nu_k\}_{k=1}^{N_T}$, reprice the market Caps/Floors and check the difference with respect to their corresponding market prices, to estimate and monitor the repricing error deriving from the numerical procedures used for volatility stripping and SABR calibration.
\item \textbf{Pricing}: estimate the implied volatility $\sigma(T,K)$ for any off-grid maturity $T$ and strike $K$ using a two-dimensional interpolation scheme based on the Hagan et al. approximate formula in the strike dimension and on interpolation of variances in the maturity dimension using formula \eqref{eq:VarianceInterpolation}.  
Finally, price Caplets/Floorlets and Caps/Floors with any maturity $T$ and strike $K$ from the corresponding implied volatilities using the normal or shifted-lognormal pricing formulas.
\end{enumerate}
We note that the Hagan et al. approximation is used both in the calibration and pricing phases. 
An extensive analysis of SABR performance with Caps/Floors can be found in \cite{Wu2012}.

\subsection{Derivation of the Scaled Shifted-SABR Model} 
\label{app:ScaledSABRmodel}
We derive here the scaled shifted-SABR model reported in sec. \ref{sec:ScaledSABRmodel}, eqs. \ref{eq:XSABRFwdDynamics}-\ref{eq:XSABRcorrelation}.
We start from the forward rate dynamics in eq. \eqref{eq:SABRFwdDynamics} and we introduce the scaled stochastic process defined as  
\begin{equation}
X(t) := \frac{\bar{F}(t)}{\bar{F}_0}.
\end{equation}
The corresponding dynamics of $X(t)$ and $\hat{\sigma}(t)$ can be derived by applying Itô's Lemma,
\begin{align}
&dX(t) = \hat{\sigma}(t) X(t)^\beta dW(t), \quad X(0) = 1,\\
&\hat{\sigma}(t) := \sigma(t) \bar{F}_0^{\beta-1},\\
&d\hat{\sigma}(t) 
    = \bar{F}_0^{\beta-1} \nu \sigma(t) dZ(t) 
    = \nu \hat{\sigma}(t) dZ(t), \quad \hat{\sigma}(0) = \alpha \bar{F}_0^{\beta-1} :=\hat{\alpha}.
\end{align}  
We observe that the scaled volatility dynamics has the same form of the original volatility dynamics in the shifted-SABR model in eq. \eqref{eq:SABRVolDynamics}. 
Consequently, the shifted-lognormal implied volatility of a Caplet/Floorlet with contract parameters 
$\theta^\textit{CF}=\left\{T,\bar{K}\right\}$ and model parameters 
$\theta^\textit{SABR}=\left\{\bar{F}_0,\alpha,\beta,\rho,\nu\right\}$ 
coincides with the shifted-lognormal implied volatility of a Caplet/Floorlet with contract parameters 
$\theta^\textit{CF}=\left\{T,\hat{K}=\frac{\bar{K}}{\bar{F}_0}\right\}$ and model parameters 
$\theta^\textit{SABR}=\left\{1,\hat{\alpha} = \alpha \bar{F}_0 ^ {\beta -1},\beta,\rho,\nu\right\}$. 
The prices of the two options differ by a multiplicative deterministic factor $\bar{F}_0$, but the associated shifted-lognormal implied volatilities coincide, since they are related to the price through the shifted-Black formula.

\section{Dataset Details} 
\label{app:DatasetDetails}
In this appendix, we report some results that can serve as benchmarks for future research, as well as details regarding the analysis we conducted on the Monte Carlo pricing error. 

\subsection{Benchmark Caplet/Floorlet Prices}
\label{app:Benchmarks}
In this section we provide two samples of Caplet/Floorlet prices\footnote{For the sake of simplicity, we omit the year fraction $\tau(T_{i-1},T_i)$ in eq. \eqref{eq:cfPrice}.} computed by Monte Carlo simulation as described in sec.~\ref{sec:MC} and using the SABR dynamics defined in \eqref{eq:MCXSABRFwdDynamics} and \eqref{eq:MCXSABRVolDynamics}. 
In tab. \ref{tab:params} we show the two parameter sets used in the simulation, and in the two tables \ref{tab:caseI} and \ref{tab:caseII} we report the corresponding option prices.  
These results can serve as benchmarks for future research in this field. 
\begin{table}[H]
\centering
\begin{tabular}{ccccccccc}
\toprule 
Case & $F_0$ & $\alpha$ & $\beta$ & $\rho$ & $\nu$ & $\lambda$ & $N_{MC}$ & $\Delta_{MC}$ (days)\\
\midrule
I   & 1 & 0.1178 & 0.8738 & -0.0702 & 0.5010 & 3\% & $2^{20}$ & 0.5 \\
II  & 1 & 0.1822 & 0.3044 &  0.1243 & 0.3127 & 3\% & $2^{20}$ & 0.5 \\
\bottomrule  
\end{tabular}
\caption{Shifted-SABR parameters used to compute the MC prices reported in the following tabs. \ref{tab:caseI} and \ref{tab:caseII}.}
\label{tab:params}
\end{table}

\begin{table}[H]
\centering
\begin{tabular}{cccc}
\toprule 
\centering Maturity (y) & Strike (K) & Floorlet price & Caplet price\\
\midrule
\centering \multirow{10}{*}{2} & 0.5 & 0.00063 $\pm$ 0.00006 & 0.50045 $\pm$ 0.00057\\
& 0.6 & 0.00177 $\pm$ 0.00005 & 0.40159 $\pm$ 0.00056\\
& 0.7 & 0.00476 $\pm$ 0.00009 & 0.30458 $\pm$ 0.00054\\
& 0.8 & 0.01249 $\pm$ 0.00014 & 0.21231 $\pm$ 0.00051\\
& 0.9 & 0.03132 $\pm$ 0.00022 & 0.13114 $\pm$ 0.00045\\
& 1.0 & 0.07070 $\pm$ 0.00031 & 0.07052 $\pm$ 0.00038\\
& 1.1 & 0.13494 $\pm$ 0.00040 & 0.03476 $\pm$ 0.00030\\
& 1.2 & 0.21742 $\pm$ 0.00046 & 0.01724 $\pm$ 0.00023\\
& 1.3 & 0.30925 $\pm$ 0.00050 & 0.00907 $\pm$ 0.00018\\
& 1.4 & 0.40530 $\pm$ 0.00052 & 0.00511 $\pm$ 0.00015\\
\midrule 
\centering \multirow{10}{*}{10} & 0.5 & 0.02865 $\pm$ 0.00030 & 0.52680 $\pm$ 0.00349\\
& 0.6 & 0.04096 $\pm$ 0.00038 & 0.43911 $\pm$ 0.00347\\
& 0.7 & 0.05776 $\pm$ 0.00046 & 0.35590 $\pm$ 0.00346\\
& 0.8 & 0.08130 $\pm$ 0.00055 & 0.27944 $\pm$ 0.00344\\
& 0.9 & 0.11498 $\pm$ 0.00064 & 0.21312 $\pm$ 0.00342\\
& 1.0 & 0.16260 $\pm$ 0.00073 & 0.16074 $\pm$ 0.00340\\
& 1.1 & 0.22549 $\pm$ 0.00081 & 0.12357 $\pm$ 0.00337\\
& 1.2 & 0.30054 $\pm$ 0.00088 & 0.09868 $\pm$ 0.00335\\
& 1.3 & 0.38368 $\pm$ 0.00093 & 0.08183 $\pm$ 0.00333\\
& 1.4 & 0.47181 $\pm$ 0.00098 & 0.06995 $\pm$ 0.00332\\
\bottomrule  
\end{tabular}
\caption{Monte Carlo Caplet/Floorlet prices (undiscounted and without the year fraction in eq. \eqref{eq:cfPrice}) obtained using case I parameters in tab. \ref{tab:params}. The Monte Carlo errors are computed as three standard deviations. We note that the strike $K$ is equal to the moneyness $K/F_0$ since in this case $F_0=1$.}
\label{tab:caseI}
\end{table}

\begin{table}[H]
\centering
\begin{tabular}{cccc}
\toprule
\centering Maturity (y) & Strike (K) & Floorlet price & Caplet price\\
\midrule
\centering \multirow{10}{*}{2} & 0.5 & 0.00256 $\pm$ 0.00007 & 0.50234 $\pm$ 0.00079\\
& 0.6 & 0.00643 $\pm$ 0.00011 & 0.40620 $\pm$ 0.00077\\
& 0.7 & 0.01493 $\pm$ 0.00016 & 0.31470 $\pm$ 0.00073\\
& 0.8 & 0.03166 $\pm$ 0.00024 & 0.23143 $\pm$ 0.00068\\
& 0.9 & 0.06081 $\pm$ 0.00034 & 0.16059 $\pm$ 0.00060\\
& 1.0 & 0.10538 $\pm$ 0.00044 & 0.10515 $\pm$ 0.00051\\
& 1.1 & 0.16571 $\pm$ 0.00053 & 0.06549 $\pm$ 0.00042\\
& 1.2 & 0.23953 $\pm$ 0.00061 & 0.03930 $\pm$ 0.00034\\
& 1.3 & 0.32327 $\pm$ 0.00067 & 0.02304 $\pm$ 0.00026 \\
& 1.4 & 0.41359 $\pm$ 0.00071 & 0.01336 $\pm$ 0.00020\\
\midrule
\centering \multirow{10}{*}{10} & 0.5 & 0.05966 $\pm$ 0.00044 & 0.56085 $\pm$ 0.00198\\
& 0.6 & 0.08183 $\pm$ 0.00053 & 0.48303 $\pm$ 0.00193\\
& 0.7 & 0.11008 $\pm$ 0.00063 & 0.41128 $\pm$ 0.00187\\
& 0.8 & 0.14556 $\pm$ 0.00074 & 0.34677 $\pm$ 0.00181\\
& 0.9 & 0.18917 $\pm$ 0.00084 & 0.29036 $\pm$ 0.00174\\
& 1.0 & 0.24118 $\pm$ 0.00094 & 0.24237 $\pm$ 0.00167\\
& 1.1 & 0.30127 $\pm$ 0.00103 & 0.20246 $\pm$ 0.00160\\
& 1.2 & 0.36860 $\pm$ 0.00112 & 0.16979 $\pm$ 0.00154\\
& 1.3 & 0.44208 $\pm$ 0.00120 & 0.14327 $\pm$ 0.00147\\
& 1.4 & 0.52059 $\pm$ 0.00127 & 0.12178 $\pm$ 0.00141\\
\bottomrule 
\end{tabular}
\caption{As in tab. \ref{tab:caseI}, using case II parameters in tab. \ref{tab:params}.}
\label{tab:caseII}
\end{table}

\subsection{Monte Carlo Error Analysis}
\label{app:MCErrorAnalysis}
As seen in the benchmark prices in tabs. \ref{tab:caseI} and \ref{tab:caseII}, we observe that Monte Carlo pricing errors are systematically smaller for out–of–the–money (OTM) short term (2Y) options. This is explained by the fact that, when an option is OTM, most simulated payoffs are zero, with only a small fraction being positive and close to zero, leading to a low variance in the payoff distribution. Since the MC error scales with the payoff’s standard deviation, OTM options exhibit smaller errors than in–the–money (ITM) options.
However, this observation does not hold for longer maturity options (10Y), where ITM Floorlet options systematically exhibit lower Monte Carlo errors. This effect occurs when volatility of volatility $\nu$, initial volatility $\alpha$, correlation $\rho$, and/or time to maturity assume higher values, i.e. in higher-uncertainty scenarios, and can be explained by the asymmetric nature of MC price paths: upward price movements are theoretically unlimited, as the forward rate can rise without bound, whereas downward price movements are limited, since the forward rate cannot fall below the rate shift $\lambda$. As a consequence, we have the following more complex behaviour of the MC price simulation.
\begin{itemize}
\item Caplets: some MC price paths of OTM Caplets may occasionally end up deep ITM, when the forward rate rises significantly. These rare but very large payoffs create a long right tail in the payoff distribution, increasing its MC error.
\item Floorlets: their payoff is capped by the (shifted) strike, because the forward rate cannot fall below the shift $\lambda$. Even in higher-uncertainty scenarios, MC price paths of OTM Floorlets are limited, leading to lower MC errors compared to the corresponding Caplets.
\end{itemize}
\par 
This behaviour can be analytically examined in the limit case of the shifted-Black model, where the Caplet/Floorlet variances (undiscounted and omitting the year fraction in eq. \eqref{eq:cfPrice}) are given by
\begin{multline} 
\textit{Var}(t;T,\bar{F},\bar{K},v,\omega) 
= \mathbb{E}_t^{\mathbb{Q}_T}\left[\max\left[\omega\left(\bar{F}(T) -\bar{K}\right);0\right]^2\right] -
\mathbb{E}_t^{\mathbb{Q}_T}\left[\max\left[\omega\left(\bar{F}(T) -\bar{K}\right);0\right]\right]^2 \\
= \bar{F}^2(t) e^{v}\Phi(\omega d_0) + \bar{K}^2 \Phi(\omega d_{-}) - 2\bar{F}(t)\bar{K}\Phi(\omega d_{+}) 
-\left[\bar{F}(t) \Phi(\omega d_{+}) - \bar{K} \Phi(\omega d_{-})\right]^2 \\
= \bar{F}^2(t) e^{v}\Phi(\omega d_0) - \bar{F}(t)\bar{K}\Phi(\omega d_{+}) 
-\bar{K}\omega V(t;T,\bar{F},\bar{K},v,\omega) - V(t;T,\bar{F},\bar{K},v,\omega)^2,
\label{eq:shiftedBlackVar}
\end{multline}
where $\mathbb{Q}_T$, is the $T-$forward probability measure, $\sigma$ is the shifted-lognormal volatility, $v = \sigma^2(T-t)$ its associated variance, $\Phi(x)$ is the standard normal cumulated distribution function, $\omega=\pm 1$ for Caplets/Floorlets, respectively, and 
\begin{equation}
V(t;T,\bar{F},\bar{K},v,\omega) = \omega\left[\bar{F}(t) \Phi(\omega d_{+}) - \bar{K} \Phi(\omega d_{-})\right],\quad 
d_\pm = \frac{\ln\left(\frac{\bar{F}(t)}{\bar{K}}\right)\pm \frac{1}{2}v}{\sqrt{v}}, \quad 
d_0 = d_- +2\sqrt{v}. 
\end{equation}
From eq. \eqref{eq:shiftedBlackVar} we observe that, for large variance $v$, some terms cancel each other and we are left with
\begin{align}
\textit{Var}(t;T,\bar{F},\bar{K},v,\omega) \underset{v\rightarrow \infty}{\approx} 
\bar{F}(t)^2e^v \Phi\left(\omega\frac{3}{2}\sqrt{v}\right),
\end{align}
which is always larger for Caplets ($\omega=1$) than for Floorlets ($\omega=-1$).
\par 
We show in tab. \ref{tab:BS} the Monte Carlo Caplet/Floorlet prices computed as described in sec. \ref{sec:MC} in the limit of shifted-Black case using the parameters listed in tab. \ref{tab:BlackParameters}, along with their analytical variances obtained from eq. \eqref{eq:shiftedBlackVar}.

\begin{table}[H]
\centering
\begin{tabular}{ccccccccc}
\toprule 
T (y) & $F_0$ & $\lambda$ & $\alpha$ & $\beta$ & $\rho$ & $\nu$ & $N_{MC}$ & $\Delta_{MC}$ (days)\\
\midrule
2 & 1 & 3\% & see tab. \ref{tab:BS} & 1 & 0 & 0 & $2^{20}$ & 0.5 \\
\bottomrule  
\end{tabular}
\caption{Shifted-SABR parameters corresponding to the limit case of shifted-Black, used to compute the MC prices reported in tab. \ref{tab:BS}. The parameter $\alpha$ corresponds to $\sigma$ in eq. \eqref{eq:shiftedBlackVar}, its values are reported directly in tab. \ref{tab:BS}.}
\label{tab:BlackParameters}
\end{table}

As discussed above, we observe that for lower values of $\alpha$ the OTM options always show lower MC errors and analytical variances. As the value of $\alpha$ increases, the Caplet variances increase more than the Floorlet variances, leading to a situation where the Floorlet MC error and variance are always lower with respect to the Caplet ones. We note that, in this limit case, the effect increases with option's time to maturity, since the variance depends on $\alpha^2(T-t)$.
\begin{table}[H]
\centering
\small
\begin{tabular}{cccccc}
\toprule
$\alpha$ & Strike (K) & MC Floorlet price & MC Caplet price & Floorlet variance & Caplet variance \\
\midrule
\multirow{7}{*}{0.1} & 0.7 & $0.00030 \pm 0.00001$ & $0.30026 \pm 0.00043$ & 0.00002 & 0.02123\\
& 0.8  & $0.00361 \pm 0.00005$ & $0.20357 \pm 0.00041$ & 0.00029 & 0.01967 \\
& 0.9  & $0.01904 \pm 0.00013$ & $0.11900 \pm 0.00036$ & 0.00190 & 0.01501 \\
& 1.0  & $0.05806 \pm 0.00023$ & $0.05802 \pm 0.00027$ & 0.00614 & 0.00855 \\
& 1.1  & $0.12344 \pm 0.00032$ & $0.02340 \pm 0.00018$ & 0.01202 & 0.00362 \\
& 1.2  & $0.20796 \pm 0.00038$ & $0.00792 \pm 0.00010$ & 0.01693 & 0.00120 \\
& 1.3  & $0.30235 \pm 0.00041$ & $0.00231 \pm 0.00005$ & 0.01970 & 0.00033 \\
\hline
\multirow{7}{*}{0.3} & 0.7  & $0.04281 \pm 0.00026$ & $0.34263 \pm 0.00121$ & 0.00811 & 0.17180 \\
& 0.8  & $0.07566 \pm 0.00037$ & $0.27548 \pm 0.00114$ & 0.01588 & 0.15165 \\
& 0.9  & $0.11929 \pm 0.00048$ & $0.21912 \pm 0.00106$ & 0.02668 & 0.13021 \\
& 1.0  & $0.17298 \pm 0.00059$ & $0.17280 \pm 0.00097$ & 0.04005 & 0.10929 \\
& 1.1  & $0.23559 \pm 0.00069$ & $0.13541 \pm 0.00088$ & 0.05518 & 0.09009 \\
& 1.2  & $0.30581 \pm 0.00078$ & $0.10563 \pm 0.00079$ & 0.07119 & 0.07324 \\
& 1.3  & $0.38232 \pm 0.00087$ & $0.08214 \pm 0.00071$ & 0.08727 & 0.05893 \\
\hline
\multirow{7}{*}{0.5} & 0.7  & $0.11993 \pm 0.00050$ & $0.41948 \pm 0.00218$ & 0.02901 & 0.55848 \\
& 0.8  & $0.16827 \pm 0.00061$ & $0.36783 \pm 0.00211$ & 0.04376 & 0.52047 \\
& 0.9  & $0.22340 \pm 0.00073$ & $0.32295 \pm 0.00203$ & 0.06122 & 0.48240 \\
& 1.0  & $0.28448 \pm 0.00083$ & $0.28403 \pm 0.00195$ & 0.08089 & 0.44533 \\
& 1.1  & $0.35074 \pm 0.00094$ & $0.25029 \pm 0.00187$ & 0.10225 & 0.40992 \\
& 1.2  & $0.42144 \pm 0.00104$ & $0.22100 \pm 0.00179$ & 0.12480 & 0.37658 \\
& 1.3  & $0.49598 \pm 0.00113$ & $0.19553 \pm 0.00172$ & 0.14810 & 0.34550 \\
\bottomrule
\end{tabular}
\caption{Monte Carlo Caplet/Floorlet prices (undiscounted and without the year fraction in eq. \eqref{eq:cfPrice}) obtained using the parameters reported in tab. \ref{tab:BlackParameters}. The Monte Carlo errors are computed as three standard deviations. The last two columns show the analytical variances computed using eq. \eqref{eq:shiftedBlackVar}, which (dividing by the number of MC scenarios in tab. \ref{tab:BlackParameters} and taking the square root) agree to high accuracy with the estimated MC errors in cols. Caplet and Floorlet price.}
\label{tab:BS}
\end{table}

\section{DNN Details} 
\label{app:DNNdetails}
\begin{figure}[H]
\begin{subfigure}{.5\textwidth}
  \centering
  \centerline{\includegraphics[scale=0.51]{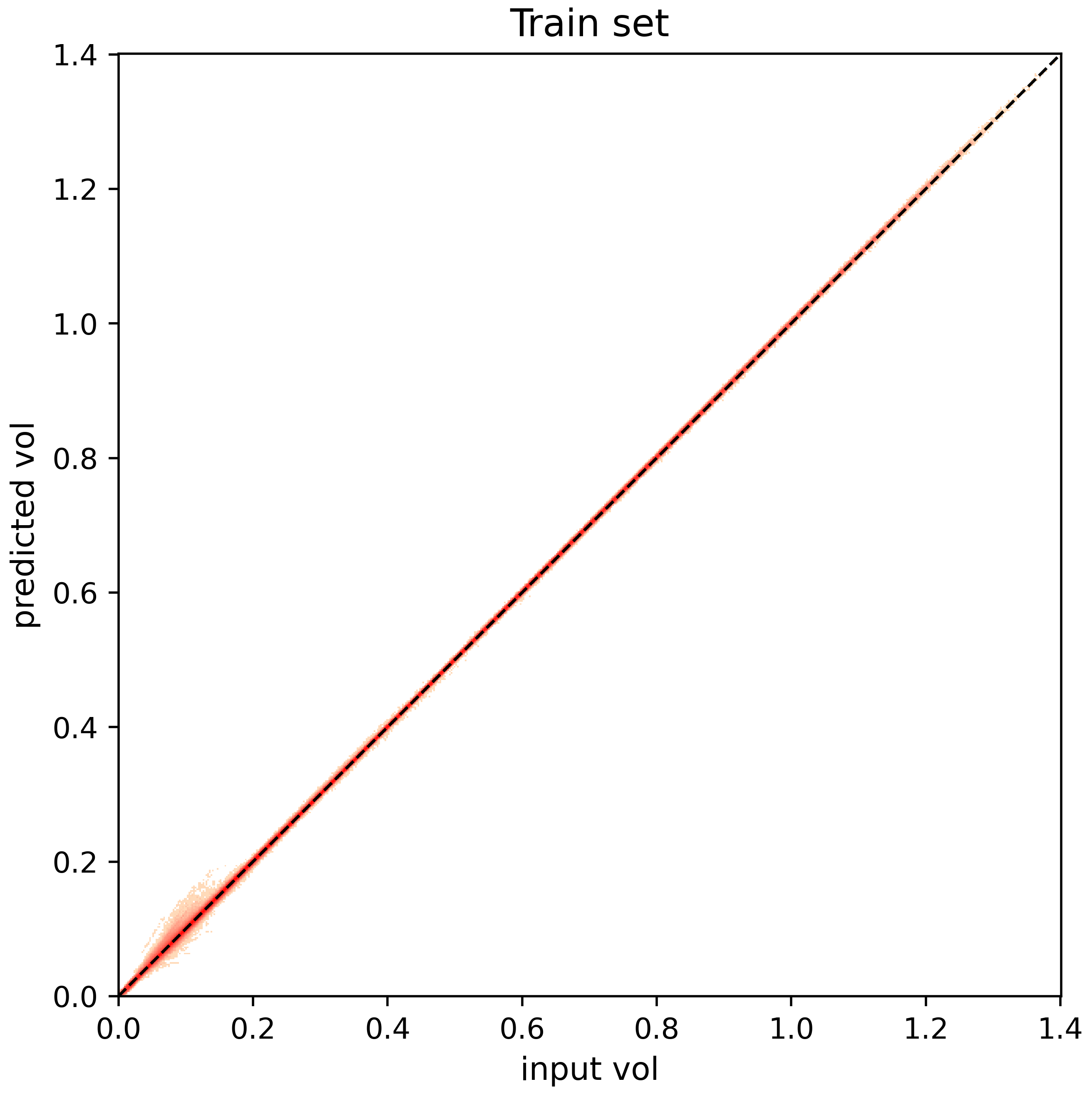}}
\end{subfigure}
\hfill
\begin{subfigure}{.5\textwidth}
  \centering
  \centerline{\includegraphics[scale=0.51]{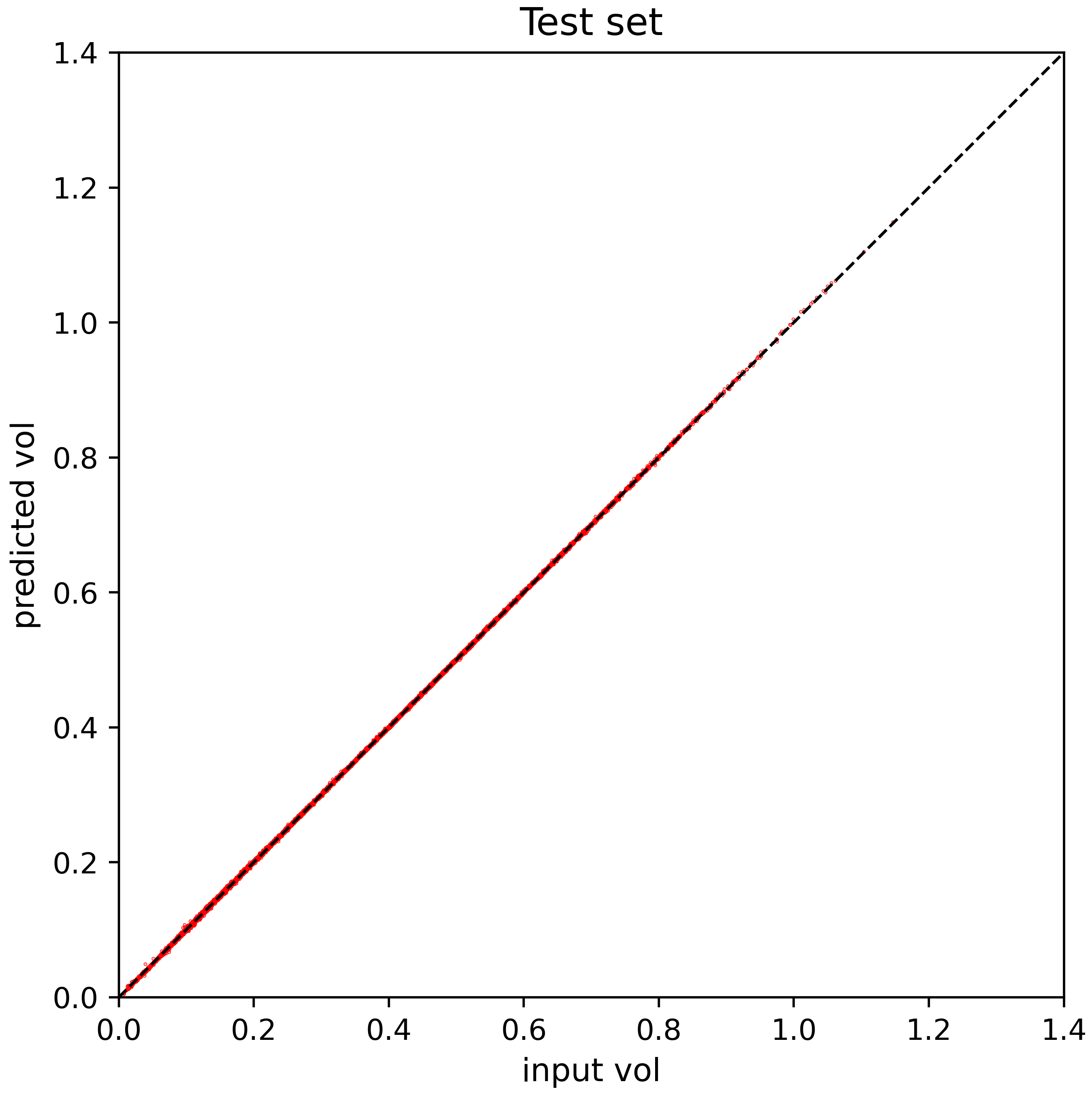}}
\end{subfigure}
\caption{Scatter plots for the training and test sets of the long term DNN.}
\label{fig:DNN3scatter}
\end{figure}

In this appendix we report more details about the three DNNs discussed in sec. \ref{sec:DNNSetup}. The following figs. \ref{fig:DNN1scatter}, \ref{fig:DNN2scatter}, and \ref{fig:DNN3scatter} show the scatter plots for training and test sets of the three DNNs.
The graphs below confirm the findings discussed in sec. \ref{sec:DNNSetup}. In particular, fig. \ref{fig:DNN1scatter} shows that, for the training set, there is a larger and more concentrated region where the discrepancies between the DNN-implied volatilities and those in the training data are more pronounced. By contrast, in figs. \ref{fig:DNN2scatter} and \ref{fig:DNN3scatter} this region is noticeably smaller. These results support the conclusions of sec. \ref{sec:DNNSetup}, namely that the most significant deviations were associated with noisier data points, which predominantly correspond to short maturities and extreme strikes.

\begin{figure}[H]
\begin{subfigure}{.5\textwidth}
  \centering
  \centerline{\includegraphics[scale=0.51]{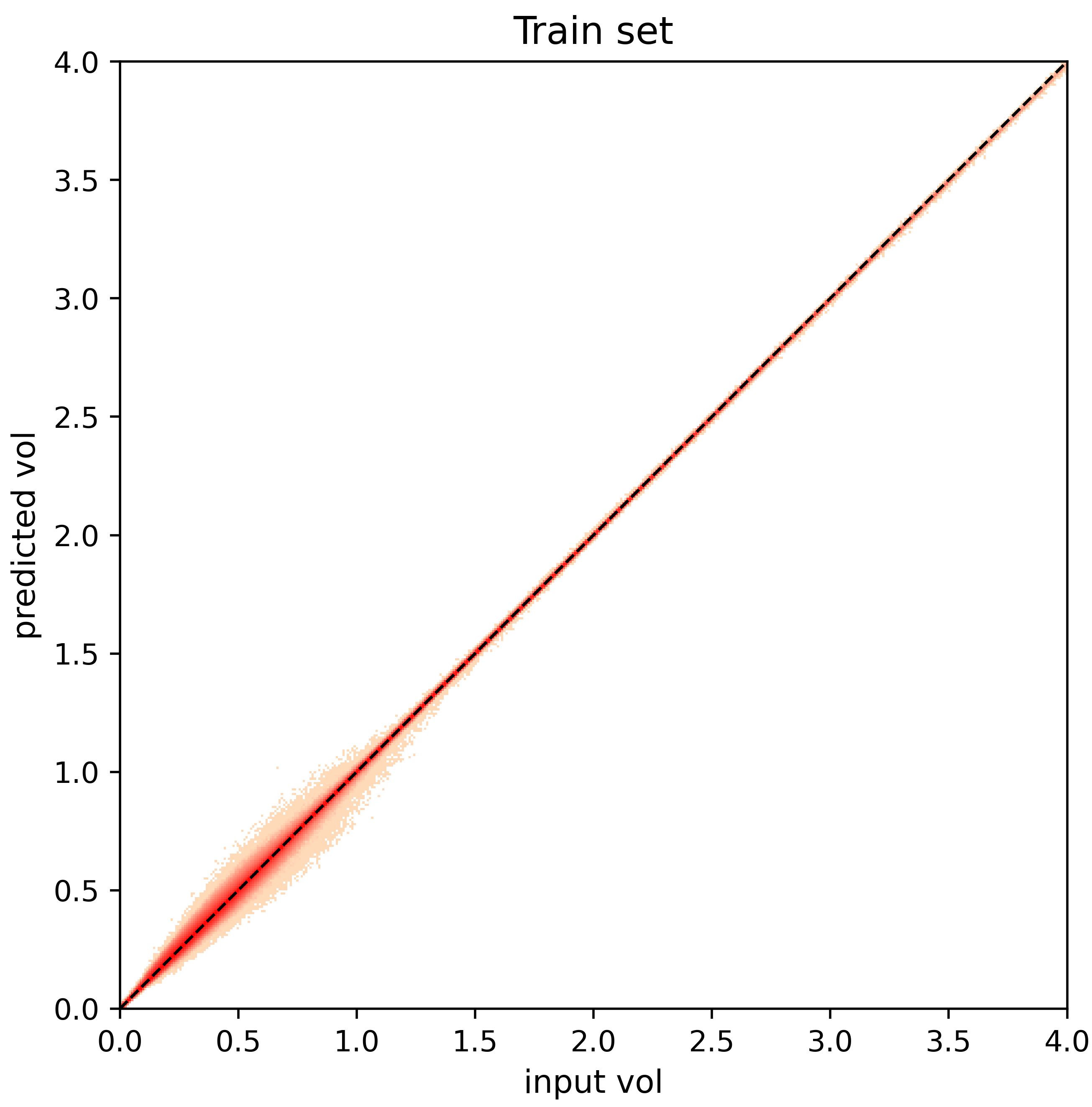}}
\end{subfigure}
\hfill
\begin{subfigure}{.5\textwidth}
  \centering
  \centerline{\includegraphics[scale=0.51]{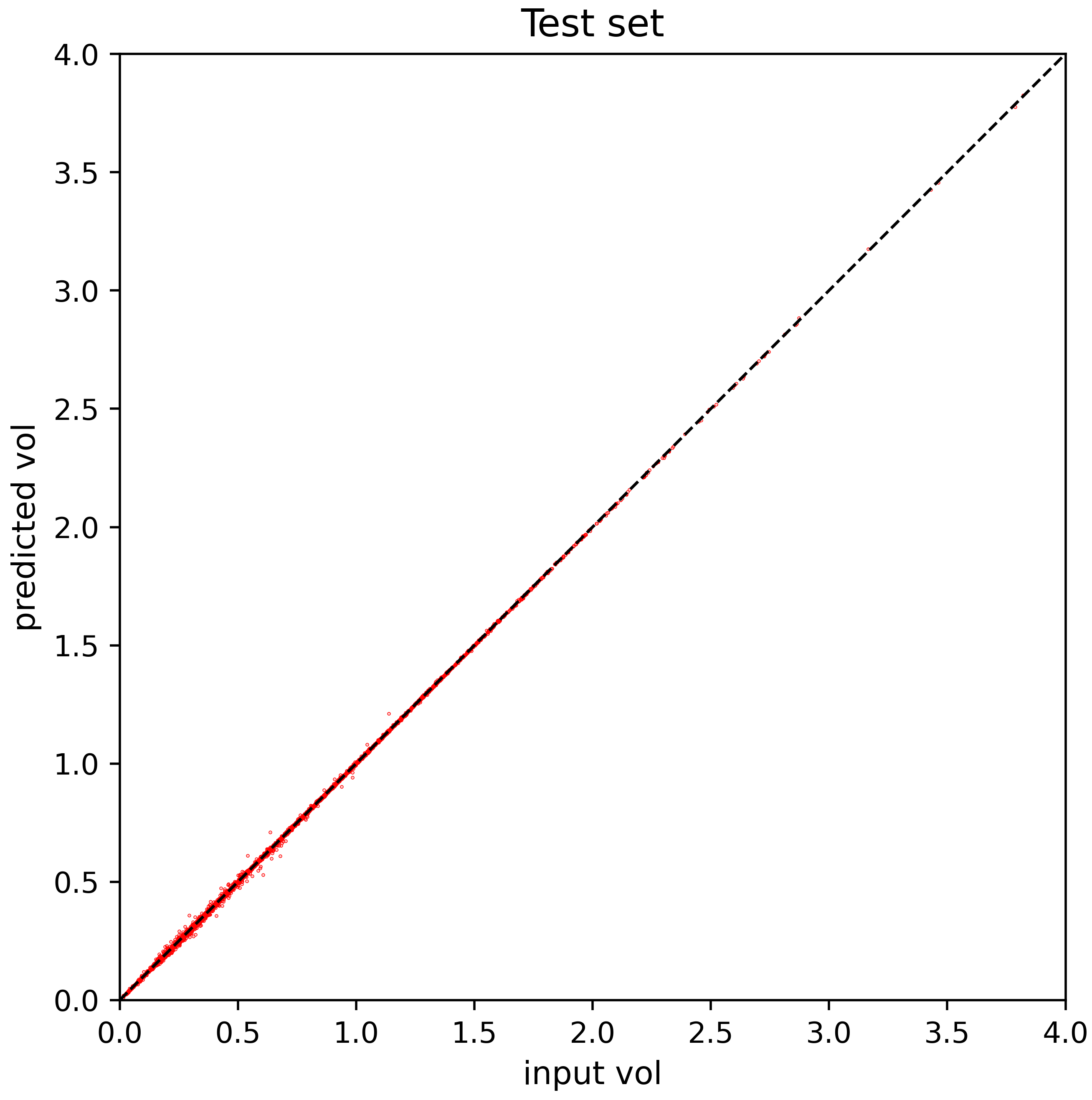}}
\end{subfigure}
\caption{Scatter plots for the training and test sets of the short term DNN.}
\label{fig:DNN1scatter}
\end{figure}

\begin{figure}[H]
\begin{subfigure}{.5\textwidth}
  \centering
  \centerline{\includegraphics[scale=0.51]{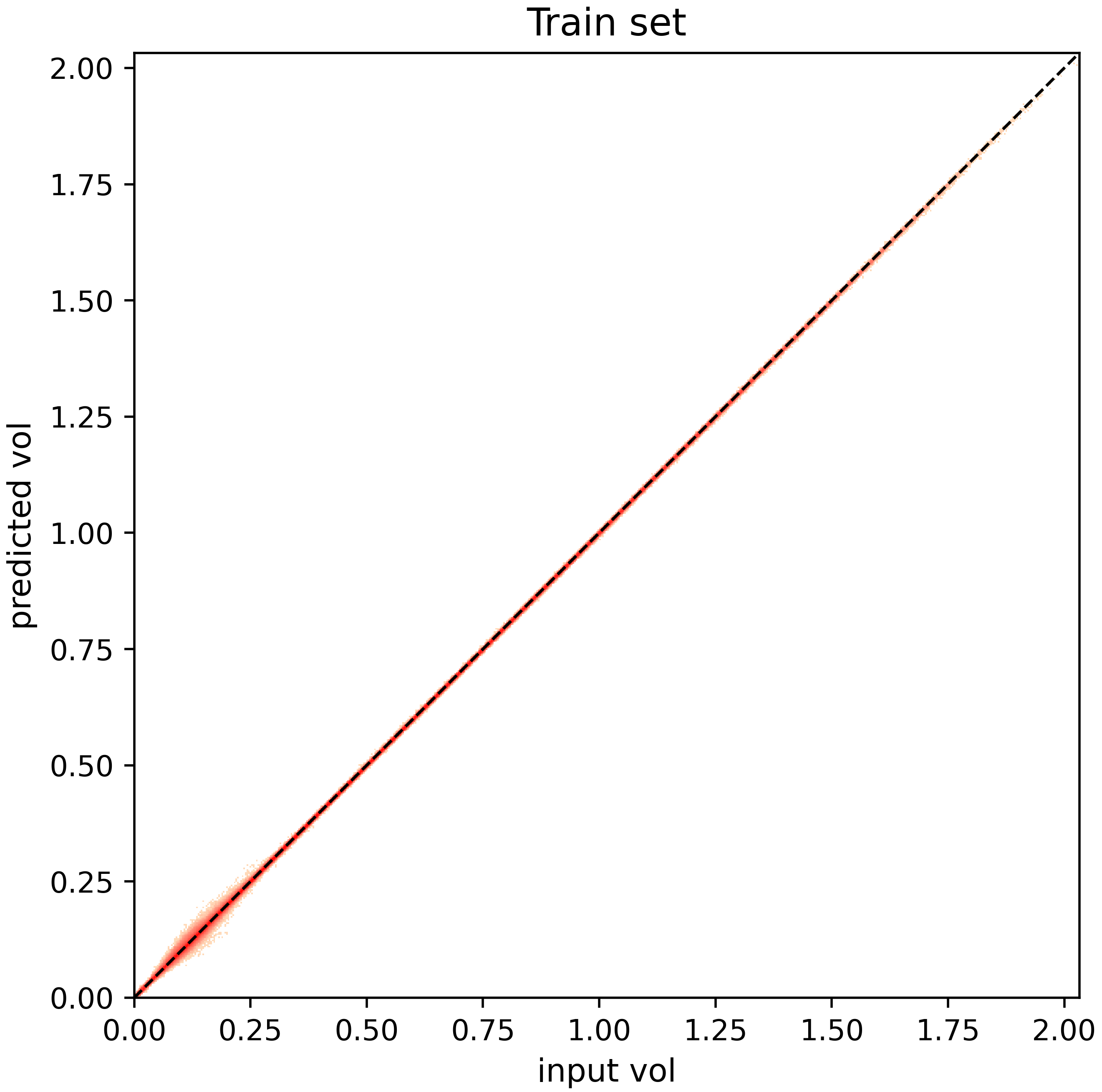}}
\end{subfigure}
\hfill
\begin{subfigure}{.5\textwidth}
  \centering
  \centerline{\includegraphics[scale=0.51]{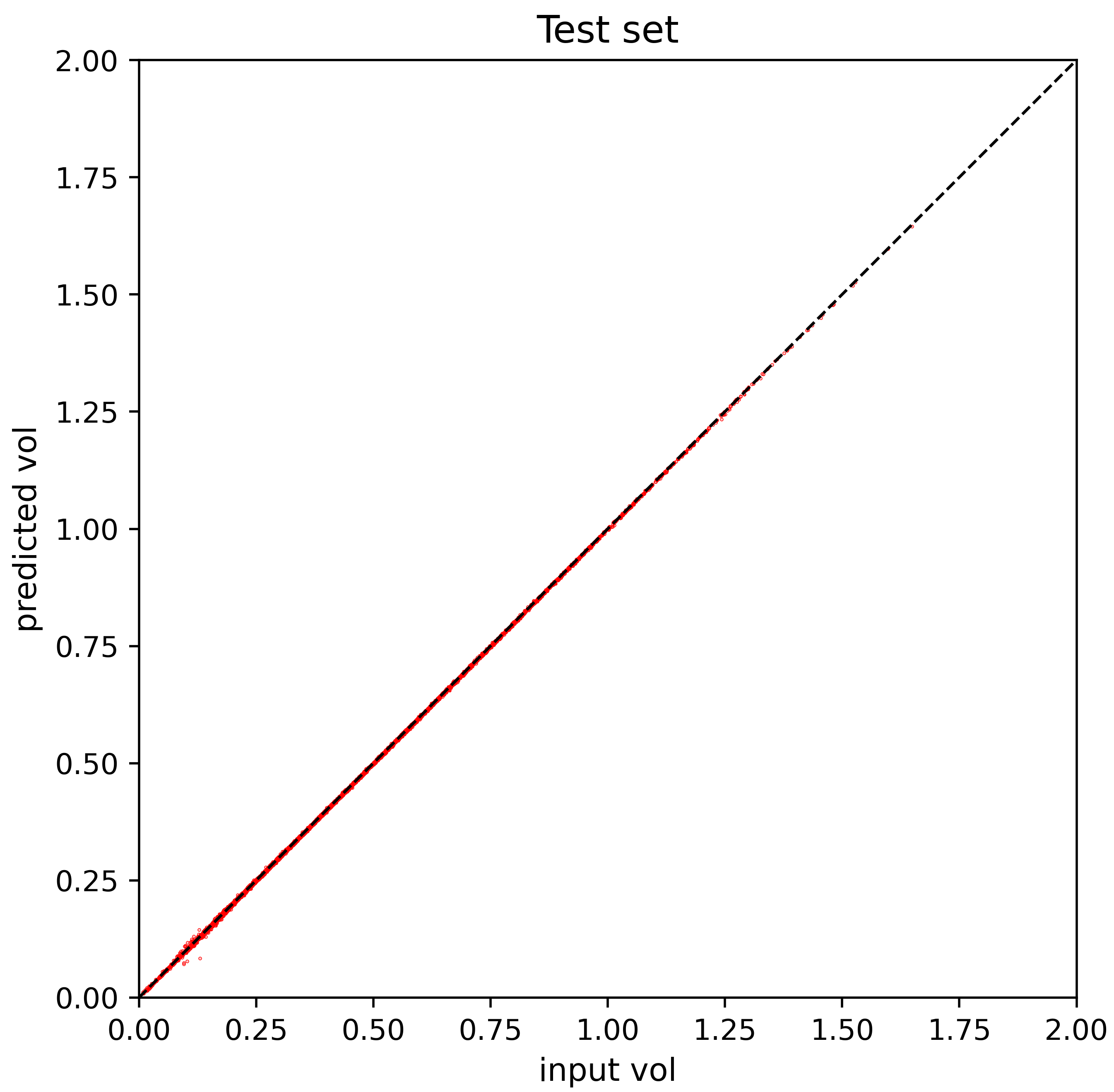}}
\end{subfigure}
\caption{Scatter plots for the training and test sets of the medium term DNN.}
\label{fig:DNN2scatter}
\end{figure}

\section{Calibration Details}
\label{app:CalibrationDetails}
In this appendix we report additional details on smile calibration discussed in sec. \ref{sec:CalibrationResults}.
We show in tab. \ref{tab:params_calibrated} the shifted-SABR parameters calibrated using both our DNNs and the \textquote{classic} Hagan et al. approximation and for the three market smiles displayed in fig. \ref{fig:smiles}. 
\par 
Furthermore, we show in tab. \ref{tab:MCvols} the numerical results associated with these three smiles. The implied volatility were calculated as discussed in sec. \ref{sec:LearningSABRwithDNN}. In particular, the MC volatility error is defined as the ratio between the MC pricing error (three standard deviations) and the shifted–Black vega sensitivity.
\par
\begin{figure}[H]
\begin{subfigure}{.5\textwidth}
    \centering
    \centerline{\includegraphics[scale=0.45]{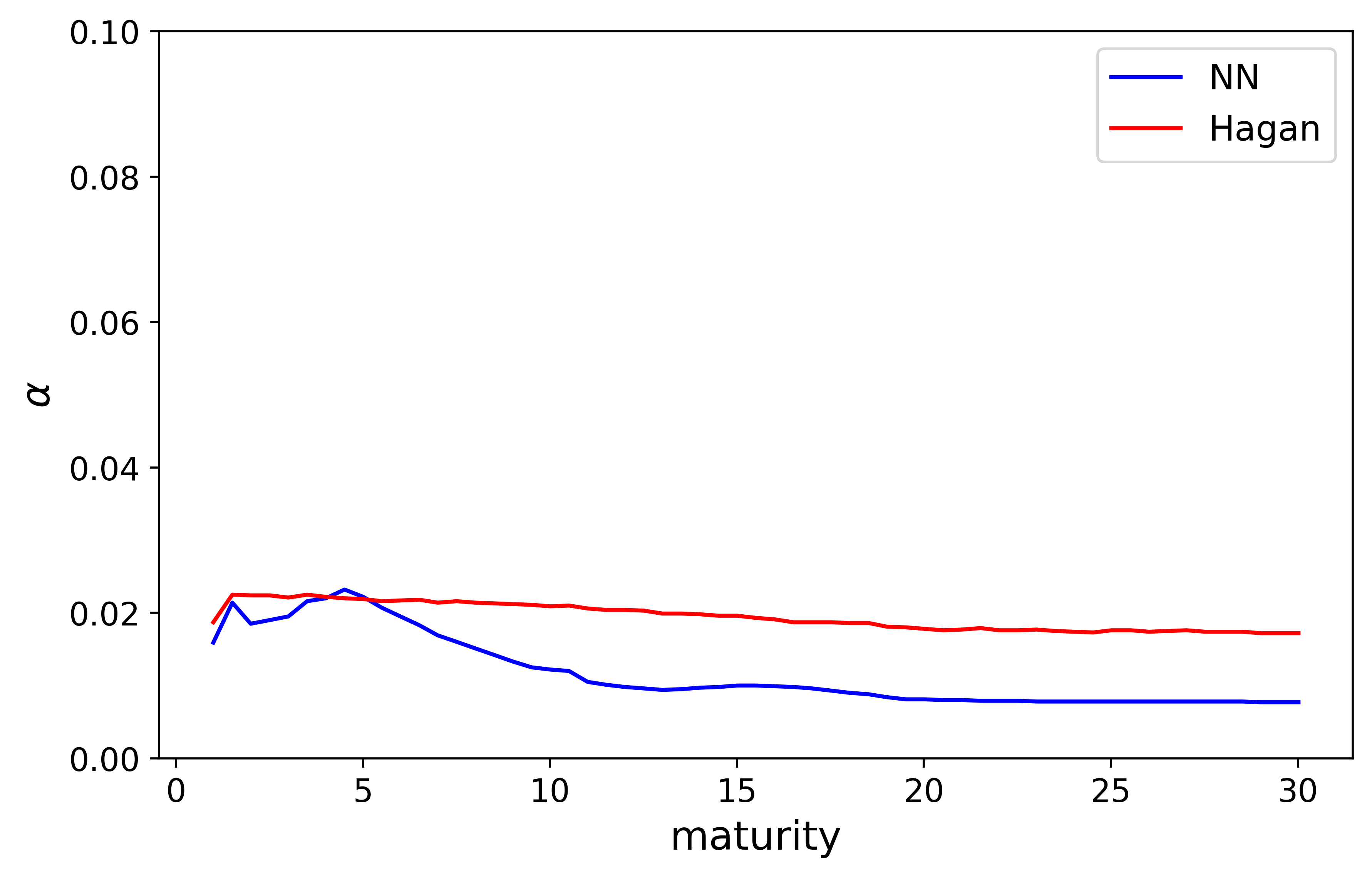}}
    \caption{$\alpha$ parameter}
\end{subfigure}
\hfill
\begin{subfigure}{.5\textwidth}
    \centering
    \centerline{\includegraphics[scale=0.45]{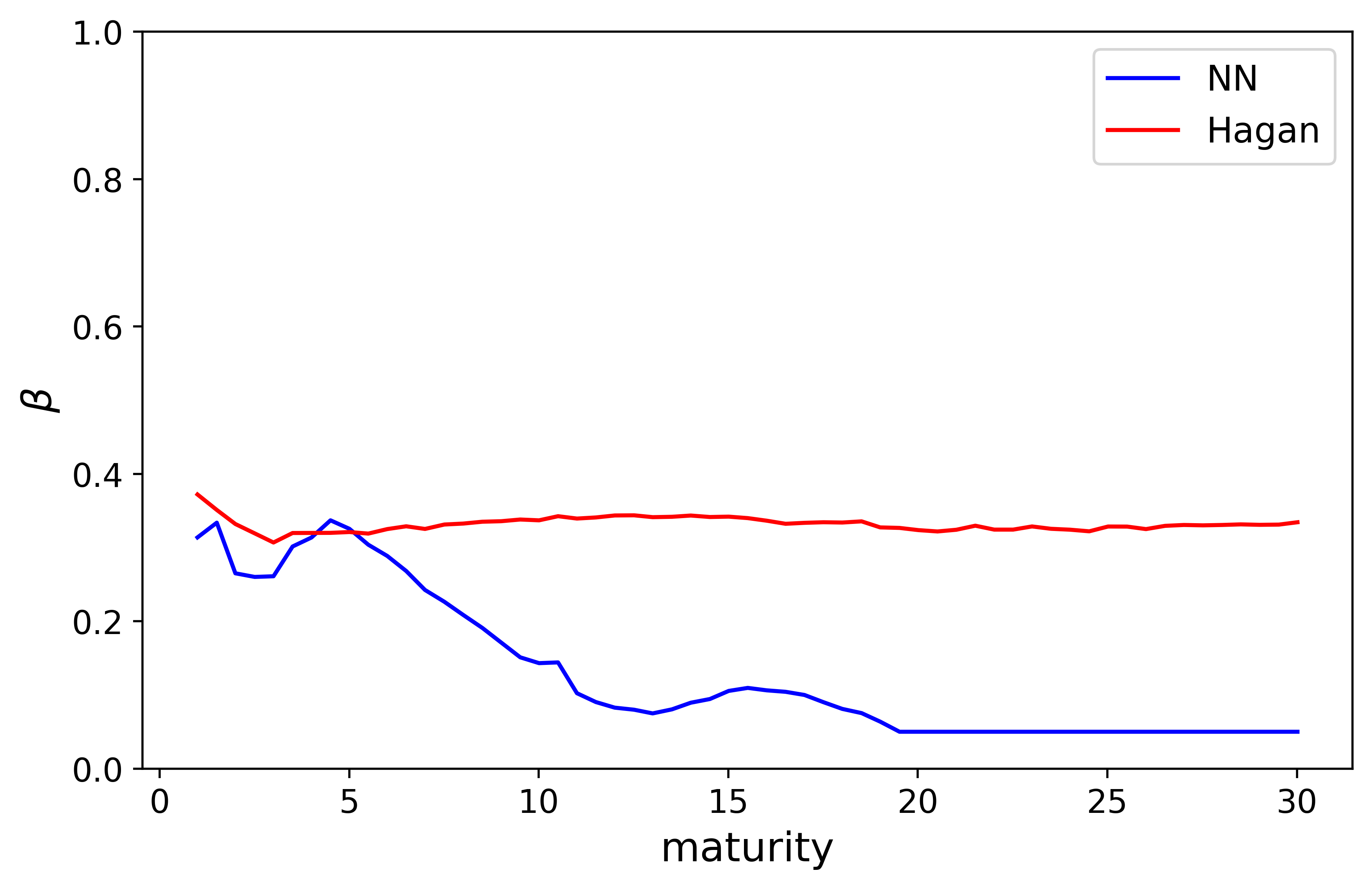}}
    \caption{$\beta$ parameter}
\end{subfigure}
\hfill
\begin{subfigure}{.5\textwidth}
    \centering
    \centerline{\includegraphics[scale=0.45]{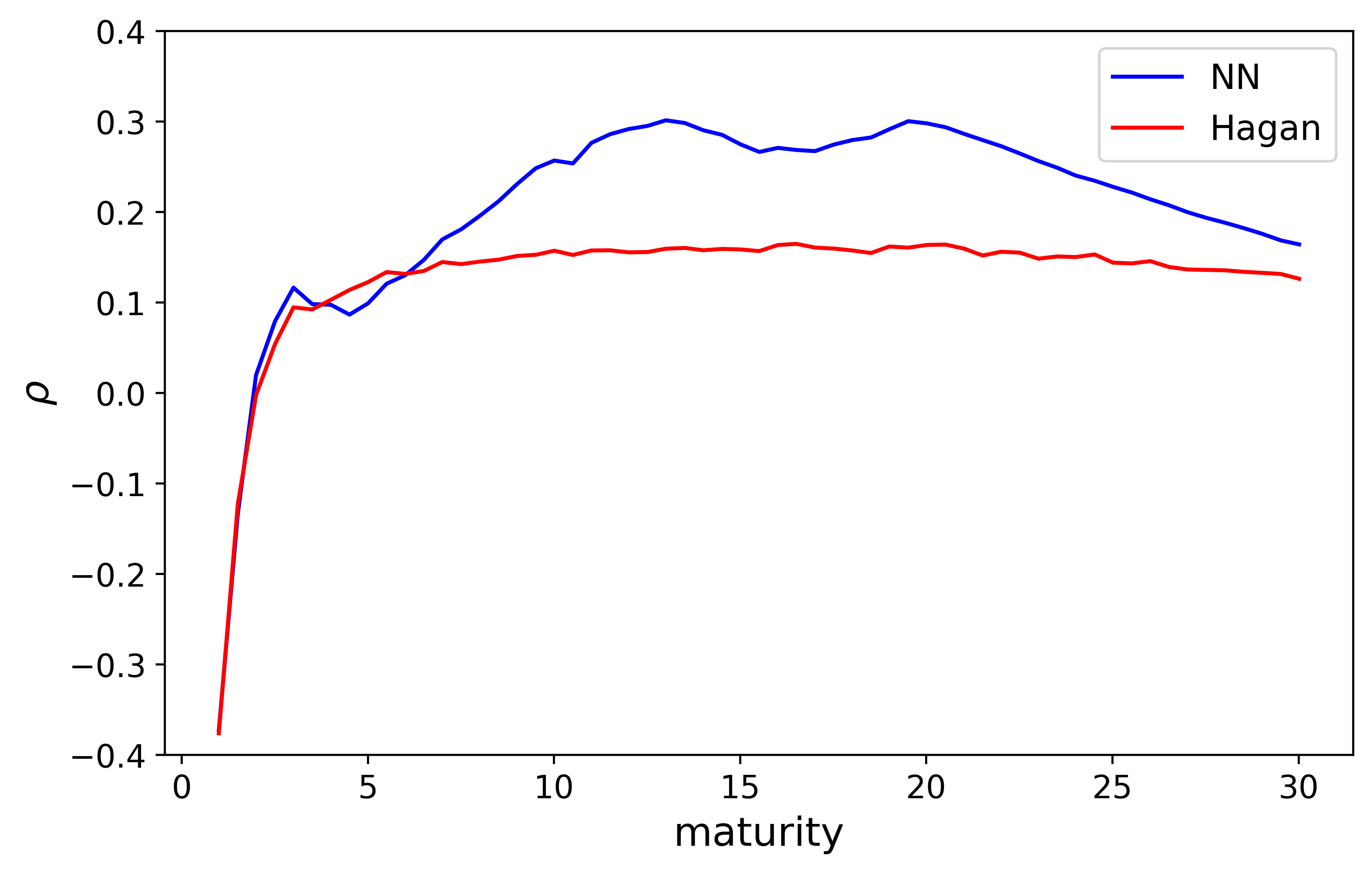}}
    \caption{$\rho$ parameter}
\end{subfigure}
\hfill 
\begin{subfigure}{.5\textwidth}
    \centering
    \centerline{\includegraphics[scale=0.45]{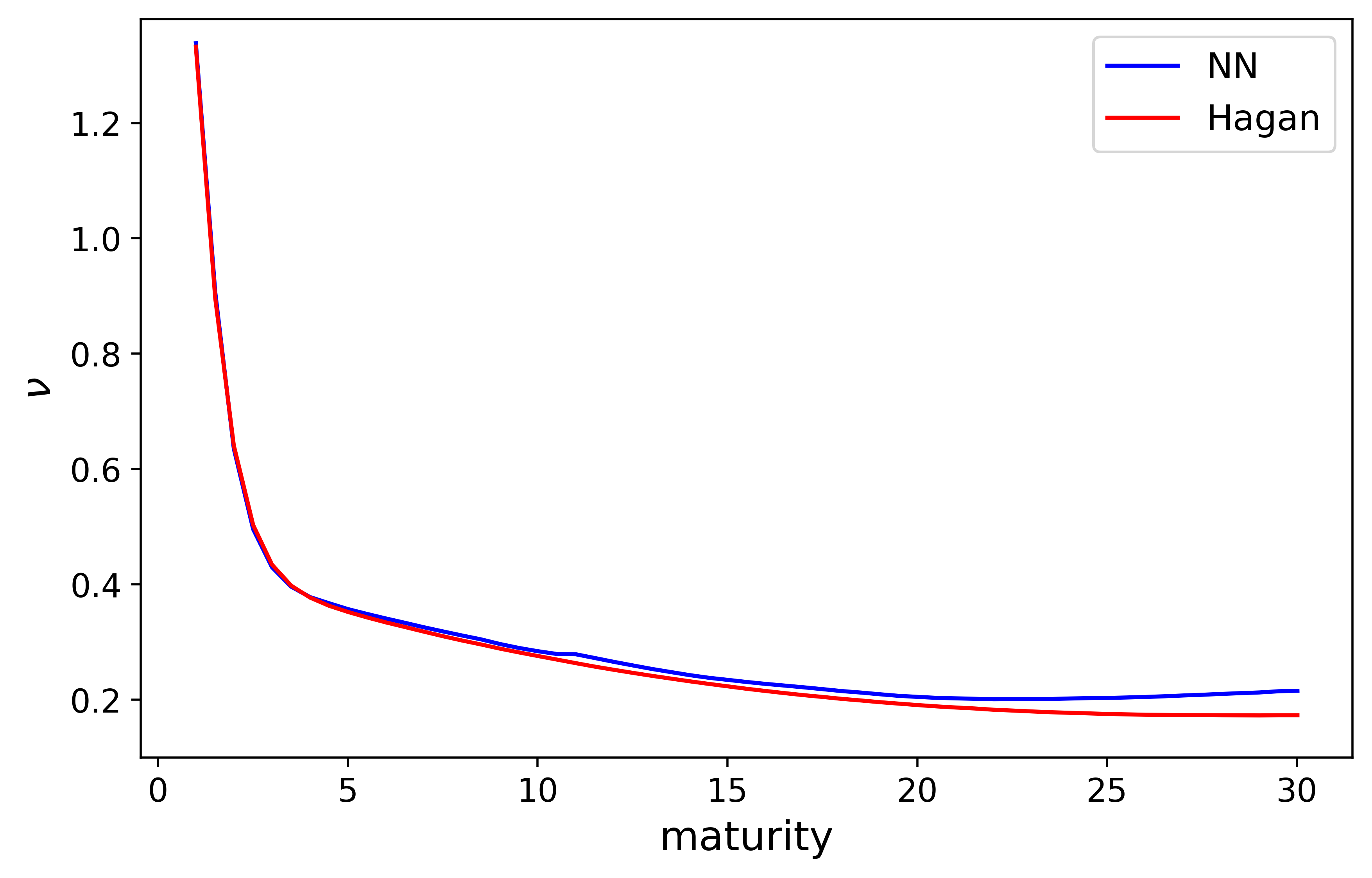}}
    \caption{$\nu$ parameter}
\end{subfigure}
\caption{Term structures of the four SABR parameters calibrated according to the two methodologies.}
\label{fig:params_calibrated}
\end{figure}

 
Finally, we show in fig. \ref{fig:params_calibrated} the term structure of the calibrated values of the shifted-SABR model parameters. As discussed in sec. \ref{sec:CalibrationResults}, to ensure the robustness of the calibration, we repeated the calibration procedure using many random initializations of the SABR parameters and verified that the resulting term structure is stable. 
We observe relatively smooth term structures for all the parameters, with no significant discontinuities. This feature is desirable to deal smoothly with Caplets/Floorlets insisting on adjacent forwards. 
Interestingly, the term structures obtained by the DNN calibration display richer shapes than those associated with the Hagan et al. approximation, reflecting the ability of the DNN to learn the more complex \textquote{exact} SABR implied volatility function.
\begin{table}[H]
\centering
\begin{tabular}{cccccccc}
\toprule  
Maturity (y) & Methodology & $F_0$ & $\alpha$ & $\beta$ & $\rho$ & $\nu$ & $\lambda$ \\
\midrule
\multirow{2}{*}{1.5}    & SABR Hagan & 2.28\% & 0.0225 & 0.3510 & -0.1232 & 0.8969 & 3\% \\
                        & SABR DNN   & 2.28\% & 0.0214 & 0.3337 & -0.1339 & 0.9074 & 3\% \\
\midrule
\multirow{2}{*}{10}     & SABR Hagan & 2.66\% & 0.0209 & 0.3369 &  0.1572 & 0.2758 & 3\% \\
                        & SABR DNN   & 2.66\% & 0.0122 & 0.1431 &  0.2569 & 0.2841 & 3\% \\
\midrule
\multirow{2}{*}{30}     & SABR Hagan & 1.56\% & 0.0172 & 0.3343 &  0.1262 & 0.1730 & 3\% \\
                        & SABR DNN   & 1.56\% & 0.0077 & 0.0500 &  0.1642 & 0.2155 & 3\% \\
\bottomrule 
\end{tabular}
\caption{Calibrated SABR parameters $\theta^\textit{SABR}=\{\alpha,\beta,\rho,\nu\}$ using both methodologies for the three market smiles shown in fig. \ref{fig:smiles}. The table also reports the (fixed) values for the forward rate $F_0$ and the shift $\lambda$.}
\label{tab:params_calibrated}
\end{table}

\begin{table}[H]
\centering
\begin{tabular}{ccccccc}
\toprule 
Maturity (y) & Strike (K) & DNN & MC\_DNN & Market & MC\_Hagan & Hagan \\
\midrule
\multirow{12}{*}{1.5} & -0.015& 50.78\%  & 50.80\% $\pm$ 0.09\% & 50.71\%  & 50.12\% $\pm$ 0.09\%  & 51.12\% \\
 & -0.01 & 42.64\%  & 42.63\% $\pm$ 0.07\% & 42.63\%  & 42.09\% $\pm$ 0.07\%  & 42.64\% \\
 & 0.00  & 30.85\%  & 30.93\% $\pm$ 0.04\% & 30.87\%  & 30.58\% $\pm$ 0.04\%  & 30.76\% \\
 & 0.005 & 26.30\%  & 26.43\% $\pm$ 0.03\% & 26.33\%  & 26.16\% $\pm$ 0.03\%  & 26.25\% \\
 & 0.01  & 22.44\%  & 22.57\% $\pm$ 0.03\% & 22.43\%  & 22.36\% $\pm$ 0.02\%  & 22.41\% \\
 & 0.015 & 19.21\%  & 19.33\% $\pm$ 0.02\% & 19.20\%  & 19.19\% $\pm$ 0.02\%  & 19.22\% \\
 & 0.02  & 16.85\%  & 16.92\% $\pm$ 0.02\% & 16.86\%  & 16.86\% $\pm$ 0.02\%  & 16.90\% \\
 & 0.03  & 16.15\%  & 16.07\% $\pm$ 0.01\% & 16.14\%  & 16.08\% $\pm$ 0.01\%  & 16.11\% \\
 & 0.04  & 18.27\%  & 18.23\% $\pm$ 0.01\% & 18.29\%  & 18.24\% $\pm$ 0.01\%  & 18.25\% \\
 & 0.05  & 20.62\%  & 20.60\% $\pm$ 0.01\% & 20.64\%  & 20.61\% $\pm$ 0.01\%  & 20.63\% \\
 & 0.06  & 22.71\%  & 22.73\% $\pm$ 0.02\% & 22.72\%  & 22.73\% $\pm$ 0.02\%  & 22.77\% \\
 & 0.07  & 24.53\%  & 24.58\% $\pm$ 0.03\% & 24.51\%  & 24.57\% $\pm$ 0.03\%  & 24.65\% \\
 & 0.10  & 28.89\%  & 28.92\% $\pm$ 0.06\% & 28.66\%  & 28.90\% $\pm$ 0.06\%  & 29.07\% \\
\midrule
\multirow{12}{*}{10} & -0.015& 27.68\% & 27.72\% $\pm$ 0.02\% & 27.67\%  & 26.46\% $\pm$ 0.02\%  & 27.78\% \\
 & -0.01 & 24.29\%  & 24.41\% $\pm$ 0.02\% & 24.27\%  & 23.42\% $\pm$ 0.02\%  & 24.22\% \\
 & 0.0   & 19.73\%  & 19.93\% $\pm$ 0.02\% & 19.76\%  & 19.30\% $\pm$ 0.02\%  & 19.70\% \\
 & 0.005 & 18.20\%  & 18.39\% $\pm$ 0.01\% & 18.22\%  & 17.88\% $\pm$ 0.01\%  & 18.19\% \\
 & 0.01  & 17.02\%  & 17.19\% $\pm$ 0.01\% & 17.03\%  & 16.76\% $\pm$ 0.01\%  & 17.03\% \\
 & 0.015 & 16.16\%  & 16.29\% $\pm$ 0.01\% & 16.14\%  & 15.92\% $\pm$ 0.01\%  & 16.16\% \\
 & 0.02  & 15.54\%  & 15.64\% $\pm$ 0.01\% & 15.51\%  & 15.32\% $\pm$ 0.01\%  & 15.54\% \\
 & 0.03  & 14.85\%  & 14.98\% $\pm$ 0.01\% & 14.85\%  & 14.69\% $\pm$ 0.01\%  & 14.87\% \\
 & 0.04  & 14.72\%  & 14.84\% $\pm$ 0.01\% & 14.73\%  & 14.58\% $\pm$ 0.01\%  & 14.72\% \\
 & 0.05  & 14.85\%  & 14.96\% $\pm$ 0.02\% & 14.87\%  & 14.74\% $\pm$ 0.02\%  & 14.85\% \\
 & 0.06  & 15.11\%  & 15.20\% $\pm$ 0.02\% & 15.12\%  & 15.02\% $\pm$ 0.02\%  & 15.10\% \\
 & 0.07  & 15.41\%  & 15.47\% $\pm$ 0.01\% & 15.41\%  & 15.33\% $\pm$ 0.01\%  & 15.40\% \\
 & 0.10  & 16.28\%  & 16.29\% $\pm$ 0.01\% & 16.24\%  & 16.28\% $\pm$ 0.01\%  & 16.30\% \\
\midrule
\multirow{12}{*}{30} & -0.015& 22.80\%  & 22.88\% $\pm$ 0.01\% & 23.03\%  & 20.44\% $\pm$ 0.01\%  & 23.07\% \\
 & -0.01 & 20.41\%  & 20.55\% $\pm$ 0.01\% & 20.37\%  & 18.49\% $\pm$ 0.01\%  & 20.33\% \\
 & 0.00  & 17.19\%  & 17.39\% $\pm$ 0.01\% & 17.09\%  & 15.97\% $\pm$ 0.01\%  & 17.07\% \\
 & 0.005 & 16.10\%  & 16.32\% $\pm$ 0.01\% & 16.06\%  & 15.14\% $\pm$ 0.01\%  & 16.06\% \\
 & 0.01  & 15.31\%  & 15.49\% $\pm$ 0.01\% & 15.30\%  & 14.52\% $\pm$ 0.01\%  & 15.31\% \\
 & 0.015 & 14.75\%  & 14.86\% $\pm$ 0.01\% & 14.74\%  & 14.06\% $\pm$ 0.01\%  & 14.75\% \\
 & 0.02  & 14.31\%  & 14.41\% $\pm$ 0.01\% & 14.33\%  & 13.73\% $\pm$ 0.01\%  & 14.35\% \\
 & 0.03  & 13.79\%  & 13.86\% $\pm$ 0.01\% & 13.84\%  & 13.34\% $\pm$ 0.01\%  & 13.84\% \\
 & 0.04  & 13.60\%  & 13.61\% $\pm$ 0.01\% & 13.61\%  & 13.19\% $\pm$ 0.01\%  & 13.60\% \\
 & 0.05  & 13.53\%  & 13.53\% $\pm$ 0.01\% & 13.53\%  & 13.15\% $\pm$ 0.01\%  & 13.51\% \\
 & 0.06  & 13.51\%  & 13.54\% $\pm$ 0.02\% & 13.51\%  & 13.19\% $\pm$ 0.02\%  & 13.50\% \\
 & 0.07  & 13.54\%  & 13.59\% $\pm$ 0.01\% & 13.54\%  & 13.26\% $\pm$ 0.02\%  & 13.53\% \\
 & 0.10  & 13.73\%  & 13.81\% $\pm$ 0.01\% & 13.69\%  & 13.52\% $\pm$ 0.01\%  & 13.72\% \\
\bottomrule 
\end{tabular}
\caption{Shifted-lognormal implied volatility smiles for the maturities displayed in fig. \ref{fig:smiles}.}
\label{tab:MCvols}
\end{table}

\end{appendices}

\end{document}